\pdfoutput=1
\newif\ifplainstyle
\plainstyletrue
\plainstylefalse

\newif\ifjhepstyle
\jhepstyletrue

\newif\ifprstyle
\prstyletrue
\prstylefalse

\ifprstyle
	\documentclass[twocolumn,nofootinbib]{revtex4-1}
\else
	\documentclass[11pt,a4paper]{article}
\fi

\ifjhepstyle
	\usepackage{jheppub}
	\usepackage{amsfonts}
	\usepackage{verbatim}
	\usepackage{float}
	\usepackage{color}
	\usepackage{array}
	\newcolumntype{C}[1]{>{\centering\arraybackslash$}p{#1}<{$}}
	\makeatletter
	\def\@fpheader{\phantom{:-)}}
	\makeatother
\else	
 	\ifprstyle
		\usepackage{verbatim}
		\usepackage{amsmath,amsfonts,amssymb}
		\usepackage[colorlinks=true
                	,urlcolor=blue
                	,anchorcolor=blue
                	,citecolor=blue
                	,filecolor=blue
                	,linkcolor=blue
                	,menucolor=blue
                	]{hyperref}
	\else
            	\usepackage{verbatim}
            	\usepackage{cite}
            	\usepackage{setspace}
            	\usepackage[top=2.5cm, bottom=2.75cm, left=2.5cm, right=2.5cm]{geometry}
            	\usepackage{amsmath,amsfonts,amssymb}
            	\usepackage[colorlinks=true
            	,urlcolor=blue
            	,anchorcolor=blue
            	,citecolor=blue
            	,filecolor=blue
            	,linkcolor=blue
            	,menucolor=blue
            	,linktoc=page
            	]{hyperref}
            	\usepackage{float}
            	\restylefloat{table}
            	\renewcommand{\arraystretch}{1.5}
            	\numberwithin{equation}{section}
	\fi
\fi

\usepackage{subfig}

\usepackage{etoolbox}

\makeatletter
\def\hlinewd#1{%
\noalign{\ifnum0=`}\fi\hrule \@height #1 %
\futurelet\reserved@a\@xhline}
\makeatother

\newcolumntype{?}[1]{!{\vrule width #1}}

\allowdisplaybreaks

\usepackage{multirow}

\usepackage[normalem]{ulem}
\usepackage{mathtools}
\usepackage{bbold}
\usepackage{transparent}
\usepackage{bm}
\usepackage{enumitem}

\usepackage[textsize=scriptsize,textwidth=2.5cm]{todonotes}

\makeatletter
\let\save@mathaccent\mathaccent
\newcommand*\if@single[3]{%
  \setbox0\hbox{${\mathaccent"0362{#1}}^H$}%
  \setbox2\hbox{${\mathaccent"0362{\kern0pt#1}}^H$}%
  \ifdim\ht0=\ht2 #3\else #2\fi
  }
\newcommand*\rel@kern[1]{\kern#1\dimexpr\macc@kerna}
\newcommand*\widebar[1]{\@ifnextchar^{{\wide@bar{#1}{0}}}{\wide@bar{#1}{1}}}
\newcommand*\wide@bar[2]{\if@single{#1}{\wide@bar@{#1}{#2}{1}}{\wide@bar@{#1}{#2}{2}}}
\newcommand*\wide@bar@[3]{%
  \begingroup
  \def\mathaccent##1##2{%
    \let\mathaccent\save@mathaccent
    \if#32 \let\macc@nucleus\first@char \fi
    \setbox\z@\hbox{$\macc@style{\macc@nucleus}_{}$}%
    \setbox\tw@\hbox{$\macc@style{\macc@nucleus}{}_{}$}%
    \dimen@\wd\tw@
    \advance\dimen@-\wd\z@
    \divide\dimen@ 3
    \@tempdima\wd\tw@
    \advance\@tempdima-\scriptspace
    \divide\@tempdima 10
    \advance\dimen@-\@tempdima
    \ifdim\dimen@>\z@ \dimen@0pt\fi
    \rel@kern{0.6}\kern-\dimen@
    \if#31
      \overline{\rel@kern{-0.6}\kern\dimen@\macc@nucleus\rel@kern{0.4}\kern\dimen@}%
      \advance\dimen@0.4\dimexpr\macc@kerna
      \let\final@kern#2%
      \ifdim\dimen@<\z@ \let\final@kern1\fi
      \if\final@kern1 \kern-\dimen@\fi
    \else
      \overline{\rel@kern{-0.6}\kern\dimen@#1}%
    \fi
  }%
  \macc@depth\@ne
  \let\math@bgroup\@empty \let\math@egroup\macc@set@skewchar
  \mathsurround\z@ \frozen@everymath{\mathgroup\macc@group\relax}%
  \macc@set@skewchar\relax
  \let\mathaccentV\macc@nested@a
  \if#31
    \macc@nested@a\relax111{#1}%
  \else
    \def\gobble@till@marker##1\endmarker{}%
    \futurelet\first@char\gobble@till@marker#1\endmarker
    \ifcat\noexpand\first@char A\else
      \def\first@char{}%
    \fi
    \macc@nested@a\relax111{\first@char}%
  \fi
  \endgroup
}
\makeatother

\newcommand{\ThisIsTheTitle}{TsT, black holes, and $\bm{\mathsf{T\bar T+J\bar T+T\bar J}}$} 
\newcommand{\ThisIsAuthorOne}{Luis Apolo}
\newcommand{\ThisIsEmailOne}{l.a.apolo@uva.nl}
\newcommand{\ThisIsAuthorTwo}{and Wei Song}
\newcommand{\ThisIsEmailTwo}{wsong2014@mail.tsinghua.edu.cn}
\newcommand{\TheseAreTheKeywords}{}

\newcommand{\ThisIsTheAbstract}{
We generate a class of string backgrounds by a sequence of TsT transformations of the NS1-NS5 system that we argue are holographically dual to states in a single-trace $T\bar T+J\bar T+T\bar J$-deformed CFT$_2$. The new string backgrounds include general rotating black hole solutions with two $U(1)$ charges as well as a smooth solution without a horizon that is interpreted as the ground state. As evidence for the correspondence we ($i$) derive the long string spectrum and relate it to the spectrum of the dual field theory; ($ii$) show that the black hole thermodynamics can be reproduced from single-trace $T\bar T+J\bar T+T\bar J$-deformed CFTs;  and ($iii$) show that the energy of the ground state matches the energy of the vacuum in the dual theory. We also study geometric properties of these new spacetimes and find that for some choices of the parameters the three-dimensional Ricci scalar in the Einstein frame can become positive in a region outside the horizon before reaching closed timelike curves and singularities. 
}

\ifjhepstyle
\title{\ThisIsTheTitle}

\author[ a, b ]{\ThisIsAuthorOne}
\author[ c, d ]{\ThisIsAuthorTwo}

\affiliation[a]{Institute for Theoretical Physics, University of Amsterdam, 1090GL Amsterdam, The Netherlands}
\affiliation[b]{Beijing Institute of Mathematical Sciences and Applications, Beijing 101408, China}

\affiliation[c]{Department of Mathematical Sciences, Tsinghua University, Beijing 100084, China}

\affiliation[d]{Peng Huanwu Center for Fundamental Theory, Hefei, Anhui 230026, China}

\emailAdd{\ThisIsEmailOne}
\emailAdd{\ThisIsEmailTwo}

\abstract{\ThisIsTheAbstract} 

\fi

\begin{document}

\ifjhepstyle
\maketitle
\flushbottom
\fi

\long\def\symfootnote[#1]#2{\begingroup%
\def\thefootnote{\fnsymbol{footnote}}\footnote[#1]{#2}\endgroup} 

\def\rednote#1{{\color{red} #1}}
\def\bluenote#1{{\color{blue} #1}}
\def\magnote#1{{\color{magenta} #1}}
\def\rout#1{{\color{red} \sout{#1}}}

\newcommand{\wei}[2]{\textcolor{magenta}{#1}\todo[color=green]{\scriptsize{W: #2}}}
\newcommand{\lui}[2]{\textcolor{red}{#1}\todo[color=yellow]{\scriptsize{L: #2}}}

\def\({\left (}
\def\){\right )}
\def\lb{\left [}
\def\rb{\right ]}
\def\lB{\left \{}
\def\rB{\right \}}

\def\Int#1#2{\int \textrm{d}^{#1} x \sqrt{|#2|}}
\def\Bra#1{\left\langle#1\right|} 
\def\Ket#1{\left|#1\right\rangle}
\def\BraKet#1#2{\left\langle#1|#2\right\rangle} 
\def\Vev#1{\left\langle#1\right\rangle}
\def\Vevm#1{\left\langle \Phi |#1| \Phi \right\rangle}\def\bbox{\bar{\Box}}
\def\til#1{\tilde{#1}}
\def\wtil#1{\widetilde{#1}}
\def\ph#1{\phantom{#1}}

\def\ra{\rightarrow}
\def\la{\leftarrow}
\def\lra{\leftrightarrow}
\def\p{\partial}
\def\barp{\bar{\partial}}
\def\diff{\mathrm{d}}

\def\sinh{\mathrm{sinh}}
\def\cosh{\mathrm{cosh}}
\def\tanh{\mathrm{tanh}}
\def\coth{\mathrm{coth}}
\def\sech{\mathrm{sech}}
\def\csch{\mathrm{csch}}

\def\a{\alpha}
\def\b{\beta}
\def\g{\gamma}
\def\d{\delta}
\def\e{\epsilon}
\def\ve{\varepsilon}
\def\k{\kappa}
\def\l{\lambda}
\def\n{\nabla}
\def\om{\omega}
\def\s{\sigma}
\def\t{\theta}
\def\z{\zeta}
\def\vp{\varphi}

\def\ss{\Sigma}
\def\dd{\Delta}
\def\GG{\Gamma}
\def\LL{\Lambda}
\def\tt{\Theta}

\def\A{{\cal A}}
\def\B{{\cal B}}
\def\C{{\cal C}}
\def\cE{{\cal E}}
\def\D{{\cal D}}
\def\F{{\cal F}}
\def\H{{\cal H}}
\def\I{{\cal I}}
\def\J{{\cal J}}
\def\K{{\cal K}}
\def\L{{\cal L}}
\def\M{{\cal M}}
\def\N{{\cal N}}
\def\O{{\cal O}}
\def\Q{{\cal Q}}
\def\P{{\cal P}}
\def\cS{{\cal S}}
\def\T{{\cal T}}
\def\W{{\cal W}}
\def\X{{\cal X}}
\def\Z{{\cal Z}}

\def\mfa{\mathfrak{a}}
\def\mfb{\mathfrak{b}}
\def\mfc{\mathfrak{c}}
\def\mfd{\mathfrak{d}}

\def\we{\wedge}
\def\re{\textrm{Re}}

\def\tilw{\tilde{w}}
\def\tile{\tilde{e}}

\def\tilL{\tilde{L}}
\def\tilJ{\tilde{J}}

\def\zz{\bar z}
\def\xx{\bar x}
\def\yy{\bar y}
\def\xp{x^{+}}
\def\xm{x^{-}}

\def\bp{\bar{\p}}
\def\note#1{{\color{red}#1}}
\def\notebf#1{{\bf\color{red}#1}}

\def\VirU1{Vir \times U(1)}
\def\VirSL2R{\mathrm{Vir}\otimes\widehat{\mathrm{SL}}(2,\mathbb{R})}
\def\U1{U(1)}
\def\u1{U(1)}
\def\SL2R{\widehat{\mathrm{SL}}(2,\mathbb{R})}
\def\sl2r{\mathrm{SL}(2,\mathbb{R})}
\def\by{\mathrm{BY}}

\def\RR{\mathbb{R}}

\def\tr{\mathrm{Tr}}
\def\bnabla{\overline{\nabla}}

\def\sint{\int_{\ss}}
\def\dsint{\int_{\p\ss}}
\def\hint{\int_{H}}

\newcommand{\eq}[1]{\begin{align}#1\end{align}}
\newcommand{\eqst}[1]{\begin{align*}#1\end{align*}}
\newcommand{\eqsp}[1]{\begin{equation}\begin{split}#1\end{split}\end{equation}}

\newcommand{\absq}[1]{{\textstyle\sqrt{\left |#1\right |}}}



\newcommand{\pp}{{\partial_+}{}}
\newcommand{\pmm}{{\partial_-}{}}


\ifprstyle
\title{\ThisIsTheTitle}

\author{\ThisIsAuthorOne}
\email{\ThisIsEmailOne}

\author{\ThisIsAuthorTwo}
\email{\ThisIsEmailTwo}

\affiliation{\ThisIsTheAffiliation}


\begin{abstract}
\ThisIsTheAbstract
\end{abstract}


\maketitle

\fi

\ifplainstyle
\begin{titlepage}
\begin{center}

\ph{.}

\vskip 4 cm

{\Large \bf \ThisIsTheTitle}

\vskip 1 cm

\renewcommand*{\thefootnote}{\fnsymbol{footnote}}

{{\ThisIsAuthorOne}\footnote{\ThisIsEmailOne} } and {\ThisIsAuthorTwo}\footnote{\ThisIsEmailTwo}

\renewcommand*{\thefootnote}{\arabic{footnote}}

\setcounter{footnote}{0}

\vskip .75 cm


\end{center}

\vskip 1.25 cm
\date{}


\end{titlepage}

\newpage

\fi

\ifplainstyle
\tableofcontents
\noindent\hrulefill
\bigskip
\fi

\section{Introduction} \label{se:introduction}

Solvable irrelevant deformations of two-dimensional CFTs \cite{Smirnov:2016lqw,Cavaglia:2016oda,Guica:2017lia} and their holographic dualities \cite{McGough:2016lol,Giveon:2017nie,Bzowski:2018pcy,Apolo:2018qpq,Chakraborty:2018vja,Guica:2019nzm} have recently attracted a lot of attention. Irrelevant deformations are in general difficult to deal with due to their nonrenormalizability. Nevertheless, in \cite{Smirnov:2016lqw,Cavaglia:2016oda} a class of irrelevant deformations built from the bilinear product of the stress tensor, namely the $T\bar T$ deformation \cite{Zamolodchikov:2004ce}, was found to be solvable in the sense that the spectrum on the cylinder can be expressed in terms of the undeformed spectrum. Remarkably, the S-matrix of $T\bar T$-deformed QFTs is related to the undeformed S-matrix by an energy-dependent phase factor which remains well defined at arbitrarily high energies, suggesting that $T\bar T$-deformed QFTs are UV complete \cite{Dubovsky:2012wk,Dubovsky:2013ira}.  The $T\bar T$ deformation can be reformulated in various ways, e.g.~as coupling the undeformed theory to random geometry \cite{Cardy:2018sdv}, as flat Jackiw-Teitelboim gravity \cite{Dubovsky:2017cnj,Dubovsky:2018bmo}, or, when the undeformed theory is a CFT, as a (non)critical string \cite{Dubovsky:2012wk, Callebaut:2019omt}. 

These features of $T\bar T$-deformed QFTs are not exclusive to this class of irrelevant deformations. Similar results have been found for the $J\bar T$ deformation \cite{Guica:2017lia} which is generated by the product of a $U(1)$ current and the stress tensor. In particular, the spectrum of $J\bar T$-deformed QFTs was worked out in~\cite{Guica:2017lia} while the deformed S-matrix was obtained in \cite{Anous:2019osb}, where $J\bar T$-deformed QFTs were also reformulated as QFTs coupled to dynamical gauge fields and gravity. Both the $T\bar T$ and $J\bar T$ deformations are generated by bilinear products of Noether currents, which we refer to as \emph{bi-current} deformations.  More generally, a linear combination of the $T\bar T$, $J\bar T$, and $T\bar J$ deformations dubbed the $T\bar T + J\bar T + T\bar J$ deformation has been studied in \cite{LeFloch:2019rut,Chakraborty:2019mdf,Frolov:2019xzi,Chakraborty:2020xyz}.  For other  studies on irrelevant deformations see e.g.~\cite{Giribet:2017imm,Datta:2018thy,Aharony:2018bad, Aharony:2018ics,Hartman:2018tkw,Taylor:2018xcy,Gross:2019ach,Cardy:2019qao}.

The aforementioned  bi-current deformations are by definition double-trace deformations. A single-trace version of these deformations can be obtained from the product of double-trace deformed theories. For instance, the single-trace $T\bar T$ deformation of a symmetric orbifold CFT is a symmetric orbifold whose seed theory is the $T\bar T$ deformation of the seed CFT~\cite{Giveon:2017nie}. Consequently, there are two kinds of holographic dualities for irrelevant bi-current deformations depending on the double or single-trace nature of the deformation.

In the double-trace version, $T\bar T$-deformed CFTs are holographically dual to Einstein gravity in asymptotically AdS$_3$ spacetimes with a finite cutoff \cite{McGough:2016lol}, or alternatively, to AdS$_3$ gravity with modified boundary conditions \cite{Guica:2019nzm}; whereas $J\bar T$-deformed CFTs are holographically dual to Einstein gravity with Chern-Simons matter fields in asymptotically AdS$_3$ spacetimes with modified boundary conditions \cite{Bzowski:2018pcy}. On the other hand, in the single-trace case,  $T\bar T$-deformed CFTs are holographically dual to (the long string sector of) string theory on a background interpolating between AdS$_3$ in the IR and a flat spacetime with a linear dilaton in the UV~\cite{Giveon:2017nie}; while $J\bar T$-deformed CFTs are holographically dual to (the long string sector of) string theory on warped AdS$_3$ spacetimes~\cite{Apolo:2018qpq,Chakraborty:2018vja}. 
 
 There are several  differences between the double and single-trace versions of holography for bi-current irrelevant deformations: ($i$) the double-trace deformation is universal and can be defined for any two-dimensional CFT whereas the single trace version has so far only been defined for symmetric orbifold CFTs; ($ii$) the cutoff AdS$_3$ proposal  applies only to one sign of the $T\bar T$ deformation parameter whereas the single-trace version applies to both signs; and ($iii$) the double-trace deformation does not change the local bulk geometry (instead it can be understood as changing the boundary conditions) while the single-trace version deforms the bulk geometry itself. The last property makes single-trace holography particularly interesting, as it provides a way to construct holographic dualities for spacetimes that are not asymptotically AdS$_3$. In fact, a special example of the $T\bar T$ black string backgrounds found in \cite{Apolo:2019zai} was studied previously in \cite{Horne:1991gn} with the motivation of understanding black holes in asymptotically flat spacetimes; while the warped AdS$_3$ backgrounds dual to the $J\bar T$ deformation \cite{Apolo:2018qpq,Apolo:2019yfj} have also been studied in \cite{Azeyanagi:2012zd} with the goal of understanding quantum gravity near the horizon of extremal Kerr black holes \cite{Guica:2008mu}. 

In the examples of single-trace holography described above, a solution-generating technique known as a TsT transformation (T-duality, shift, and T-duality) \cite{Lunin:2005jy}, has been used to generate the aforementioned non-AdS$_3$ backgrounds \cite{Apolo:2018qpq,Araujo:2018rho,Apolo:2019yfj,Apolo:2019zai}. This suggests a deep connection between TsT transformations, non-AdS holography, and irrelevant bi-current deformations. Let us assume that the long string sector of type IIB string theory on AdS$_3\times M^7$ supported by NS-NS flux is holographically dual to a symmetric orbifold CFT~\cite{Yu:1998qw,Hosomichi:1998be,Hikida:2000ry,Argurio:2000tb,Giveon:2005mi,Eberhardt:2019qcl}, and let $X^m$ and $X^{\bar m}$ denote two bulk directions with translational or rotational symmetries. It was conjectured in \cite{Apolo:2019zai} that performing the following TsT transformation on the string theory is holographically equivalent to deforming the seed of the dual symmetric orbifold,
 \eq{
\textrm{TsT}_{(X^m,X^{\bar m};\check\mu)} \quad\Longleftrightarrow\quad  \frac{\p S_{\mathcal M_\mu}}{\p \mu} = -\frac{1}{\pi} \int J_{(m)} \wedge J_{(\bar{m})}, \label{conjecture}
}
where $\textrm{TsT}_{(X^m,X^{\bar m};\check\mu)}$ denotes T-duality on  $X^m$, shift $ X^{\bar m} \to X^{\bar m} - 2\check\mu X^m$,  and T-duality on $ X^m$, $S_{\mathcal M_\mu}$ is the action of the deformed seed, $\mu$($\check\mu$) is the dimensionful(less) deformation parameter, and $J_{(m)}$, $J_{(\bar{m})}$ are the Noether currents generating translations along $X^m$, $X^{\bar m}$. It is interesting to note that from a purely worldsheet perspective, a TsT transformation is equivalent to a \emph{marginal} bi-current deformation of the worldsheet action, the latter of which can be shown to be equivalent to a gauged $(SL(2,R) \times U(1))/ U(1)$ WZW model~\cite{Apolo:2019zai}.

For individual  $T\bar T$ or $J\bar T$ deformations, evidence supporting the conjecture \eqref{conjecture} has been given in \cite{Apolo:2019yfj,Apolo:2019zai}. In this paper, we find that this connection still holds between multiple TsT transformations and multiple bi-current deformations. More concretely, we use three successive TsT transformations to construct backgrounds of type IIB string theory that are holographically dual to states in single-trace $T\bar T + J\bar T + T\bar J$-deformed CFTs. While the background dual to the Ramond vacuum has been studied in \cite{Chakraborty:2019mdf,Chakraborty:2020cgo} using other methods, our construction enables us to obtain more general backgrounds including black holes with arbitrary mass, angular momentum, and two $U(1)$ charges, as well as the spacetime dual to the Neveu-Schwarz (NS) vacuum of the dual field theory.

We refer to the finite-temperature backgrounds obtained by the aforementioned sequence of TsT transformations as \emph{tri-TsT black holes}.\footnote{The black holes reported in this paper were referred to as ``black strings'' in \cite{Apolo:2019zai} since the topology of the spacetime at infinity is $R^{(1,1)} \times S^1$ instead of $R^{(1,2)}$, as appropriate for asymptotically flat spacetimes.} These black holes are 7-parameter solutions of type IIB supergravity with NS-NS flux that describe the most general class of black holes in the space of single-trace $T\bar T$, $J\bar T$, and $T\bar J$ deformations. From the seven parameters characterizing these solutions, three parameters are identified with the action of the $T\bar T$, $J\bar T$, and $T\bar J$ deformations, while the other four correspond to phase space variables controlling the energy, angular momentum, and two $U(1)$ charges. As a result, the space of tri-TsT solutions constructed herein contains the $T\bar T$ and $J\bar T$ black holes found previously in \cite{Apolo:2019zai} and \cite{Apolo:2019yfj}, as well as generalizations of the latter that include additional $U(1)$ charges. 

Evidence for the correspondence between a sequence of TsT transformations and single-trace $T\bar T + J \bar T + T\bar J$ deformations given in this paper consists of:
\renewcommand\labelitemi{\small$\bullet$}
\begin{itemize}[leftmargin=\parindent]
\item The spectrum of long strings winding on tri-TsT transformed backgrounds. In particular, we show that the spectrum of strings with one unit of winding matches the spectrum of untwisted states in a single-trace $T\bar T + J \bar T + T\bar J$-deformed CFTs. Furthermore,  we argue that the spectrum of strings with more than one unit of winding corresponds to the spectrum of twisted states in the deformed symmetric orbifold.
\item The entropy (and thermodynamics) of tri-TsT black holes which matches the density of high-energy states in single-trace $T\bar T + J \bar T + T\bar J$-deformed CFTs.
\item The ground state geometry obtained by requiring a smooth tri-TsT background without a horizon. In particular, we show that the conserved charges of this background match the corresponding quantities of the NS vacuum in the dual theory. 
\end{itemize}
In addition, we study geometric properties of the tri-TsT backgrounds in different regions of parameter space. We find that there exists a region of parameter space that yields geometries free of CTCs and curvature singularities. For solutions featuring curvature singularities,  it is possible to find tri-TsT backgrounds where the three-dimensional Ricci scalar in the Einstein frame is positive in a smooth region outside the horizon before reaching any pathologies.

This paper is organized as follows. In section \ref{se:triTsT} we briefly review the relationship between single TsT transformations and single bi-current deformations and propose a sequence of three TsT transformations that are holographically dual to the single-trace $T\bar T + J\bar T + T\bar J$ deformation in the dual CFT. In section \ref{se:tstbackgrounds} we use this sequence of TsT transformations to construct neutral and charged tri-TsT black holes, as well as the ground state geometry. The geometric properties of the tri-TsT backgrounds are studied in section~\ref{se:tstbtzproperties} where we discuss possible singularities, CTCs, and regions of positive curvature. In section \ref{se:spectrum} we derive the spectrum of winding strings on arbitrary tri-TsT backgrounds and discuss its relationship to the spectrum of single-trace $T\bar T + J\bar T + T\bar J$-deformed CFTs. The charges and thermodynamics of the tri-TsT black holes are derived in section~\ref{se:thermodynamics} where we match the entropy and ground state energy to the corresponding quantities in the dual field theory. Additional results are collected in appendices \ref{ap:spectrum} -- \ref{ap:btzU1}. In appendix \ref{ap:spectrum} we derive a general formula for the spectrum of winding strings on backgrounds obtained from multiple TsT transformations; in appendix \ref{ap:tstbackgrounds} we construct additional tri-TsT backgrounds and derive the spectrum of strings winding on these backgrounds; in appendix \ref{ap:charges} we collect the formulae used for the computation of gravitational charges; and in appendix \ref{ap:btzU1} we comment on properties of the charged BTZ black hole that is used in the construction of the charged tri-TsT black holes. 
 

\section{TsT and irrelevant deformations} \label{se:triTsT}
In this section we describe the triality \cite{Apolo:2019yfj,Apolo:2019zai} between TsT transformations in IIB supergravity with NS-NS flux, marginal deformations on the string worldsheet, and irrelevant deformations of two-dimensional CFTs. In \cite{Apolo:2019yfj} we have demonstrated how a specific single-trace irrelevant deformation --- for instance the $T\bar T$, $J \bar T$, or $T\bar J$ deformation --- is related to a specific TsT transformation. We further propose that a linear combination of the aforementioned three deformations corresponds to a sequence of TsT transformations in string theory.

\subsection{Irrelevant deformations from TsT} \label{se:tst}

\begin{figure}[ht!]
\centering
 \includegraphics[scale=0.45]{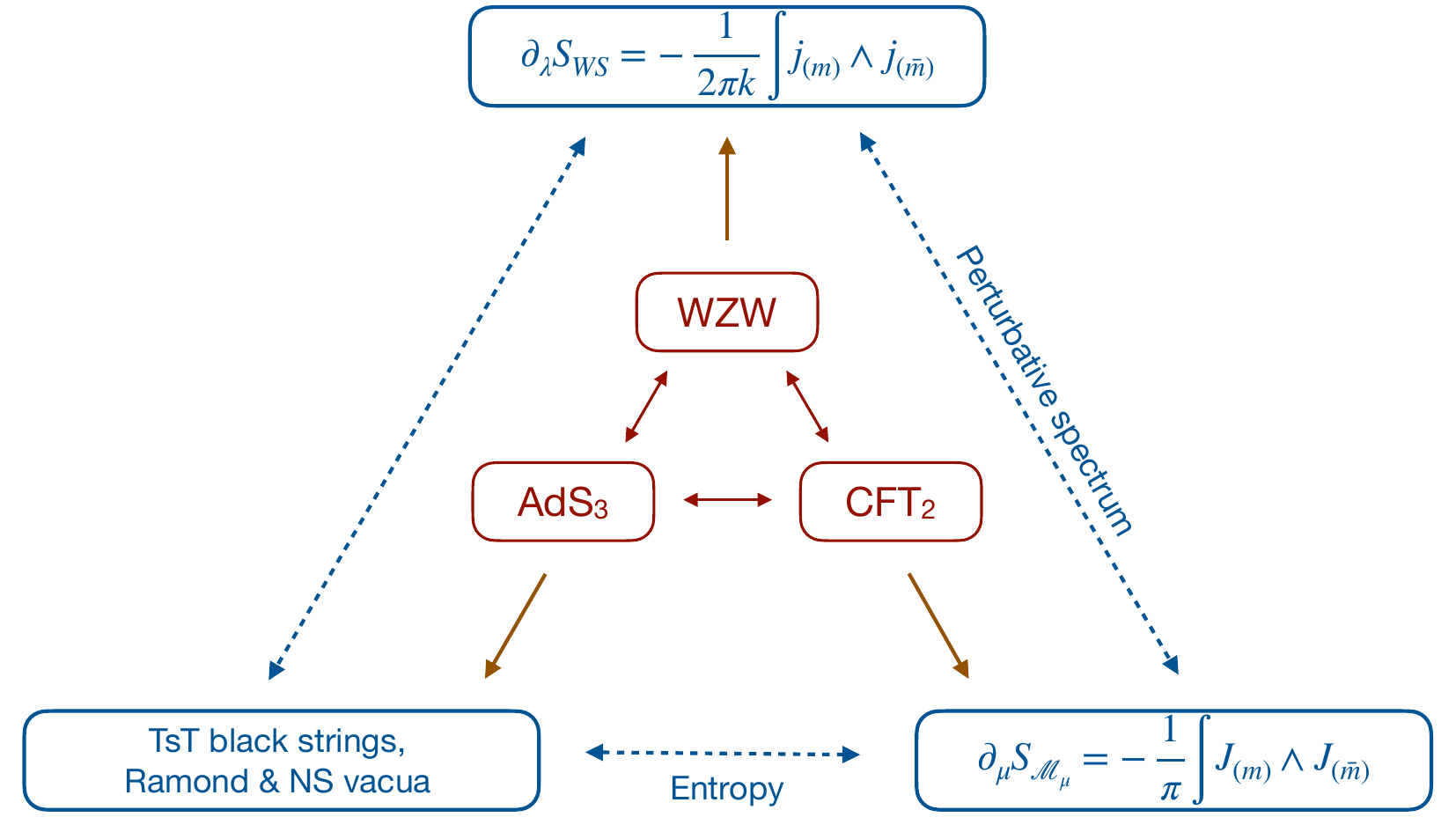}
\caption{The holographic correspondence between string theory on AdS$_3 \times M^7$ backgrounds and two-dimensional CFTs (inner triangle), and its relationship to the conjectured duality between TsT transformations of AdS$_3 \times M^7$ spacetimes, bi-current deformations of the worldsheet action, and single-trace irrelevant deformations of two-dimensional CFTs (outer triangle).}
\label{fig:triangle}
\end{figure}

Let us begin by reviewing the relationship between TsT transformations and single-trace irrelevant deformations of two-dimensional CFTs described in \cite{Apolo:2019yfj,Apolo:2019zai}, which can be summarized by the triangle diagram in fig.~\ref{fig:triangle}.

\subsubsection{The inner layer: triality before the deformations}  \label{se:innerlayer}

The inner triangle of fig.~\ref{fig:triangle} describes three equivalent descriptions of IIB string theory prior to any deformations. String theory on AdS$_3 \times M^7$ backgrounds supported by NS-NS flux admits a spacetime description in terms of supergravity (the left vertex), a weakly-coupled worldsheet description in terms of a WZW model (the top vertex), and a holographic description as a symmetric orbifold CFT~\cite{Yu:1998qw,Hosomichi:1998be,Hikida:2000ry,Argurio:2000tb,Giveon:2005mi,Eberhardt:2019qcl}. 

In this paper we will focus on the bosonic sector of string theory on AdS$_3 \times S^3 \times T^4$ spacetimes.  In this case the CFT dual to the long string sector is a symmetric orbifold $\textrm{Sym}^p\, {\M}$, where the seed CFT $\M$ has central charge $6k$ and $(p,k)$ are the number of coincident (NS1, NS5) branes sourcing the background. We expect the symmetric orbifold structure, as well as the relationship between TsT transformations and irrelevant deformations we are about to describe, to hold for generic choices of the internal manifold $M^7$ and not to depend on the details of the seed CFT.

\subsubsection{The outer layer: triality after the deformations} \label{se:outerlayer}

The outer layer of fig.~\ref{fig:triangle} describes deformations on the spacetime geometry, the worldsheet theory, and the symmetric orbifold CFT that are generated by two $U(1)$ symmetries.\footnote{Here $U(1)$ may refer to a compact (internal) $U(1)$ symmetry or a non-compact (translational) one.}

The right vertex of the outer layer describes the deformation on the dual CFT side: a single-trace irrelevant deformation of the symmetric orbifold $\textrm{Sym}^p\, \mathcal M\to \textrm{Sym}^p\, \mathcal M_\mu$ that is obtained by deforming each copy of the seed CFT $\M \to \M_\mu$. The action of the deformed seed satisfies the following differential equation
\eq{
 \frac{\p S_{\mathcal M_\mu}}{\p \mu} = - \frac{1}{\pi} \int J_{(m)} \wedge J_{(\bar{m})}, \label{cftdef}
}
where $\mu$ is a dimensionful parameter and the Noether currents $J_{(m)}$ and $J_{(\bar m)}$ generate two $U(1)$ transformations on the deformed seed $\mathcal M_\mu$. 
The simplest examples of such irrelevant deformations are the $T\bar T$ \cite{Zamolodchikov:2004ce,Smirnov:2016lqw,Cavaglia:2016oda} and the $J\bar T$ \cite{Guica:2017lia} deformations where the Noether currents are given by
 \eqsp{
T\bar T&:\qquad  \,J_{(x)} = T_{\mu x} dx^\mu, \quad J_{(\bar{x})} = T_{\mu \bar x } dx^\mu, \\
  J\bar T&: \qquad J_{(m)} = J_\mu dx^\mu, \quad \,\,   J_{({\bar{x}})} = T_{\mu \bar x } dx^\mu, \label{irrelevantcurrents}
  }  %
where $(x,\bar x)$ are lightcone coordinates, $T_{\mu\nu}$ is the stress tensor, and $J_{\mu}$ is a $U(1)$ current of the deformed seed $\M_\mu$.

The top vertex in the outer layer of fig.~\ref{fig:triangle} describes the deformation of the string worldsheet theory. Generalizing previous studies \cite{Giveon:2017nie,Apolo:2018qpq,Chakraborty:2018vja}, the authors of~\cite{Apolo:2019zai} proposed that the holographic description of single-trace deformations generated by two $U(1)$ currents~\eqref{cftdef} corresponds to an instantaneous deformation of the worldsheet action that is given by
  \eq{
  \frac{\p S_{\textrm{WS}}}{\p {\check \mu}} = -\frac{1}{\pi} \int  { j}_{(m)} \wedge {j}_{({\bar m})}, 
  \label{wsdef}
  }
where $\check \mu$ is a dimensionless parameter, while ${j}_{(m)}$ and ${j}_{({\bar{m}})}$ are the Noether currents that generate the $U(1)$ symmetries on the deformed worldsheet action $S_{\textrm{WS}}$.\footnote{When written as 1-forms, we adopt the convention that the worldsheet Noether currents are given by $j_{(m)} \equiv 2  j^a_{(m)} \gamma_{ab} dz^b$ where $\gamma_{ab}$ and $z^a$ are the worldsheet metric and coordinates, and $j^a_{(m)}$ is given in \eqref{noethervector}.} The worldsheet currents ${j}_{(m)}$ and ${j}_{({\bar{m}})}$ are in one-to-one correspondence with the Noether currents $J_{(m)}$ and $J_{(\bar m)}$ of the deformed field theory, since they generate the same $U(1)$ symmetries. 

For example, in the $T\bar T$ deformation, the Noether current $J_{(x)}$ generates a translation along the boundary coordinate $x$. Using the holographic dictionary, $x$ corresponds to a target space coordinate $u$ in the AdS$_3$ part of the bulk spacetime. Hence, the worldsheet Noether current $j_{(u)}$ generates shifts of the target space coordinate $u$. Similarly, the boundary lightcone coordinate $\bar x$ is identified with the $\bar u$  coordinate of AdS$_3$ such that $J_{(\bar x)}$ in \eqref{cftdef} corresponds to  $j_{(\bar u)}$ in \eqref{wsdef}. We can also consider a shift along one of the coordinates $X^n$ of the internal manifold $S^3\times T^4$ which is generated by the worldsheet current $j_{(n)}$. In the dual field theory side, this worldsheet current corresponds to a Noether current $J_{(n)}$ that generates a global $U(1)$ transformation. The dictionary relating the field theory \eqref{cftdef} and worldsheet  \eqref{wsdef} deformations is thus
\eqsp{ 
T\bar T&: \qquad \ell J_{(x)} \leftrightarrow  j_{(u)},\qquad \ell J_{(\bar x)} \leftrightarrow j_{(\bar u)},\qquad  \mu = \ell^2 \check\mu ,\\
J\bar T&: \qquad  \phantom{\ell} J_{(n)} \leftrightarrow j_{(n)},\qquad   \ell J_{(\bar x)} \leftrightarrow j_{(\bar u)},\qquad \mu = \ell \check\mu,    \label{dictionary}
}
where $\ell$ is the AdS scale. Note that in this identification the normalization of the currents should be fixed appropriately. In particular, the $U(1)$ current $J_{(n)}$ in the boundary field theory is assumed to be canonically normalized such that the coordinate $X^n$ in the bulk satisfies $X^n \sim X^n + 2\pi$. Furthermore, we note that the factor of $\ell$ between the  field theory and worldsheet currents in \eqref{dictionary} comes from our conventions where the boundary coordinates $(x,\bar x)$ are dimensionful while those in the bulk $(u, \bar u)$ are dimensionless so that $(x, \bar x) = (\ell u, \ell \bar u)$. 

Finally, the left vertex on the outer layer of fig.~\ref{fig:triangle} describes a new solution of IIB supergravity with NS-NS flux that is generated by a TsT transformation along two $U(1)$ isometries generated by $\p_{X^m}$ and $\p_{X^{\bar m}}$. This follows from the fact that the differential equation \eqref{wsdef} corresponds to an exactly marginal deformation of the worldsheet action to leading order in $\alpha'$ that is equivalent to a TsT transformation of the undeformed background. In order to see this let $M_{\a\b}  = G_{\a\b} + B_{\a\b}$ where $G_{\a\b}$ and $B_{\a\b}$ are the ten-dimensional deformed string frame metric and Kalb-Ramond field. The differential equation \eqref{wsdef} can then be written as
\eq{
  \frac{\p M}{\p{{\check\mu}}}  = -2\ell_s^{-2} M  \GG M, \qquad \GG_{\a\b} = \d^{m}_{\a} \d^{{\bar{m}}}_{\b} - \d^{{\bar{m}}}_{\a} \d^{m}_{\b},   \label{Mdef}
  }
  where $\ell_s$ is the string length. The solution of \eqref{Mdef} is given in terms of the undeformed background field $\hat M_{\a\b}$ by
  \eq{
  M = \hat M \big (I+ {2\check \mu \ell_s^{-2}} \GG \hat M \big)^{-1}. \label{tst} 
  }
We recognize \eqref{tst} as a TsT transformation of $\hat M_{\alpha\beta}$ that yields the deformed background $M_{\alpha\beta}$ and consists of
\eq{
\!\!\!\! \textrm{TsT}_{(X^m,X^{\bar m}; \,\check\mu)}: \,\,\,\, \textrm{T-duality on } X^m, \quad \textrm{shift } X^{\bar m} \to X^{\bar m} - 2\check\mu  X^m, \quad \textrm{T-duality on } X^m.\label{tstdef}
}
In particular, if one of the $U(1)$ symmetries corresponds to an isometry of AdS$_3$, the resulting background will no longer have a locally AdS$_3$ factor. We refer to the solutions obtained from \eqref{tstdef} as TsT backgrounds.

In this section we have described the relationship between deformations generated by a single pair of $U(1)$ symmetries and a single TsT transformation of the background geometry. Evidence for the correspondence between TsT transformations and irrelevant deformations has been provided in \cite{Apolo:2019yfj,Apolo:2019zai} and includes: ($i$) matching  the worldsheet spectrum of long strings to the (untwisted sector) spectrum of single-trace irrelevant deformations of the symmetric orbifold $\textrm{Sym}^p\, \mathcal M$; and ($ii$) matching the thermodynamics of TsT black holes/strings with the thermodynamics of the dual field theory. In the next subsection we generalize this relationship and propose that linear combinations of several pairs of $U(1)$ symmetries --- specifically a linear combination of the $T\bar T$, $J\bar T$, and $T\bar J$ deformations --- corresponds to a sequence of TsT transformations of the background geometry.


\subsection{$T\bar T + J \bar T + T\bar J$ from multiple TsT transformations} \label{se:multipletst}

We now deform the symmetric orbifold $\textrm{Sym}^p\, \mathcal M$ by a linear combination of the single-trace $T\bar T$, $J\bar T$, and $T\bar J$ deformations, namely by the so-called $T\bar T + J \bar T + T\bar J$ deformation. 

Let $\mu_0$, $\mu_+$, and $\mu_-$ parametrize the $T\bar T$, $J\bar T$ and $T\bar J$ deformations such that $\mu_0$ has dimensions of length squared, while $\mu_\pm$ have dimensions of length. A single-trace $T\bar T + J \bar T + T\bar J$ deformation of the symmetric orbifold $\textrm{Sym}^p\, \mathcal M$ amounts to deforming the seed CFT $\M \to \M_{\mu}$ such that
\eq{
\!\!\!\! \frac{\p S_{\M_{\mu}}}{\p \mu_0} & = -\frac{1}{\pi} \int J_{(x)} \wedge J_{(\bar x)}, \,\,\, \frac{\p S_{\M_{\mu}}}{\p \mu_+} = -\frac{1}{\pi} \int J_{(n)} \wedge J_{(\bar x)}, \,\,\,  \frac{\p S_{\M_{\mu}}}{\p \mu_-} = -\frac{1}{\pi} \int J_{(x)} \wedge J_{(\bar{n})}, \label{cftdef2}
}
where $J_{(\bar n)}$ and $J_{(n)}$ are Noether currents generating internal $U(1)$ transformations in the deformed seed $\mathcal M_\mu$, while $J_{(x)}$ and $J_{(\bar x)}$ generate spacetime translations and can be written in terms of the stress tensor $T_{\mu\nu}$ as in \eqref{irrelevantcurrents}.

In order to discuss the deformation on the string theory side, we first need to find the corresponding Noether currents. Before the deformation, translations along $x$ and $\bar x$  in the boundary CFT correspond to isometries of the locally AdS$_3$ spacetime in the bulk, while internal $U(1)$ transformations correspond to isometries of the internal manifold $S^3 \times T^4$. Following the discussion in section \ref{se:tst}, we propose that the string theory description of single-trace $T\bar T + J\bar T + T\bar J$ deformations corresponds to an instantaneous deformation of the worldsheet action $S_{\textrm{WS}}$ such that
\eq{
\frac{\p S_{\textrm{WS}}}{\p \check\mu_0} &= - \frac{1}{\pi} \int  {j}_{(u)} \wedge {j}_{(\bar u)},  \,\,\,\,\, \frac{\p S_{\textrm{WS}}}{\p \check\mu_+} = - \frac{1}{\pi} \int {j}_{(n)} \wedge  {j}_{(\bar u)},  \,\,\,\,\, \frac{\p S_{\textrm{WS}}}{\p \check\mu_-} = - \frac{1}{\pi} \int {j}_{(u)} \wedge {j}_{({\bar{n}})}. \label{wsdef2}
  }
The worldsheet Noether currents in \eqref{wsdef2} are in one-to-one correspondence with the Noether currents of the field theory via the dictionary \eqref{dictionary}; similarly, the $\check\mu_i$ parameters of the worldsheet are related to the $\mu_i$ parameters of the field theory via \eqref{dictionary}.  As a consistency check, we note that when the three-dimensional part of the background is the zero mass BTZ black hole, \eqref{wsdef2} reduces to chiral current-current deformations of the Ramond vacuum~\cite{Giveon:2017nie}. 

The instantaneous deformations of the worldsheet action \eqref{wsdef2} can also be interpreted geometrically as a sequence of TsT transformations of the AdS$_3\times S^3 \times T^4$ background. This can be made manifest by writing each of the differential equations in \eqref{wsdef2} in terms of the deformed background field $M_{\a\b}$, namely
\eq{
\frac{\p M}{\p{{\check\mu}_i}}  = -2\ell_s^{-2} M  \GG_{i} M, \label{Mdef2}
 }
where the $\GG_{i}$ matrices read
\eq{
(\GG_{0})_{\a\b} = \d^{u}_{\a} \d^{{\bar{u}}}_{\b} - \d^{{\bar{u}}}_{\a} \d^{u}_{\b}, \qquad  (\GG_{+})_{\a\b} = \d^{n}_{\a} \d^{{\bar{u}}}_{\b} - \d^{{\bar{u}}}_{\a} \d^{n}_{\b}, \qquad (\GG_{-})_{\a\b} = \d^{u}_{\a} \d^{{\bar{n}}}_{\b} - \d^{{\bar{n}}}_{\a} \d^{u}_{\b}. \label{Gamma}
 }
The general solution to the differential equation \eqref{Mdef2} can be succinctly written as %
 \eq{
 M = \hat M \big[ I + 2\ell_s^{-2} \textstyle\sum_i \check\mu_i \GG_i \hat M\big]^{-1}. \label{tst2}
 }
Using \eqref{tst} it is not difficult to show that \eqref{tst2} corresponds to a sequence of three TsT transformations that consist of
\eq{
 \textrm{TsT}_{(u,\bar u ;\, \check\mu_0)}, \qquad \textrm{TsT}_{(X^n,\bar u ;\, \check\mu_+)},\qquad \textrm{TsT}_{(u,X^{\bar n};\, \check\mu_-)}.  \label{tstrecipe2}
}
In particular, it is clear from \eqref{tst2} that the we can perform this (or any other) sequence of TsT transformations in any order and obtain the same deformed background.

We thus propose that states in $T\bar T + J \bar T + T\bar J$-deformed CFTs are dual to backgrounds obtained from a sequence of TsT transformations of AdS$_3\times S^3 \times T^4$ spacetimes. This sequence of TsT transformations is described in \eqref{tstrecipe2} and is compactly written in \eqref{tst2}. In particular, depending on the choice of TsT coordinates on the internal manifold, we can obtain different backgrounds that are characterized by the kinds of the internal $J$ and $\bar J$ currents featured in the $T\bar T + J \bar T + T\bar J$ deformation, as summarized in Table~\ref{table1}.

\renewcommand{\arraystretch}{1.175}
   \begin{table}[h!]
  \begin{center}
  \begin{tabular}{?{.75pt}c|c|c|c|c|c|c|c|c|c?{.75pt}c?{.75pt}}
\hlinewd{.75pt} 
    \multicolumn{3}{?{.75pt}c|}{AdS$_3$}  &   \multicolumn{3}{c|}{$S^3$}  & \multicolumn{4}{c?{.75pt}}{$T^4$}   & \multirow{2}{*}{internal worldsheet currents} \\  \cline{1-10}
  $u$  &  $\bar u$  &  $r$  &  $\theta$  &  $\psi$  &  $\phi$  &  $y_7$  &  $y_8$  &  $y_9$  &  $y_{10}$  &    \\  \hlinewd{.75pt}
  $\times$  &  $\times$  &  &  &  &  &  $\times$  &  &  &  &                  $j_{(n)} = j_{(\bar n)} = j_{(y_7)}$ \\  \hline
  $\times$  &  $\times$  &  &  &  &  &  $\times$  &  $\times$  &  &  &  $j_{(n)} = j_{(y_7)}, \quad  j_{(\bar n)} = j_{(y_8)}$  \\  \hline
  $\times$  &  $\times$  &  &  &  $\times$  &  &  $\times$  &  &  &  &  $j_{(n)} = j_{(\psi)}, \quad  j_{(\bar n)} = j_{(y_7)}$  \\  \hline
  $\times$  &  $\times$  &  &  &  $\times$  &  $\times$  &  &  &  &  &  $j_{(n)} = j_{(\psi)}, \quad  j_{(\bar n)} = j_{(\phi)}$  \\  \hlinewd{.75pt} 
  \end{tabular}
  \caption{Different choices of TsT coordinates corresponding to the single-trace $T\bar T + J\bar T + T\bar J$ deformation. Two of the TsT coordinates are always taken along the lightcone $u$ and $\bar u$ coordinates of AdS$_3$ as these correspond to the $T$ and $\bar T$ parts of the deformation.}
  \label{table1}
  \end{center}
  \end{table}
\vspace{-15pt}

In this paper we will focus on the first case in table \ref{table1} where the internal $U(1)$ currents of the $T\bar T + J \bar T + T\bar J$ deformation are associated with the same isometry of $T^4$. In the next section we will construct black hole backgrounds that are generically dual to thermal states in this class of $T\bar T + J \bar T + T\bar J$-deformed CFTs but include other states like the Ramond and Neveu-Schwarz vacua after an appropriate choice of parameters. We refer to these backgrounds as \emph{tri-TsT black holes}. The other cases listed in table \ref{table1} are also interesting. In appendix \ref{ap:tstbackgrounds} we discuss the TsT-transformed backgrounds corresponding to the second and fourth cases, and derive the spectrum of strings winding on these backgrounds. 


\section{Tri-TsT black holes} \label{se:tstbackgrounds}

In this section we construct black hole solutions that are generically dual to thermal states in single-trace $T\bar T + J\bar T + T\bar J$-deformed CFTs. We classify these solutions by the absence or presence of $U(1)$ charges before any TsT transformations. When these charges vanish, we obtain special black hole solutions that are dual to thermal states whose \emph{deformed} $U(1)$ charges also vanish. On the other hand, when the undeformed background has additional $U(1)$ charges, we obtain the most general class of charged tri-TsT black holes with arbitrary values of the deformed $U(1)$ charges. Finally, we construct the background that is dual to the NS vacuum in the dual $T\bar T + J\bar T + T\bar J$-deformed CFT.

\subsection{Neutral tri-TsT black holes} \label{se:specialtstbtz}

Let us first consider a simpler class of black hole solutions that are dual to neutral thermal states in single-trace $T\bar T + J\bar T + T\bar J$-deformed CFTs. We will accomplish this by performing a sequence of TsT transformations on a generic BTZ$\,\times S^3 \times T^4$ background where the TsT coordinates are identified as shown in the first entry of table~\ref{table1}.
 
We begin by describing the BTZ$\,\times S^3 \times T^4$ background and setting up our conventions. For convenience, we write this background in terms of the three-dimensional fields of the dimensionally reduced theory which are related to the ten-dimensional fields via (see appendix \ref{ap:charges} for more details)
\eqsp{
ds^2_{(10)}  &=  ds^2  + \ell_4^2 e^{-\omega } \big(  dy + A^{(1)}\big)^2 + d \Omega_3^2+ \ell^2_4 {\textstyle \sum_{i=8}^{10}} dy_i^2 , \\
B_{(10)}  &= B    + \ell_4^2 A^{(2)} \!  \wedge dy + B_{\Omega_3} ,  \\
e^{2\Phi_{(10)} } &= k^2_4 e^{2\Phi }.
\label{10dbackground}
}
Here $ds^2$, $B$, and $\Phi$ are the three-dimensional line element, Kalb-Ramond field, and dilaton; while $A^{(1)}$, $A^{(2)}$, and $\omega$ are the additional three-dimensional gauge fields and dilaton that result from the dimensional reduction. The constant $k_4$ in the definition of the ten-dimensional dilaton is given by $k_4 = \ell^2_4/\ell_s^2$ where $\ell_4$ is the scale of $T^4$. We use $x^\mu \in (u,v,r)$ for the coordinates of the three-dimensional fields, $(\theta, \psi, \phi)$ for the coordinates of the 3-sphere, and $(y, y_8, y_9, y_{10})$ for the coordinates of the four-torus. For convenience, we have introduced $v \equiv \bar u$ and $y \equiv y_7$, as these coordinates are featured prominently in what follows.

The BTZ black hole is characterized by the following three-dimensional fields
\eqsp{
\begin{gathered}
ds^2 =  \ell^2 \bigg[ \frac{dr^2}{4 \big (r^2 - 4 T_u^2 T_{v}^2 \big )} + r du {dv} + T_u^2 du^2 + T_{v}^2 d{v}^2 \bigg] , \\
B =  \frac{ \ell^2 r}{2}  {dv} \we du,  \qquad A^{(1)} = A^{(2)} =0, \qquad e^{2{\Phi}} = \frac{k}{p}, \qquad  \omega = 0,  \label{btz}
 \end{gathered}
}
where $k = \ell^2/\ell_s^2$ and $\ell$ denotes the scale of AdS. The contributions from the three-sphere to the ten-dimensional metric and the $B$-field can be written as
\eq{
d\Omega_3^2 =  \frac{\ell^2}{4} \big (d \t^2 + d\psi^2 + 2\cos\t d\phi d\psi + d\phi^2  \big), \qquad B_{\Omega_3} = \frac{\ell^2}{4} \cos\t\, d \phi \we d\psi, \label{S3}
}
where the $(\theta, \psi, \phi)$ coordinates satisfy $\theta \in [0,\pi)$ together with
\eq{
\psi \sim \psi + 4\pi, \qquad (\psi,\phi) \sim (\psi + 2\pi, \phi + 2\pi). \label{coordsim1}
}
In our conventions, each of the coordinates of $T^4$ are identified as
\eq{
y_i \sim y_i + 2\pi, \label{coordsim2}
}
while the $(u,v)$ coordinates of AdS$_3$  satisfy
  \eq{
  (u,v) \sim (u + 2\pi, v+ 2\pi), \label{coordsim3}
  }
such that the dual CFT lives on a cylinder of size $2\pi\ell$. We refer to each of the circles in \eqref{coordsim2} as an internal circle while \eqref{coordsim3} corresponds to the so-called spatial circle.

The constants $k$ and $p$ in the definition of the dilaton \eqref{btz} correspond to the background magnetic and electric charges that count the number of coincident NS5 and NS1 branes supporting the background 
  \eq{
  Q_m \equiv\frac{1}{(2\pi \ell_s)^2} \int_{S^3} dB_{(10)} = k, \qquad Q_e \equiv \frac{1}{(2 \pi \ell_s)^6} \int_{S^3 \times T^4} e^{-2\Phi_{(10)}} \star_{10} dB_{(10)} = p, \label{QeQmBTZ}
}
 where $\star_{10}$ denotes the Hodge dual in the ten-dimensional string frame. It is also convenient to spell out the relationships between $k$, $p$, the dilaton $e^{\Phi_{(10)}}$, and the ten and three-dimensional Newton's constants $G^{(10)}_N$ and $G^{(3)}_N$, which read
\eq{
 G_N^{(10)} = 8\pi^6 g_s^2 \ell_s^8, \qquad  {G^{(3)}_N} = \frac{G^{(10)}_N e^{2\Phi_{(10)}}}{V_{S^3} V_{T^4} g_s^2},
}
where $V_{S^3}$ is the volume of $S^3$, $V_{T^4}$ is the volume of $T^4$. In addition, the central charge $c$ of the dual CFT is given by 
\eq{
c = 6 k p = \frac{3\ell}{2 G_N^{(3)}}.
}
The black hole \eqref{btz} features left and right-moving energies that are given by
\eq{
\Q_{u}  =  \frac{c}{6\ell} T_u^2, \qquad  \Q_{v}   = \frac{c}{6\ell} T_{v}^2,
}
where $\Q_{u} $ and $\Q_{v} $ are the gravitational charges associated with the Killing vectors $\ell^{-1} \p_u$ and $-\ell^{-1} \p_v$. Furthermore, note that $T_u$ and $T_v$ parameters are related to the left and right-moving temperatures of the BTZ black hole via
\eq{
T_L = \frac{T_u}{\ell \pi}, \qquad T_R = \frac{T_{v}}{\ell \pi}.
}

Let us now perform the sequence of TsT transformations \eqref{tstrecipe2} on the BTZ black hole described by \eqref{10dbackground} and \eqref{btz}. This allows us to generate backgrounds that are generically dual to neutral thermal states in single-trace $T\bar T + J\bar T + T\bar J$-deformed CFTs. We are interested in the case where both the $J$ and $\bar J$ currents are associated with the same isometry of $T^4$ (see appendix~\ref{ap:tstbackgrounds} for other choices). Consequently, we perform the following TsT transformations,  
\eq{
& \textrm{TsT}_{(u,v;\, \check\mu_0)}, \qquad \textrm{TsT}_{(y,v;\, \check\mu_+)},  \qquad \textrm{TsT}_{(u,y;\, \check\mu_-)}.  \label{tstrecipe3}
}
At this stage, it is convenient to introduce another set of dimensionless deformation parameters that are related to the $\check\mu_i$ and $\mu_i$ parameters by
\eq{
\l_0 & \equiv 2 k \check\mu_0 = \frac{2}{\ell_s^2} \mu_0, \qquad \l_\pm \equiv 2 \sqrt{k k_4} \,\check\mu_\pm = \frac{2 \ell_4}{\ell_s^2} \mu_\pm. \label{dictionarylambdas}
}
Using \eqref{tst2} and \eqref{dictionarylambdas} we find that the deformed background can be written as in \eqref{10dbackground} where the three-dimensional fields are given by
\eqsp{
\!\! ds^2 &=  \frac{\ell^2 dr^2}{4(r^2 - 4 T_u^2 T_{v}^2)}  + \ell^2 h \bigg [  \big( r + 2 \l_+  \l_- T_u^2  T_{v}^2 \big) du d{v}   +  \big( 1 + \l_+^2  T_{v}^2 \big) T_u^2 du^2 \\
& \hspace{15pt} + \big(1 + \l_-^2  T_u^2 \big)   T_{v}^2 d{v}^2 + \frac{(r + 2\l_0 T_u^2 T_{v}^2)^2(\l_+ du + \l_- d{v})^2}{4(1+\l_0 r +  \l_0^2 T_u^2T_{v}^2)}  \bigg],  \\
\!\!  B &= \frac{\ell^2 h}{2} \big( r + 2 \l_0 T_u^2 T_{v}^2 \big) d{v} \we du, \qquad e^{2\Phi} = \frac{k h}{ \eta p},   \qquad  e^{-\omega} =  h\big(1+\l_0 r +  \l_0^2 T_u^2T_{v}^2  \big),   \\
 \!\!   A^{(1)} & = -\frac{\ell h}{2 \ell_4}e^\omega  \big(  r  + 2 \l_0  T_{v}^2  T_u^2 \big) \big(\l_+ du + \l_-  d{v} \big ),  \\
\!\! A^{(2)} & = \frac{\ell h}{2 \ell_4} \big[   \big( \l_+ r  - 2  \l_-  T_u^2  \big ) du -   \big( \l_- r  -2 \l_+ T_{v}^2    \big) d{v} \big],   \label{neutraltstbtz} 
}  
and the coordinates satisfy the identification \eqref{coordsim3}. The dilaton in \eqref{neutraltstbtz} is determined up to a constant $\eta$ using Buscher's rule~\cite{Buscher:1987sk,Buscher:1987qj} while the function $h$ is given by
 \eqsp{
 h ^{-1} & \equiv \det (I + 2\ell_s^{-2} \textstyle\sum_i \check\mu_i \GG_i \hat M) \\
 & = 1 + (\l_0 - \l_+ \l_-)r+  \l_-^2 T_u^2 + \l_+^2 T_{v}^2 + \l_0^2 T_u^2 T_{v}^2, \label{shdef1}
 }
where the $\Gamma_i$ matrices are defined in \eqref{Gamma} and $\hat M_{\a\b}  = \hat G_{(10)\a\b} + \hat B_{(10)\a\b}$ refers to the ten-dimensional undeformed background fields \eqref{10dbackground} and \eqref{btz}. We refer to $h$ as the \emph{flow function}, as it interpolates between the locally AdS$_3$ region near the horizon at $r = 2T_uT_{v}$ in the IR and a deformed asymptotic region in the UV.  In particular, when $\lambda_0\neq\lambda_+\lambda_-$ the UV is described by a linear dilaton background.

The background described by \eqref{10dbackground} and \eqref{neutraltstbtz} is a solution to the ten-dimensional supergravity equations of motion for any constant value of $\eta$. The latter can be fixed by requiring the values of the magnetic and electric charges \eqref{QeQmBTZ} to be preserved after the deformation. Preserving the values of the magnetic and electric charges \eqref{QeQmBTZ} then yields
\eq{
\eta^{-1} = 1 - \l_0^2 T_u^2 T_{v}^2 .
}
Since $\eta$ guarantees quantization of the magnetic and electric charges, we refer to this constant as the \emph{quantization parameter} of the black hole. Note that in order for the dilaton to be real, the $\l_0$, $T_u$, and $T_{v}$ parameters are constrained to satisfy $\eta^{-1} = 1 - \l_0^2 T_u^2 T_{v}^2 > 0$. This bound only depends on the $T\bar T$ deformation parameter, and it reproduces the bound on the product of the left and right-moving temperatures in single-trace $T\bar T$-deformed CFTs, as first observed in \cite{Apolo:2019zai}. This suggests that neutral tri-TsT black holes mimic the behavior of thermal states in $T\bar T$-deformed CFTs, as described in more detail in section~\ref{se:machingthermo}. 

Let us briefly discuss the asymptotic behavior of the TsT background \eqref{neutraltstbtz} as $r\to \infty$. A more detailed analysis of a generalization of this background is given in section~\ref{se:asympt}. The leading order behavior of the metric near the asymptotic boundary depends only on the TsT parameters $\l_0$, $\l_+$, and  $\l_-$, consistent with the fact that they are identified with irrelevant deformations in the dual field theory. On the other hand, the $T_u$ and $T_v$ parameters appear at subleading order and are therefore identified with the phase space variables of the background, as is also the case before the deformations. In particular, when $T_u > 0$ and $T_{v} > 0$, the solutions \eqref{neutraltstbtz} describe black holes that are dual to a special class of thermal states in single-trace $T\bar T + J\bar T + T\bar J$-deformed CFTs, namely states where the deformed $U(1)$ charges associated with the $J$ and $\bar J$ currents vanish. We will justify these statements in section \ref{se:thermodynamics}, where we study the thermodynamics of this background and the charged generalization constructed in the next section. Finally, when $T_u = T_v = 0$, the black hole \eqref{neutraltstbtz} reduces to the Ramond vacuum studied in \cite{Chakraborty:2019mdf} where our deformation parameters are identified with the parameters used there via $(\l_0, \l_+, \l_-) = (\l, -2\e_+, -2\e_-)$.


\subsection{Charged tri-TsT black holes} \label{se:tstbtz}

The TsT backgrounds \eqref{neutraltstbtz} are characterized by the three fixed deformation parameters $(\l_0, \l_+, \l_-)$ and two free phase space variables $(T_u, T_{v})$ that parametrize the mass and angular momentum of the black hole. On the other hand, a state in a $T\bar T + J\bar T + T\bar J$-deformed CFT is characterized by two additional quantum numbers that correspond to the zero mode charges of the $J$ and $\bar J$ currents. In order to describe charged states it is necessary to turn on two arbitrary $U(1)$ charges in the bulk. This can be achieved by first adding $U(1)$ charges to the BTZ$\,\times S^3 \times T^4$ background~\eqref{btz} and then performing the sequence of TsT transformations \eqref{tstrecipe3}.

The simplest way to add $U(1)$ charges to the BTZ black hole~\eqref{btz} is to perform a spectral flow transformation to the ten-dimensional solution via the following change of coordinates and gauge transformation
  \eq{
  \!\!\! y  \to y - \frac{\ell}{\ell_4} (\a_u u +\a_{v} v), \quad B_{(10)} \! \to B_{(10)} \! + \ell^2 \a_u \a_{v} d{v} \we du +  \ell \ell_4 (\a_u du - \a_{v} d{v}) \we dy . \label{cc}
  }
  This spectral flow transformation  turns on the gravitational charges associated with momentum and winding along the $y$ coordinate. Furthermore, it guarantees that the metric and $B$-field satisfy the same condition on $\hat M_{\alpha\beta}$ as the BTZ background \eqref{btz}, namely,
\eq{
\hat M_{uy} = \hat M_{yv } = \hat M_{uv} = 0. \label{bfieldbc}
}
 As described in~\cite{Apolo:2019zai}, constant values of $ \hat M_{uy}$, $ \hat M_{y v}$, and $ \hat M_{u v}$ are necessary for the existence of chirally conserved currents in the worldsheet theory that are related to translations along $u$, $v$, and $y$. These chiral currents play an important role in the derivation of the spectrum and imply that the instantaneous deformations of the worldsheet action \eqref{wsdef2} are related to the current-current deformations associated with gauged WZW models~\cite{Apolo:2019zai}.

After the shift of coordinates and gauge transformation \eqref{cc}, the resulting background can be written in the form \eqref{10dbackground} where the three-dimensional metric, $B$-field, gauge fields, and dilatons are given by
\eq{
\begin{gathered}
\!\!\!\! ds^2 = \ell^2 \bigg[   \frac{dr^2}{4 \big (r^2 - 4 T_u^2 T_v^2 \big )} + r du dv + T_u^2 du^2 + T_v^2 dv^2 \bigg] , \quad B =  \ell^2 \Big( \frac{r}{2} + \a_u \a_v \Big) dv \we du, \\
\!\!\!\!  A^{(1)} = - \frac{\ell}{\ell_4} (\a_u du + \a_v dv), \quad A^{(2)} =  \frac{\ell}{\ell_4} \big( \a_u du - \a_v dv \big), \quad e^{2{\Phi}} = \frac{k}{p}, \quad \omega = 0, \label{btzU1}
 \end{gathered}
}
where the identifications of the coordinates remain the same as in \eqref{coordsim3}, namely $(u,v) \sim (u + 2\pi, v + 2\pi)$. By construction, the spectral flow parameters $\a_u$ and $\a_v$ do not affect the magnetic and electric charges, which are given by \eqref{QeQmBTZ}, such that the central charge of the dual CFT is still given by $c = 6 Q_e Q_m = 6 p k$.  Although \eqref{btzU1} is locally equivalent to \eqref{btz}, these backgrounds are physically distinct solutions that carry different gravitational charges. Indeed, the energies of the charged BTZ black hole are now given by
\eq{
\Q_{u} &= \frac{c}{6 \ell }  \big( T_u^2 +  \a_u^2 \big) , \qquad \Q_{v}  =  \frac{c}{6 \ell }   \big( T_{v}^2  + \alpha_{v}^2\big). \label{ELERbtzU1}
}
The background also features additional left and right-moving $U(1)$ charges that read
\eq{
Q_L = - p \sqrt{k k_4} \a_u, \qquad Q_R = p \sqrt{k k_4} \a_{v}, \label{qLqRbtzU1}
}
where $p k_4$ is the total level of two $U(1)$ affine symmetry algebras in the dual CFT. The derivation of these charges, and their role in the thermodynamics of the undeformed background \eqref{btzU1}, is discussed in appendix \ref{ap:btzU1}. In that appendix we also provide a holographic interpretation of the charged BTZ black hole in the three-dimensional reduced theory.\footnote{From a three-dimensional perspective, the background \eqref{btzU1} can be obtained from \eqref{btz} by the gauge transformations $A^{(1)} \to A^{(1)}  -  \frac{\ell}{\ell_4} ( \a_u du + \a_{v} d{v})$, $A^{(2)} \to  A^{(2)} + \frac{\ell}{\ell_4} (\a_u du - \a_{v} d{v})$, and $B \to B + \ell^2 \a_u \a_{v} d{v} \we du$.} 

The charged BTZ black hole \eqref{btzU1} describes thermal states in the dual CFT with energies $\Q_{u,v}$ and $U(1)$ charges $Q_{L,R}$. In order to construct backgrounds dual to charged thermal states in single-trace $T\bar T+J\bar T+T\bar J$-deformed CFTs, we perform the sequence of TsT transformations~\eqref{tstrecipe3} on the background described by \eqref{10dbackground} and \eqref{btzU1}. In terms of the $\l_i$ parameters defined in \eqref{dictionarylambdas}, the  charged tri-TsT black hole is characterized by the following three-dimensional fields
\eq{
ds^2  &=  \frac{ \ell^2 dr^2}{4(r^2 - 4 T_u^2 T_v^2)}  \! + \! \ell^2 h \bigg\{ \! \big[ r + 2 ( \a_u \!-\! \l_- \a_u^2 \!-\! \l_- T_u^2 )  ( \a_v \!-\! \l_+ \a_v^2 \!-\! \l_+  T_v^2) \big] du dv  - \frac{1}{h e^\omega} \big(A^{(1)}\big)^2  \notag \\
& \hspace{15pt} + (\a_u^2 + T_u^2 ) \big[ (1 - \a_v \l_+)^2 + \l_+^2 T_v^2 \big]  du^2 +  (\a_v^2 + T_v^2) \big[(1 - \a_u \l_-)^2 + \l_-^2 T_u^2 \big] dv^2  \bigg\}, \notag \\
B &= \ell^2 h \Big[ \frac{r}{2} + \a_u \a_v -  \a_v \l_- (\a_u^2 + T_u^2) -  \a_u \l_+ (\a_v^2 + T_v^2) + \l_0 (\a_u^2 + T_u^2)(\a_v^2 + T_v^2) \Big] d{v} \we du, \notag \\
\begin{split}
A^{(1)} &= -\frac{\ell h}{2 \ell_4} e^\omega \Big\{ \big[ \l_+ r +2 \a_u (1 - \a_v \l_+) -2 \l_0 (  \a_v - \l_+ \a_v^2 - \l_+  T_v^2 )(\a_u^2 + T_u^2 )  \big] du  \label{tstbtz} \\
& \hspace{15pt} + \big[  \l_- r +2 \a_v (1 - \a_u \l_-) -2 \l_0  ( \a_u - \l_- \a_u^2 -  \l_- T_u^2  )( \a_u^2 + T_u^2)  \big] d{v} \Big\},
\end{split} \\
A^{(2)} & = \frac{\ell h}{2 \ell_4} \Big\{\big[ \l_+ r +2 \a_u (1 - \a_v \l_+) +2 (\a_v \l_0 - \l_-)  (\a_u^2 + T_u^2 )  \big] du \notag \\
& \hspace{15pt} - \big[ \l_- r +2 \a_v (1 -  \a_u \l_-) +2 ( \a_u \l_0 - \l_+)  (\a_v^2 + T_v^2)  \big] d{v} \Big\} , \notag\\
e^{2\Phi} &= \frac{k h}{ \eta p}, \qquad \qquad  e^{-\omega} = h\big[\l_0 r + (1 - \a_u \a_v \l_0)^2  +  (\a_v^2T_u^2 + \a_u^2  T_{v}^2  +T_u^2T_{v}^2) \l_0^2\big],\notag
}  
where the quantization parameter $\eta$ and the flow function $h$ read
 \eq{
   \eta^{-1} &= (1 - \a_u \l_-)(1 - \a_v \l_+) - \big[\a_u \l_+ - \l_0 (\a_u^2 + T_u^2) \big] \big[ \a_v \l_- - \l_0 (\a_v^2 + T_{v}^2) \big], \label{etadef} \\
  \begin{split}
h^{-1} &= (1 - \a_u \l_- - \a_v \l_+ + \a_u \a_v \l_0)^2 + (\l_- - \a_v \l_0)^2 T_u^2 + 
  (\l_+ - \a_u \l_0)^2 T_{v}^2\\
  & \hspace{15pt} +   (\l_0 - \l_+ \l_-)r +  \l_0^2 T_u^2 T_{v}^2. \label{hdef}
  \end{split}
}
As described earlier, the value of $\eta$ is determined by requiring that the electric charge is preserved after the TsT transformations such that \eqref{QeQmBTZ} still holds. Requiring the dilaton to be real then implies 
\eq{
\eta^{-1} > 0. \label{etaconstraint}
}
This condition imposes physical constraints on the phase space variables which translate into constraints on the thermodynamic potentials of the dual field theory. 

Note that the charged \eqref{tstbtz} and neutral \eqref{neutraltstbtz} tri-TsT black holes share the same asymptotic behavior, and hence describe states in the same $T \bar T + J\bar T + T\bar J$-deformed CFT. The charged black holes are characterized by additional phase space variables which, as described in section \ref{se:tstthermo}, lead to nonvanishing values of the deformed $U(1)$ charges.

The charged tri-TsT black hole \eqref{tstbtz} is a 7-parameter solution to the supergravity equations of motion that describes a variety of backgrounds that are dual to thermal states in two-dimensional CFTs deformed by single-trace irrelevant operators. For example, when $\l_\pm$ vanish, the black hole \eqref{tstbtz} describes a generalization of the finite-temperature backgrounds dual to thermal states in $T\bar T$-deformed CFTs described in \cite{Apolo:2019zai}. Similarly, when $\l_0$ and $\l_-$ vanish, the black hole \eqref{tstbtz} describes thermal states in single-trace $J\bar T$-deformed CFTs with extra $U(1)$ charges, generalizing the backgrounds constructed in \cite{Apolo:2019yfj}.

In this paper we will focus on the charged tri-TsT black holes with arbitrary values of the deformation and phase space parameters. As described in the next section, the background \eqref{tstbtz} can also be used to construct the spacetime dual to the NS vacuum by analytic continuation of the phase space variables. The geometric properties of the charged tri-TsT black holes, including their curvature and the asymptotic behavior of their background fields, are considered in section~\ref{se:tstbtzproperties}. The gravitational charges, entropy, and thermodynamics of these backgrounds are studied in detail in section \ref{se:thermodynamics}, where we also show that the neutral and charged tri-TsT black holes describe thermal states in $T \bar T + J\bar T + T\bar J$-deformed CFTs.


\subsection{The Neveu-Schwarz vacuum} \label{se:NSvacuum}

In this section we construct the background dual to the NS vacuum or ground state in $T\bar T + J\bar T + T\bar J$-deformed CFTs, i.e.~the analog of global AdS$_3$ in the AdS$_3$/CFT$_2$ correspondence. Generalizing the method used in \cite{Apolo:2019zai} for $T\bar T$ deformations, our strategy is to find a unique point in the parameter space of charged tri-TsT backgrounds \eqref{tstbtz} where the solution is nonrotating and smooth everywhere without the appearance of a horizon or singularities. 

We begin by noting that when the $T_u$ and $T_v$ parameters of the TsT backgrounds~\eqref{tstbtz} are real, the solutions  describe rotating black holes with at event horizon at $r = 2 T_u T_v$. In order to construct the NS vacuum, we must first analytically continue $T_u$ and $T_v$ to purely imaginary values, and then choose phase space variables that guarantee a smooth background. Let us perform the following coordinate transformation
\eq{ 
r = 2 |T_uT_v| (1 + 2 \rho^2),\quad \rho\in[0,\infty).
}  
For generic values of the phase space parameters, the aforementioned analytic continuation introduces conical singularities as well as divergences in the norms of the gauge fields at the origin $\rho=0$. In three dimensions, conical singularities can be interpreted as the result of adding point particles whose masses are proportional to the deficit angle. However, we are interested on the empty (vacuum) spacetime which is not rotating, and where both the metric and the gauge fields are regular at the origin. These conditions can be used to determine the values of the phase space variables $T_u$, $T_v$, $\a_u$ and $\a_v$. 

The NS vacuum is required to satisfy the following conditions
\eq{
g_{t\vp} |_{\rho=0} &= 0,   \qquad \frac{g_{\vp\vp}}{g_{\rho\rho}} \Big|_{\rho\to0} = \rho^2, \qquad  A^{(1)}_\vp |_{\rho=0} = A^{(2)}_\vp |_{\rho=0} = 0. \label{smoothcondition3}
} 
where $(u,v) = (\vp + t, \vp - t)$. The first condition turns off the angular momentum so that the solution is not rotating, the second condition translates into the absence of conical defects, and the third one imposes a trivial holonomy on the gauge fields. The last condition is equivalent to requiring the norm of the gauge fields to be finite at the origin. Imposing the above conditions we obtain 
\eq{
\begin{gathered}
T_u = T_v = i r_0, \qquad r_0   = \frac{1}{2} - \frac{\l_0  \big[2-\l_0 -\sqrt{4 - 4\l_0 + (\l_+ + \l_-)^2 }\, \big] }{2 \big [ (\l_+ + \l_-)^2 - \l^2_0 \big]}, \label{r0sol}  \\
\a_u = \a_{v} =\frac{ (\l_+ + \l_-) \big[2- \lambda_0- \sqrt{4 - 4\l_0 + (\l_+ + \l_-)^2 } \,\big]}{ 2\big [ (\l_+ + \l_-)^2 - \l^2_0 \big]}.
\end{gathered}
}
Note that in order for the background to remain real the values of $\a_u$ and $\a_v$ should also be real. As a result, the deformation parameters must satisfy
\eq{
\l_0 - \tfrac{1}{4}(\l_+ + \l_-)^2 \le 1.  \label{NSconstraint}
}
This condition \eqref{NSconstraint} generalizes a similar condition found for the deformation parameter in $T\bar T$-deformed backgrounds~\cite{Apolo:2019zai} and, as described in section \ref{se:NSenergies}, it also guarantees that the energy of the vacuum is real. 

The phase space parameters \eqref{r0sol} determine the bulk solution that is dual to the NS vacuum in the $T\bar T + J\bar T + T\bar J$-deformed CFT.  In particular, setting $\lambda_\pm =  0$ yields $\a_u = \a_v = 0$ and the corresponding background reduces to the NS vacuum in single-trace $T\bar T$-deformed CFTs~\cite{Apolo:2019zai}. For general $T\bar T+J\bar T+T\bar J$ deformations, the spectral flow parameters $\a_u$ and $\a_v$ are generically nonvanishing. As shown in \eqref{se:NSenergies}, the choice \eqref{r0sol} guarantees that the undeformed $U(1)$ charges of the analytically continued background vanish, which is consistent with the spectrum of the $T\bar T+J\bar T+T\bar J$-deformed CFTs. 

Let us further elaborate on this point with another special example: the $J\bar T$ deformation where $\l_0 = \l_- = 0$. From \eqref{r0sol} we obtain $ r_0 = 1/2$, which is the same value of $r_0$ found in global AdS$_3$. On the other hand, the spectral flow parameters are given by 
\eq{
J\bar T: \qquad  \alpha_u=\alpha_v =  \frac{ 2-\sqrt{4+\smash[b]{\l_+^2}} }{2 \l_+}. \label{JTbar}
} 
Hence the NS vacuum for $J\bar T$-deformed CFTs is dual to the following smooth solution
\eqsp{
ds^2_{0} &= \ell^2 \bigg\{ \frac{d\rho^2}{\rho^2 +1} + \rho^2 (\rho^2 + 1) du^2 - \frac{1}{4} \Big [ dv - \big ( 1  + \sqrt{4 + \smash[b]{\l_+^2}} \, \rho^2 \big) du \Big]^2 \bigg\}, \\
B_0 &= \frac{\ell^2}{2} \bigg( \rho^2  +  \frac{2\sqrt{4 + \smash[b]{\l_+^2}} - 4 }{\l_+^2}\bigg) dv \we du, \qquad e^{2\Phi_0} = \frac{k \sqrt{4 + \smash[b]{\l_+^2}}}{2p}, \qquad \om_0 = 0, \\
A_0^{(1)} &= - A_0^{(2)} = -\frac{\ell}{2\ell_4} \Big[ \l_+ \rho^2 du + \frac{\sqrt{4 + \smash[b]{\l_+^2}} - 2}{\l_+} \big(du - dv \big)  \Big]. \label{JTbarvacuum}
}
The non-vanishing spectral flow parameters \eqref{JTbar} play a crucial role in the construction of the NS vacuum for the $J\bar T$ deformation, since without them the TsT transformation of global AdS$_3$~\cite{Apolo:2019yfj} would not satisfy all of the requirements \eqref{smoothcondition3} that guarantee a smooth geometry. It is interesting to note that the three-dimensional metric in \eqref{JTbarvacuum} corresponds to a timelike warped AdS spacetime, in contrast to the Ramond vacuum (null warped AdS), and a generic black hole solution (spacelike warped AdS)~\cite{Anninos:2008fx}.

We also note that \eqref{r0sol} implies that the phase space variables depend on the deformation parameters after the TsT transformation. This means that the phase space variables $T_u$, $T_v$, $\a_u$, and $\a_v$ should not be interpreted as the same variables parametrizing the space of solutions before the deformation, except in the limit where all of the deformation parameters vanish. A similar behavior is observed in the backgrounds dual to both single and double-trace $T\bar T$ and $J\bar T$ deformations~\cite{Apolo:2019yfj,Apolo:2019zai,Bzowski:2018pcy,Guica:2019nzm}. Nevertheless, we note that letting the phase space parameters depend on $\l_0$, $\l_+$, and $\l_-$ does not affect the matching of the worldsheet spectrum or the microcanonical entropy to the spectrum and entropy of $T\bar T + J\bar T + T\bar J$-deformed CFTs described in sections~\ref{se:spectrum} and~\ref{se:thermodynamics}.

To summarize, the values of the phase space parameters given in \eqref{r0sol} determine the background dual to the NS vacuum in single-trace $T\bar T + J\bar T + T\bar J$-deformed CFTs, the latter of which is characterized by smooth background fields free of conical defects or singular gauge fields. We will derive the gravitational charges of this background in section \ref{se:NSenergies}, where we will show that these charges match the energies and $U(1)$ charges of the vacuum in single-trace $T\bar T + J\bar T + T\bar J$-deformed CFTs. 


\section{Geometric properties} \label{se:tstbtzproperties}

In this section we describe geometric properties of the charged tri-TsT black hole~\eqref{tstbtz} in different regions of parameter space. In particular, we describe the behavior of the Ricci scalar in the Einstein frame, the potential existence of CTCs and curvature singularities, and the asymptotic behavior of the background fields.

\subsection{Singularities and CTCs} \label{se:singctcts}

The three-dimensional Ricci scalar of the tri-TsT black holes \eqref{tstbtz} can be written in the string and Einstein frames as\footnote{In three dimensions the string $(G_{\mu\nu})$ and Einstein frame $(g_{\mu\nu})$ metrics are related by $g_{\mu\nu} = e^{-4\Phi-\om} G_{\mu\nu}$.}
\eq{
R_{(s)} = \frac{f^{(s)}_{\textrm{reg}}}{\ell^2 h^{-1} (e^{-\om} h^{-1})^3}, \qquad R_{(E)} = e^{4\Phi}  \frac{f^{(E)}_{\textrm{reg}}}{\ell^2 h^{-1} (e^{-\om} h^{-1})^3}, \label{ricci}
}
where $f^{(s)}_{\textrm{reg}}$ and $f^{(E)}_{\textrm{reg}}$ are regular but cumbersome functions of the radial coordinate, the deformation parameters, and the phase space variables. For example, for the $T\bar T$ deformation with $\a_u = \a_v = 0$, the $f^{(s)}_{\textrm{reg}}$ and $f^{(E)}_{\textrm{reg}}$ functions are given by
\eq{
\!\!\!\! \! f^{(s)}_{\textrm{reg}}\big|_{\l_{\pm} = \a_u = \a_v = 0} & = - 2h^{-2} \big[  3 - 22 \l_0^2 T_u^2 T_{v}^2 + 3 \l_0^4 T_u^4 T_{v}^4 - 4 \l_0 r (1 + \l_0^2 T_u^2 T_{v}^2)\big], \\
\!\!\!\! \!  f^{(E)}_{\textrm{reg}}\big|_{\l_{\pm} = \a_u = \a_v  = 0} & = - 2h^{-2} \big[  3 + 26  \l_0^2 T_u^2 T_{v}^2 + 3 \l_0^4 T_u^4 T_{v}^4 + 12 \l_0 r (1 + \l_0^2 T_u^2 T_{v}^2) + 4 \l_0^2 r^2 \big]. 
}
Note that for generic values of the deformation parameters the $f^{(s)}_{\textrm{reg}}$ and $f^{(E)}_{\textrm{reg}}$ functions do not feature inverse powers of the flow function $h$, in contrast to the special case considered above. In what follows we will be concerned with the factors of $e^{\om}$ and $h$ in the Ricci scalar, but we will have more to say about $f_{\textrm{reg}}^{(E)}$ in section \ref{se:signricci}.

The Ricci scalar in the string and Einstein frames \eqref{ricci} diverges at the zeroes of the denominator, namely when $e^{-\om}h^{-2} = 0$. From the expressions for $e^{-\om}$ and $h^{-1}$ given in \eqref{tstbtz} and \eqref{hdef}, we see that these functions are strictly positive for positive values of the radial coordinate provided that
\eq{
\l_0 > 0 \quad \textrm{and} \quad  \l_0 - \l_+ \l_- > 0. \label{singcondition1}
}
Consequently, the condition \eqref{singcondition1} guarantees that the tri-TsT black holes  \eqref{neutraltstbtz} and \eqref{tstbtz} are free of curvature singularities.

When $\l_0 < 0$ the equation $h^{-1} e^{-\om} = 0$ is satisfied at a positive value of the radial coordinate, namely at 
\eq{
r^*_\omega = \frac{(1 - \a_u \a_v \l_0)^2 + \a_u^2 \l_0^2 T_{v}^2 + \a_v^2 \l_0^2 T_u^2 + \l_0^2 T_u^2 T_{v}^2}{|\l_0 |}, \label{r1sing}
}
where the dilaton $\om$ becomes singular and the Ricci scalar (in either frame) diverges. On the other hand, when $\l_0 - \l_+ \l_-  < 0$, the equation $h^{-1} = 0$ has a solution at a  positive value of the radial coordinate that is given by
  \eq{
  \!\!\!  r^*_h  = \frac{ (1 - \a_u \l_-\! - \a_v \l_+ + \a_u \a_v \l_0)^2 + (\l_- \! - \a_v \l_0)^2 T_u^2 + 
  (\l_+ - \a_u \l_0)^2 T_{v}^2  +  \l_0^2 T_u^2 T_{v}^2}{|\l_0 - \l_+ \l_-|}, \label{r2sing}
  }
where the Ricci scalar and both of the dilatons diverge. In particular, we have the following special cases:
  \begin{itemize}[leftmargin=\parindent]
\item when $\l_{\pm} = 0$ and either $\a_u$ or $\a_v$ vanish, the singularities at $r_\om^*$ and $r_h^*$ merge as in the uncharged $T\bar T$-deformed background discussed in \cite{Giveon:2017nie,Apolo:2019zai} (see fig.~\ref{fig:ttbarsing1});
\item since $e^{-\omega} h^{-1} = 1$ when $\l_0$ vanishes,  only the second singularity  \eqref{r2sing} at $r_h^*$ is present in the case where $\l_0 = 0$ and $\l_+ \l_- > 0$;
\item there are no curvature singularities in the $J\bar T$ or $T\bar J$ cases where only $\l_+$ or $\l_-$ are nonvanishing.
\end{itemize}

\paragraph{CTCs} Let us now discuss the possiblity of having closed timelike curves (CTCs) in the charged tri-TsT black holes \eqref{tstbtz}. CTCs are possible in this class of spacetimes due to the identification of coordinates \eqref{coordsim3}. In order to see this, we first note that the norm of the spatial circle \eqref{coordsim3} generated by $\p_\vp = \tfrac{1}{2}(\p_u + \p_{v})$ is given in the Einstein frame by
 \eq{
\! \p_\vp \cdot \p_\vp = g_{\vp \vp} = \frac{\eta^2 \ell^2 p^2}{k^2} \Big\{ \big[\l_0 - \frac{1}{4}(\l_+ + \l_-)^2 \big] (r^2 - 4 T_u^2 T_{v}^2) + r_u r_{v}  r + r_u^2 T_{v}^2 + r_{v}^2 T_u^2 \Big\}, \label{norm}
}
where $r_u$ and $r_v$ are defined by
\eq{
r_u = 1 - \a_u( \l_+ + \l_-) + \l_0 (\a_u^2 + T_u^2), \qquad r_{v} = 1 - \a_v ( \l_+ + \l_-) + \l_0 (\a_v^2 + T_{v}^2).
}
Thus, the tri-TsT backgrounds \eqref{tstbtz} are free of CTCs when the deformation parameters satisfy
\eq{
\l_0 - \frac{1}{4} ( \l_+ + \l_-)^2 \ge 0, \label{ctccondition}
}
since this guarantees that $r_u > 0$ and $r_{v} > 0$ such that each term in \eqref{norm} is positive outside of the horizon at $r_h = 2T_u T_v$. In particular, when restricted to the Ramond vacuum, the constraint \eqref{ctccondition} reduces to that of \cite{Chakraborty:2019mdf}.

Since \eqref{ctccondition} is stronger than \eqref{singcondition1}, we find that \eqref{ctccondition}  guarantees the existence of black hole solutions without CTCs or curvature singularities outside of the event horizon. Thus, the space of tri-TsT backgrounds \eqref{tstbtz} contains a real ground state and black holes free of pathologies for the following choice of deformation parameters
\eq{
0\le \l_0 - \frac{1}{4}(\l_+ + \l_-)^2 \le 1. \label{fullconstraint}
}
The condition \eqref{fullconstraint} is also reflected on the perturbative spectrum of winding strings and, consequently, on the spectrum of single-trace $T\bar T + J\bar T + T\bar J$-deformed CFTs. Indeed, as we will see in sections \ref{se:matchingspectrum} and \ref{se:NSenergies}, the first inequality in \eqref{fullconstraint} guarantees a real spectrum for generic values of the energy above the Ramond vacuum (i.e.~for real and nonvanishing values of $T_u$ and $T_v$); while the second inequality in \eqref{fullconstraint} is necessary for the energy of the vacuum to be real. 

When the deformation parameters of the charged tri-TsT black hole satisfy
\eq{
\l_0 - \frac{1}{4}(\l_+ + \l_-)^2 < 0, \label{CTCregion} 
}
closed timelike curves appear in the region $r > r_c$ where $r_c$ is given by
\eq{
r_c = \frac{r_u r_{v} + \sqrt{\big(r_u^2 + \big|4 \l_0 - (\l_+ + \l_-)^2\big| T_u^2\big)\big(r_{v}^2 + \big|4 \l_0 - (\l_+ + \l_-)^2\big| T_{v}^2\big)}}{2\big|\l_0 - \tfrac{1}{4}(\l_+ + \l_-)^2\big|}. \label{rctc}
}
It is useful to compare the locations of the CTCs ($r_c$), the horizon ($r_h$), and the curvature singularities ($r_h^*$, $r_\om^*$). From \eqref{r1sing}, \eqref{r2sing}, and \eqref{rctc}, it is not difficult to show that 
\eq{
 r_\om^* \ge r_c \ge r_h ,  \qquad r_h^* \ge r_c \ge r_h.  \label{relpos}
}
As a result, we find that CTCs are encountered before or at the location of either one of the singularities and that all of these pathologies are located in the exterior of the black hole. Note that the relative positions of the curvature singularities is the same in the string and Einstein frames but depends on the choice of parameters. For example, for the Ramond vacuum where $T_u = T_v = \a_u = \a_v = 0$ we find that $r_\om^* \ge r_h^* \ge r_c> 0$ if $\l_+ \l_- \ge 0$ and $r_h^* > r_\om^* > r_c > 0$ if $\l_+ \l_- < 0$.

In the region of parameter space  \eqref{CTCregion}, the perturbative string spectrum becomes complex at high energies (see section~\ref{se:spectrum}), which suggests the necessity of a cutoff in the bulk at large radius. It would be interesting to directly relate the maximum value of the real energy, a cutoff in the bulk, and the location of the CTCs. 


\subsection{A region of positive curvature} \label{se:signricci}
Let us now consider the curvature of the charged tri-TsT black holes in more detail. As described in section \ref{se:singctcts}, the Ricci scalar in the Einstein frame can be written as $R_{(E)} = \ell^{-2} e^{4\Phi}  e^{3\om} h^4 f^{(E)}_{\textrm{reg}}$ where $f^{(E)}_{\textrm{reg}}$ is a quartic polynomial on the radial coordinate that depends on the deformation parameters and the phase space variables. Interestingly, we find that there are regions where the curvature of the spacetime becomes positive before we reach any singularities or CTCs. We will illustrate the emergence of these regions of positive curvature quantitatively in the case of the $T\bar T$ deformation, as well as qualitatively in the most general case where none of the deformation parameters or phase space variables vanish. We note that the string frame Ricci scalar and other curvature invariants have also been studied for the Ramond vacuum in the pure $T\bar T$ case with $\l_0 > 0$ in \cite{Giribet:2021cyy}.

\subsubsection{The $\mathbf{T\bar T}$ case}

Let us first consider the $\a_u = \a_v = 0$ case where the Ricci scalar is given by
\eq{
R_{(E)} = - 2 \ell^{-2} e^{4\Phi} h^{2} \big[  3 + 26  \l_0^2 T_u^2 T_{v}^2 + 3 \l_0^4 T_u^4 T_{v}^4 + 12 \l_0 r (1 + \l_0^2 T_u^2 T_{v}^2) + 4 \l_0^2 r^2 \big].\label{ricciTTbar}
}
We see that the Ricci scalar vanishes asymptotically such that
\eq{
R_{(E)} = -\frac{8 k^2 }{( \ell \eta p)^2} \frac{1}{\l_0^2 r^2} + \O(r^{-3}).
} 
When $\l_0 > 0$ the factor in the square bracket of \eqref{ricciTTbar} is always positive outside the horizon which results in a Ricci scalar that is always negative. The negative curvature and asymptotic behavior of the $T\bar T$-deformed backgrounds persists when we turn on the $\a_u$ and $\a_v$ parameters as illustrated in fig.~\ref{fig:ttbarhealthy}. 

Interestingly, when $\l_0 < 0$ the Ricci scalar \eqref{ricciTTbar} has positive roots at
\eq{
r_{\pm} = \frac{3 + 3 \l_0^2 T_u^2 T_{v}^2 \pm \sqrt{6 - 8 \l_0^2 T_u^2 T_{v}^2 + 6 \l_0^4 T_u^4 T_{v}^4}}{2|\l_0|}.
}
Since the curvature is negative in the IR, the Ricci scalar becomes positive in the region $r_-<r<r_+$. Furthermore, the first root at $r_{-}$ is located before the region with CTCs and singularities, that is 
\eq{
r_{-} < r_c = r_h^* = r_\om^*  = \frac{1 + \l_0^2 T_u^2 T_{v}^2}{|\l_0|}.
}
This means that the bulk solution interpolates between an AdS$_3$ region in the IR and a region of positive curvature at $r_- < r < r_c$ before reaching the threshold of CTCs and the location of the singularity (see fig.~\ref{fig:ttbarsing1}). This is similar to the interpolating two and three-dimensional geometries described in \cite{Anninos:2017hhn,Gorbenko:2018oov}. 

Note that since $r_-$ scales inversely with $|\l_0|$,  the region of positive curvature is pushed away to infinity as we take the limit $|\l_0|\to 0$. In addition, we note that the region of positive curvature can be made arbitrarily large by making $|\l_0|$ small, namely
\eq{
r_c - r_- \sim \frac{\sqrt{6} - 1}{2 |\l_0|} + \O(\l_0),
} 
such that the proper distance in the radial direction scales as $-\log |\l_0|$.

The existence of a smooth region with positive curvature outside the horizon is  a generic feature of $T\bar T$-deformed backgrounds with $\l_0 < 0$ and arbitrary values of the $\a_u$ and $\a_v$ parameters. In this more general case, the behavior of the Ricci scalar when only $\a_u$ or $\a_v$ vanishes is shown in fig.~\ref{fig:ttbarsing2} while its behavior when neither of them vanish is illustrated in fig.~\ref{fig:ttbarsing3}.

\begin{figure}[H]
{\centering
\subfloat[$\l_0 > 0$ \label{fig:ttbarhealthy}]{
 \includegraphics[scale=0.3056]{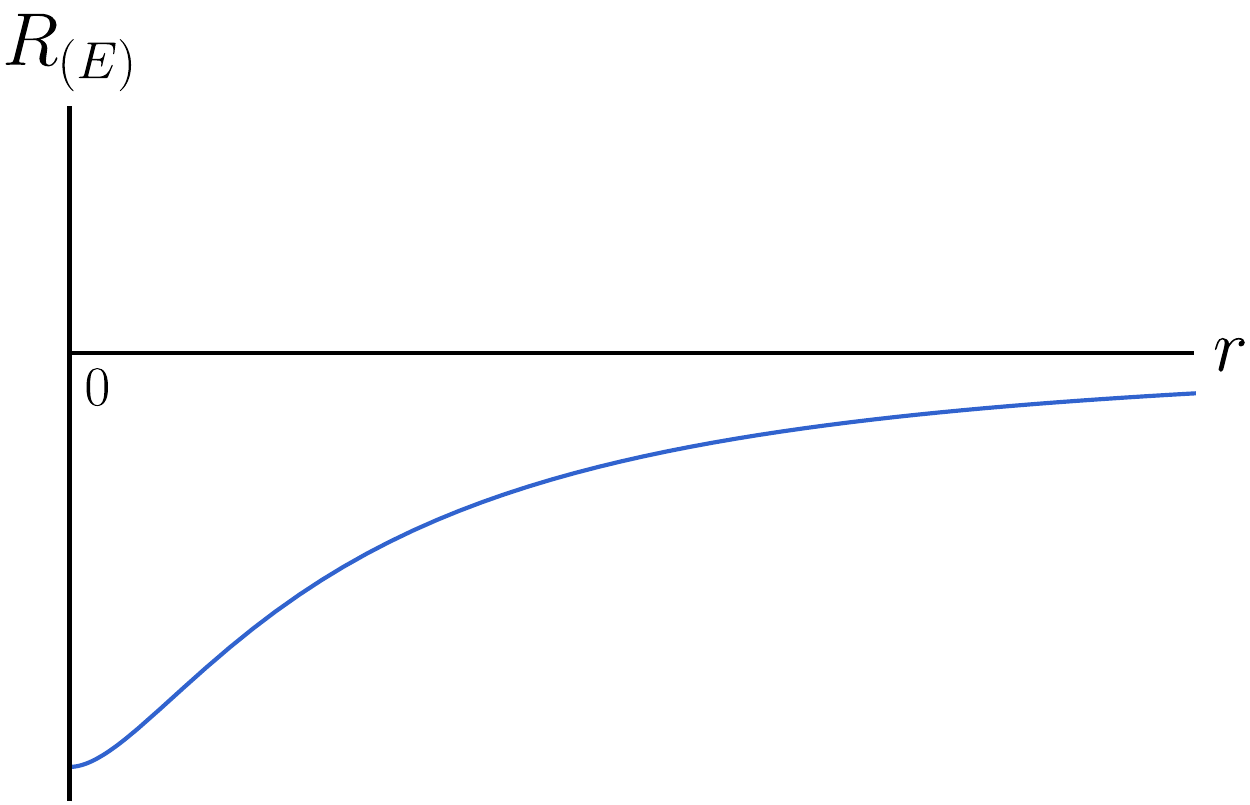}
}\hfill
\subfloat[$\l_0 < 0, \,\,\,\,  \a_u = \a_{v} = 0 $ \label{fig:ttbarsing1}]{
 \includegraphics[scale=0.3056]{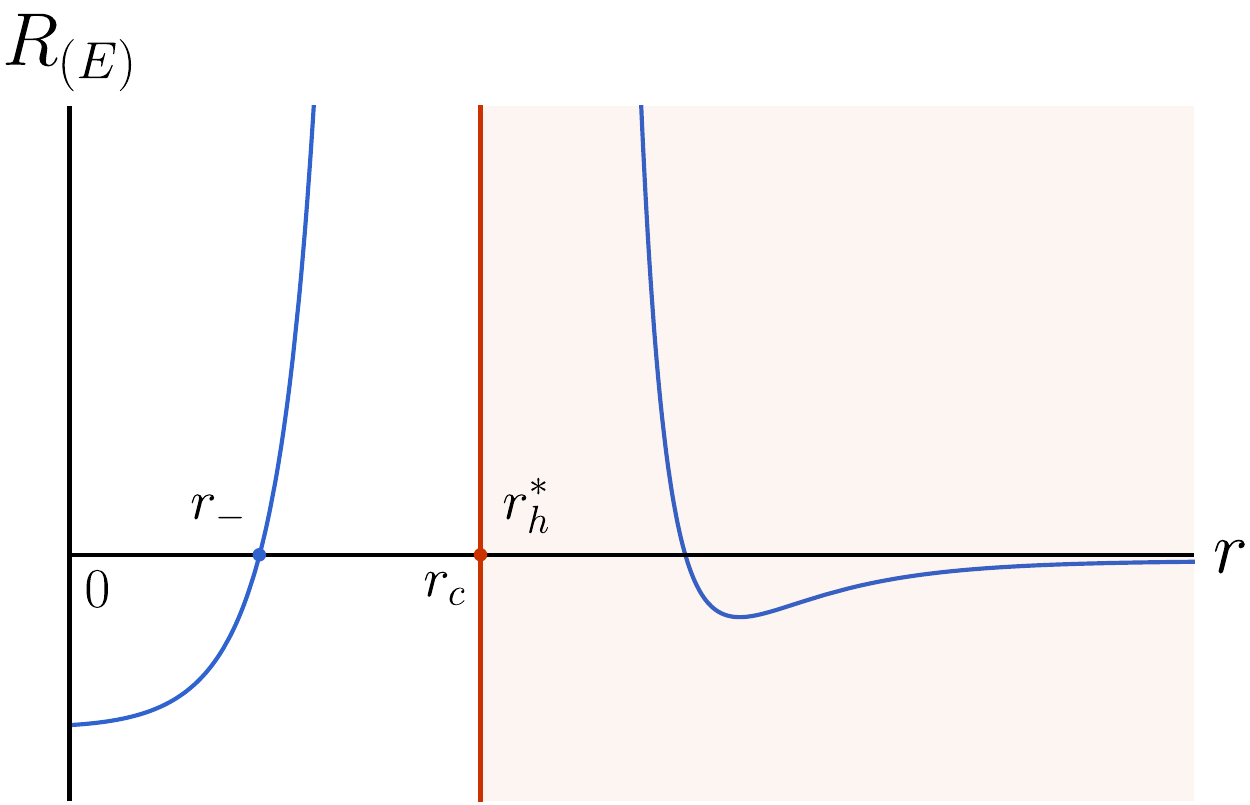}
}
}
{\centering
\subfloat[$\l_0 < 0, \,\,\,\, \a_u = 0 \,\,\,\, \textrm{or} \,\,\,\, \a_{v} = 0$ \label{fig:ttbarsing2}]{
 \includegraphics[scale=0.3056]{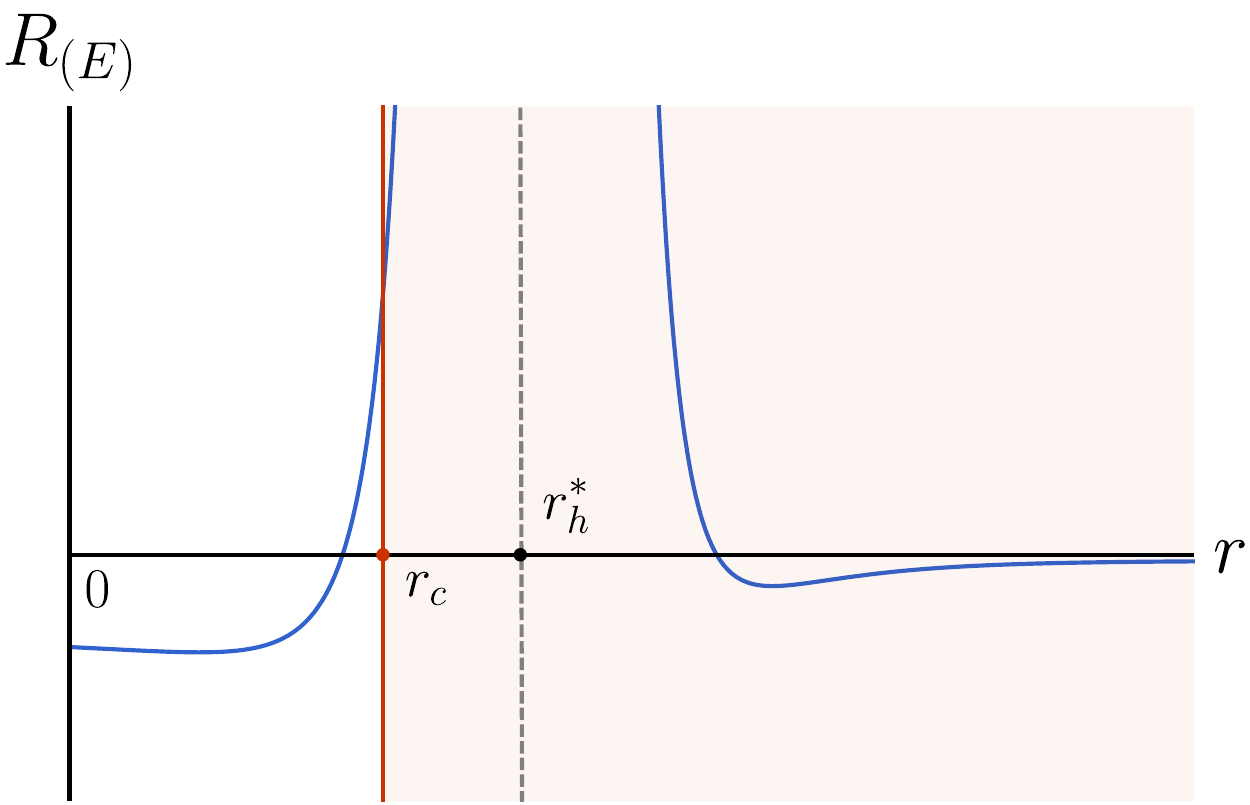}
}\hfill
\subfloat[$\l_0 < 0, \,\,\,\,  \a_u \ne 0, \,\,\,\, \a_{v} \ne 0 $ \label{fig:ttbarsing3}]{
 \includegraphics[scale=0.3056]{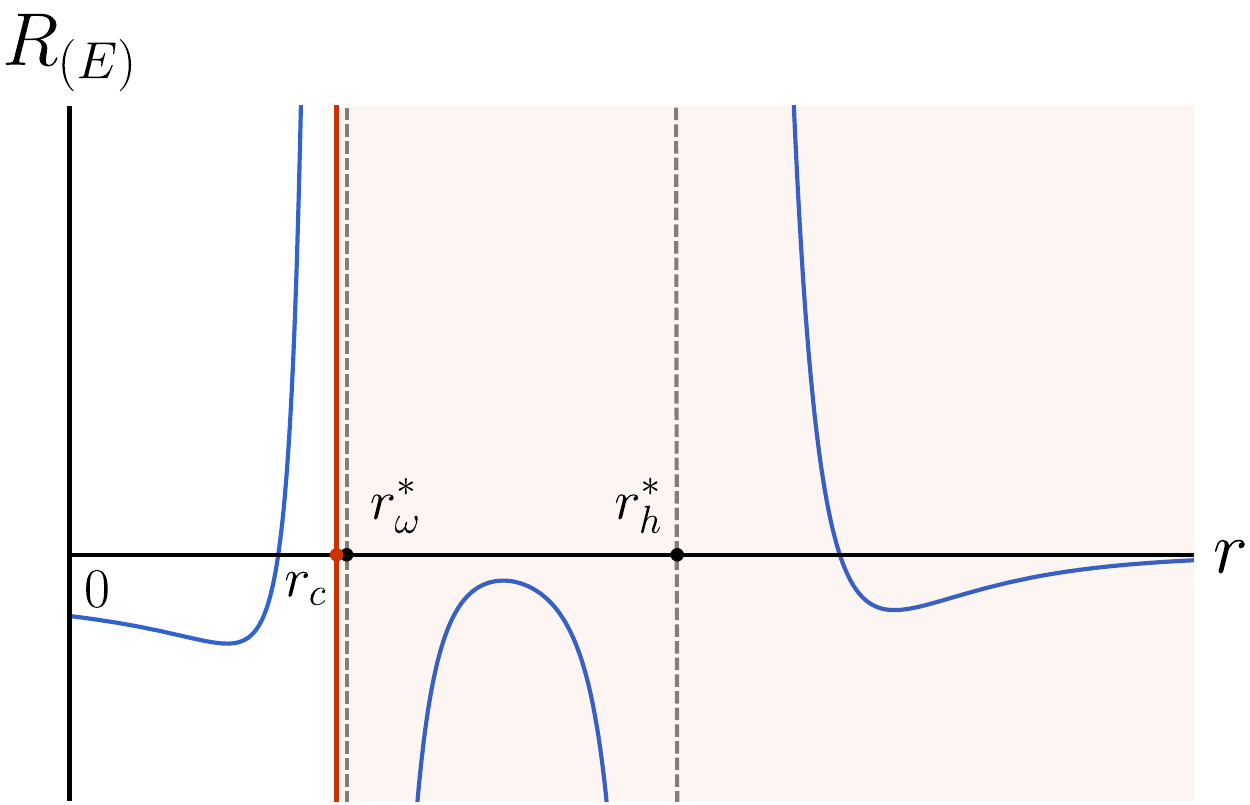}
 }
 }
\caption{The Ricci scalar for $T\bar T$ backgrounds with generic values of the $T_u$ and $T_v$ parameters and different choices of $\a_u$ and $\a_v$ (not drawn to scale). When $0 < \l_0 < 1$, the curvature is negative everywhere and the spacetime is smooth and free of CTCs. When $\l_0 < 0$ the background has a region of positive curvature that is found before any CTCs (the shaded region) or curvature singularities (dashed lines) are reached.}
\label{fig2}
\end{figure}


\subsubsection{The general case}\label{se:generalricciTTbar}

Let us now consider the most general class of charged tri-TsT black holes where none of the deformation parameters or phase space variables vanish. For these backgrounds, the Ricci scalar in the Einstein frame is asymptotically flat but approaches zero from above or below depending on the choice of the deformation parameters
\eq{
R_{(E)} = - \frac{8 k^2 }{( \ell \eta p)^2}  \frac{1}{\l_0 (\l_0 - \l_+ \l_-)r^2}  +\O(r^{-3}).  \label{Rgeneral}
}
We observe from \eqref{singcondition1} and \eqref{Rgeneral} that $R_{(E)} \to 0^+$ when the black hole has one singularity while $R_{(E)} \to 0^-$ when the black hole has two or no singularities. 

Whenever a singularity is present, namely when $\l _0< 0$ and/or $\l_0 < \l_+ \l_-$, there is always at least one region of the spacetime with positive curvature that is found before reaching the first singularity, as illustrated in figs.~\ref{fig:sing2} and~\ref{fig:sing3}. Depending on the parameters, the region of positive curvature may be found before encountering any CTCs. When the spacetime has one singularity there is an additional region of positive curvature found after the singularity, see fig.~\ref{fig:sing2}. Furthermore, when there are two singularities it is possible to find yet another region of positive curvature that lies between $r_\om^*$ and $r_h^*$, as shown in fig.~\ref{fig:sing3}.

\begin{figure}[H]
{\centering 
\subfloat[$\l_0 - \tfrac{1}{4}(\l_+ + \l_-)^2 > 0$ \label{fig:healthy}]{
 \includegraphics[scale=0.3056]{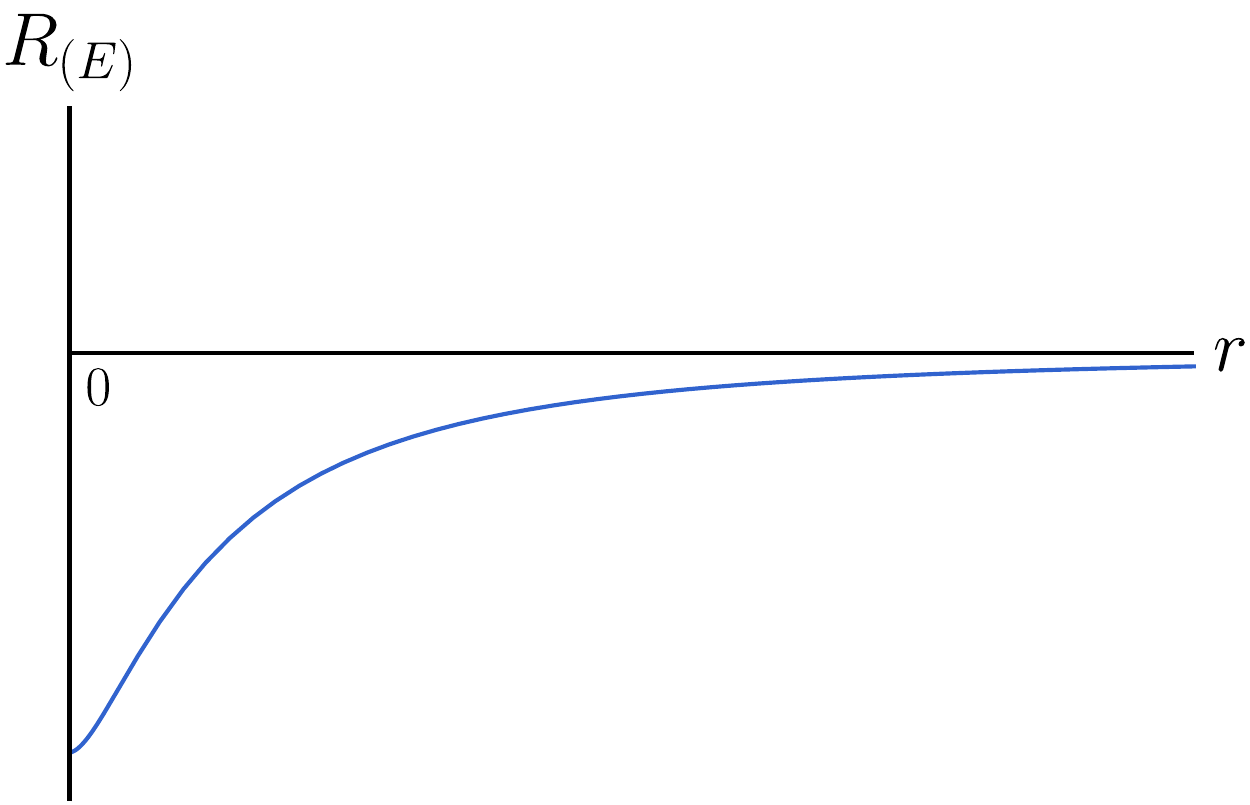}
}\hfill
\subfloat[$0 < \l_+ \l_- < \l_0 < \tfrac{1}{4}(\l_+ + \l_-)^2$ \label{fig:sing1}]{
 \includegraphics[scale=0.3056]{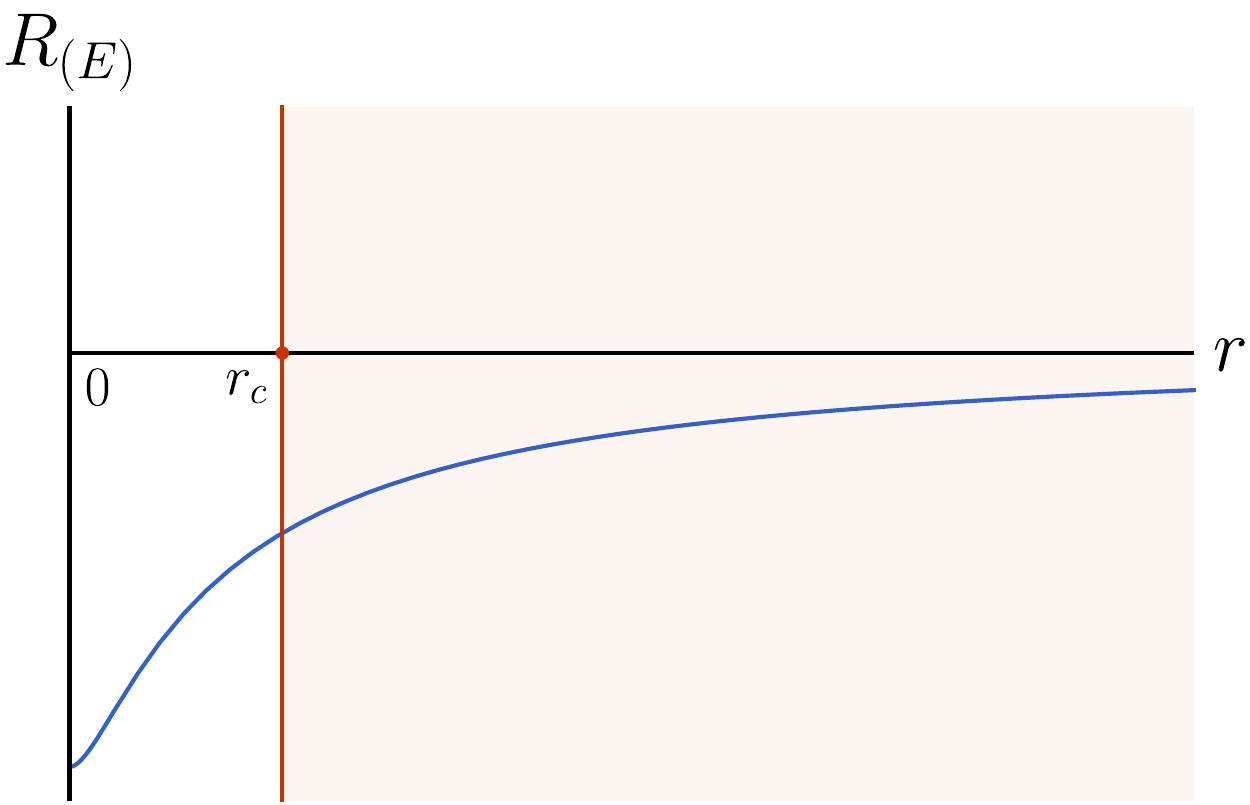}
}\\
\subfloat[$0 < \l_0 < \l_+ \l_-$ \label{fig:sing2}]{
 \includegraphics[scale=0.3056]{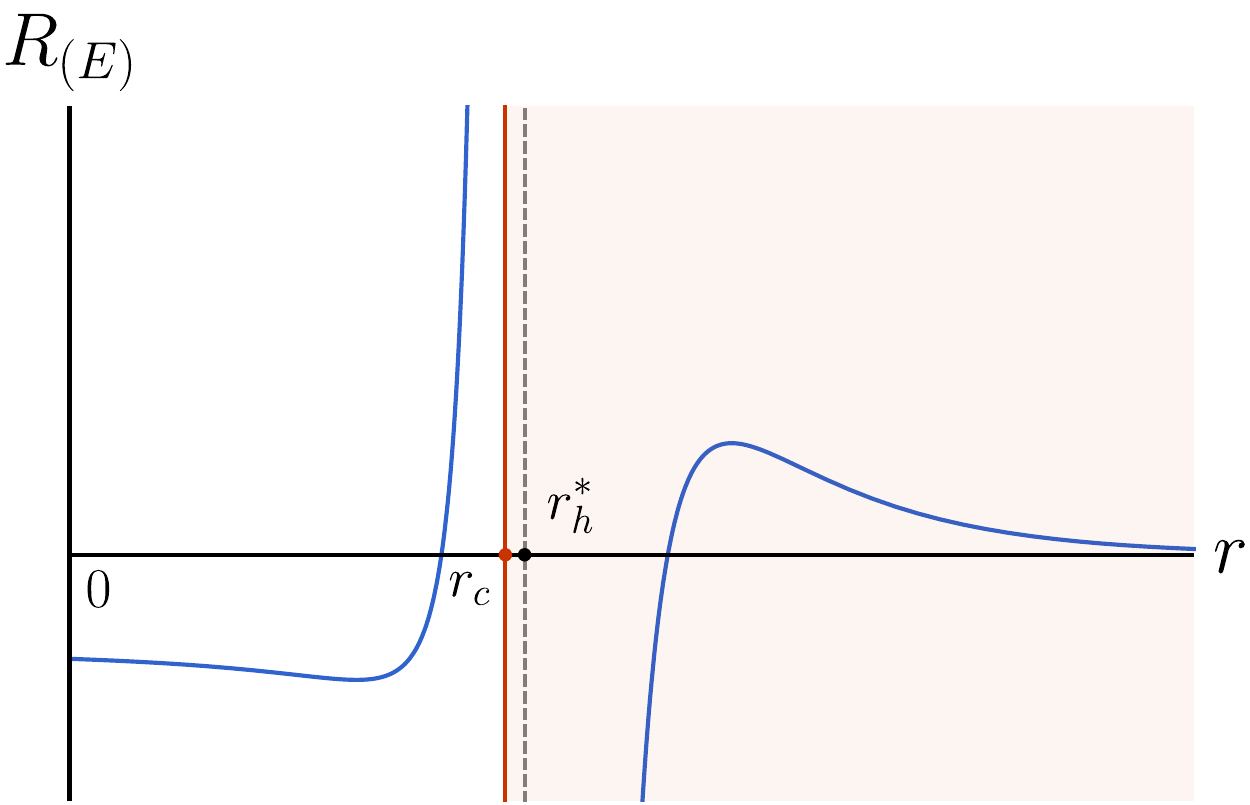}
}\hfill
\subfloat[$\l_0 < 0 < \l_+ \l_-$ \label{fig:sing3}]{
\includegraphics[scale=0.3056]{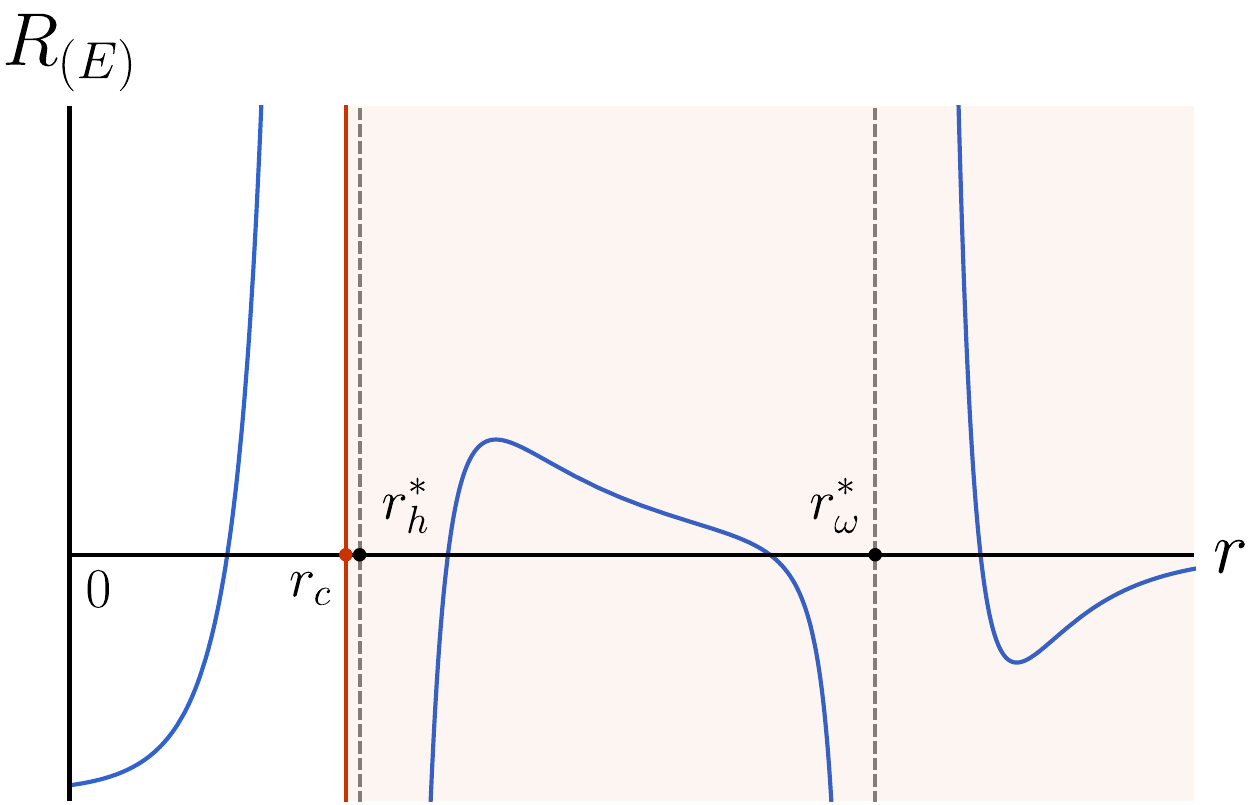}
}
}
\caption{The Ricci scalar for generic values of the phase space variables and different choices of the deformation parameters of the charged tri-TsT black hole (not drawn to scale). In these plots we have assumed that $\l_+ \l_- > 0$ but a similar behavior is found when $\l_+ \l_- < 0$ provided that the inequalities in \eqref{singcondition1} are handled appropriately.} \ContinuedFloat
\label{fig1}
\end{figure}

When $\l_0 - \frac{1}{4}(\l_+ + \l_-)^2 > 0$, the Ricci scalar is negative and the spacetime is free of CTCs and curvature singularities (see fig.~\ref{fig:healthy}). This is the same behavior observed for the $T\bar T$ backgrounds in the previous section. Similarly when $0 < |\l_+ \l_-| < \l_0 < \frac{1}{4} (\l_+ + \l_-)^2$ the curvature remains negative everywhere but the spacetime develops CTCs as shown in fig.~\ref{fig:sing1}. This situation --- CTCs but no singularities --- is not possible for $T\bar T$ backgrounds. On the other hand, this is generically the case for $J\bar T$ or $T\bar J$ backgrounds which feature CTCs and a constantly negative curvature for any choice of parameters; and is the best case scenario for $J\bar T + T\bar J$ backgrounds which also feature CTCs for all values of the parameters.


\subsection{Asymptotic behavior} \label{se:asympt}

We conclude this section by describing the asymptotic behavior of the charged tri-TsT black holes. Let us first consider the gauge fields and dilatons. In contrast to the charged BTZ black hole \eqref{btzU1}, the gauge fields $A^{(1)}$, $A^{(2)}$, $B$, and dilaton $\om$ of the TsT-transformed background \eqref{tstbtz} approach constant ratios of the deformation parameters at the asymptotic boundary,
\eqsp{
B &= \frac{\ell^2}{2(\l_0 - \l_+ \l_-)} dv \we du+ \O\big(\tfrac{1}{r} \big),  \quad\,\, A^{(1)} = - \frac{\ell}{2\ell_4} \Big( \frac{\l_+}{\l_0} du + \frac{\l_-}{\l_0} dv \Big) + \O\big(\tfrac{1}{r} \big), \\
 e^\omega &= \frac{\l_0 - \l_+ \l_-}{\l_0} + \O\big(\tfrac{1}{r} \big), \quad \,\, A^{(2)} = \frac{ \ell}{2\ell_4} \Big( \frac{\l_+}{\l_0 - \l_+ \l_-}du - \frac{\l_-}{\l_0 - \l_+ \l_-} dv \Big) + \O\big(\tfrac{1}{r} \big).
\label{asy}
}
On the other hand, the dilaton $\Phi$, which is constant for the charged BTZ black hole, becomes a linear dilaton for the tri-TsT background
\eq{
\Phi = \frac{1}{2} \log \bigg[ \frac{k}{p \eta(\l_0 - \l_+ \l_-)r}  \bigg] + \O(r^{-1}).
}

Let us now describe the behavior of the metric near the asymptotic boundary. For generic values of the deformation parameters, the three-dimensional metric in the string frame is flat near the boundary since the Ricci scalar vanishes there. More explicitly, the line element behaves as
\eq{
ds^2 &\sim  \frac{ \lambda_0  \ell^2 dr^2}{4(\lambda_0-\lambda_+\lambda_- ) r^2} + \frac{ds_b^2}{(\lambda_0-\lambda_+\lambda_-)^2}, \quad ds_b^2 = \ell^2 \big[ \l_0 du d{v} - \frac{1}{4} (\l_+ du + \l_- d{v})^2\big]. \label{boundarymetric}
}
Note that the asymptotic expansion of the metric depends only on the deformation parameters $\l_0$, $\l_+$, and $\l_-$. This is compatible with the interpretation of the sequence of TsT transformations \eqref{tstrecipe2} as describing irrelevant deformations of the dual CFT.

The determinant of the boundary metric \eqref{boundarymetric} is given by\footnote{In the Einstein frame the additional factor of $(\l_0 - \l_+ \l_-)^{-2}$ in \eqref{boundarymetric} cancels, so that the determinant of the boundary metric is still proportional to \eqref{detdsb}.}
\eq{
\textrm{det} \,g_b = - \frac{\ell^2 \l_0 (\l_0 - \l_+ \l_-)}{4} .  \label{detdsb}
}
It is interesting to note that the boundary metric becomes degenerate when $\l_0 = 0$ or $\l_0 = \l_+ \l_-$, mirroring the behavior of the metric in models of holography without Lorentz invariance, such as the flat$_3$/BMS field theory correspondence \cite{Barnich:2006av} and the (warped) AdS$_3$/warped CFT correspondence \cite{Anninos:2008fx,Detournay:2012pc}. For these choices of the deformation parameters, the bulk metric is asymptotically flat since $R_{(s)} \to 0$ at infinity, except in the $J\bar T$ and $T\bar J$ cases where the Ricci scalar is constant and negative everywhere, and is given by 
\eq{
J\bar T: \quad R_{(s)} = -6 \ell^{-2}(1 - \a_v \l_+)^2 - 8 \ell^{-2} \l_+^2 T_v^2,\\
T\bar J: \quad R_{(s)} = -6 \ell^{-2}(1 - \a_u \l_-)^2 - 8 \ell^{-2} \l_-^2 T_v^2.
}
 Note that whenever the boundary metric becomes degenerate, the bulk spacetime necessarily features CTCs, since $\lambda_0=0$ and $\l_0 = \l_+ \l_-$  are both in the region~\eqref{CTCregion}.


\subsection{Summary of the phase space}

Let us conclude this section by summarizing the phase space of charged tri-TsT black holes:
\begin{itemize}[leftmargin=\parindent]
\item in the region $ 0 \le \l_0 - \tfrac{1}{4}(\l_+ + \l_-)^2 \le 1$ the tri-TsT backgrounds have negative curvature everywhere, are free of any pathologies, and their phase space includes the NS vacuum and the charged rotating black holes described in sections \ref{se:NSvacuum} and \ref{se:tstbtz}, respectively;
\item for $\l_0 - \tfrac{1}{4}(\l_+ + \l_-)^2 > 1$ the black holes are free of CTCs and curvature singularities outside of the event horizon, the curvature remains negative, but the NS vacuum becomes complex;
\item when $0 < |\l_+ \l_-| < \l_0 < \frac{1}{4} (\l_+ + \l_-)^2$ the charged tri-TsT backgrounds have negative curvature and CTCs, but no singularities;
\item finally, when $\l_0 < 0$ and/or $\l_0 < \l_+ \l_-$ the black holes have one or two singularities as well as CTCs. In these cases the Ricci scalar is negative near the horizon and becomes positive at a finite value of the radial coordinate before reaching the first singularity. By an appropriate choice of the phase space variables it is always possible to find the region of positive curvature before encountering any CTCs.
\end{itemize}


\section{The perturbative spectrum} \label{se:spectrum}

In this section we derive the spectrum of strings winding on the TsT-transformed backgrounds described in section~\ref{se:tstbackgrounds} where two of the TsT coordinates are identified with the same isometry of $T^4$. We then interpret the worldsheet spectrum from the point of view of single-trace $T\bar T + J\bar T +T \bar J$-deformed CFTs. In particular, we show that the spectrum of strings with one unit of winding matches the spectrum of states in the untwisted sector of the symmetric orbifold $\textrm{Sym}^p\, \M_\mu$. The spectrum of strings winding on TsT-transformed backgrounds with other choices of the TsT coordinates is considered in appendix~\ref{ap:tstbackgrounds}. 

\subsection{The spectrum of winding strings}

The spectrum of strings winding on TsT-transformed backgrounds can be obtained from a spectral flow transformation that relates the values of the worldsheet stress tensor before and after the TsT transformations~\cite{Apolo:2019zai}. In this section we delineate the main steps necessary in the derivation of the spectrum of the tri-TsT backgrounds described by \eqref{10dbackground} and \eqref{tstbtz} satisfying the identification of coordinates \eqref{coordsim1} -- \eqref{coordsim3}. A more general but technical derivation of the spectrum is given in appendix \ref{ap:spectrum}.

Let us denote the target space coordinates before and after the TsT transformations respectively by $\tilde X^\a$  and $X^\a$. The $\tilde X^\a$ coordinates satisfy periodic boundary conditions, while some of the $X^\a$ coordinates satisfy winding boundary conditions where $X^\a(\sigma+2\pi)=X^\a(\sigma)+2\pi w^{(\a)}$. The main steps necessary in the derivation of the spectrum are:
 \begin{itemize}[leftmargin=37.5pt]
\item[step 1:] find a (nonlocal) field redefinition $\hat X^\a[X^\b]$ such that  $\hat X^\a$ satisfies, locally, the same equations of motion and constraints as the undeformed fields $\tilde X^\a$; 
\item[step 2:]  read off the boundary conditions $\hat  X^\a(\sigma+2\pi) = \hat  X^\a(\sigma+2\pi) + 2\pi\gamma^{(\a)}$ where $\gamma^{(\a)}$ may depend on the winding and/or the momentum along the TsT coordinates;
\item[step 3:] implement the boundary conditions by a spectral flow transformation $\hat  X^{\a} = \tilde X^\a +\gamma^{(\a)} (\zeta z + \bar \zeta \bar z )$ where $\zeta$  and $\bar \zeta$ are constants determined by the background geometry;
\item[step 4:] use the $\hat  X^\a$ found in step 3 to relate the values of the worldsheet stress tensor and other currents before and after the TsT transformations;
\item[step 5:]  impose the Virasoro constraints and express the deformed quantum numbers in terms of the undeformed ones.
\end{itemize}

Let us now go through these steps and derive the spectrum of the worldsheet theory. The first step follows from the fact that TsT transformations can be realized by a nonlocal field redefinition that is given by \cite{Frolov:2005dj,Rashkov:2005mi,Alday:2005ww}, 
  \eqsp{
  \p \hat{X}  = \p X - 2\ell^{-2}_s \textstyle \sum_i \check\mu_i \, \p X  M \GG_{i},\\
   \bp \hat{X}  = \bp X - 2\ell^{-2}_s \textstyle \sum_i \check\mu_i \,\GG_{i} M \bp X, \label{nonlocalbd}
  }
where $M_{\alpha\beta}=G_{(10)\alpha\beta}+B_{(10)\alpha\beta}$ and the $\Gamma_{i}$ matrices are defined in \eqref{Gamma}. For the TsT-transformed backgrounds considered in section \ref{se:tstbackgrounds}, the TsT coordinates $\hat X^{u} = \hat u$, $\hat X^{v} = \hat{v}$, and $\hat X^{y} = \hat y$ mix with each other after the TsT transformations while the other coordinates are identified directly without any mixing, i.e.~$\hat X^i = X^i$ for $i \not \in \{u,v,y\}$.  

One of the crucial ingredients in the derivation of the spectrum are the nonlocal boundary conditions satisfied by the target space coordinates $\hat X^\a$. These boundary conditions, which are described in step 2, are captured by the so-called $\gamma$ parameters that are sensitive to the winding boundary conditions of the strings under consideration. We are interested in strings that wind $|w|$ times along the spatial circle \eqref{coordsim3} of the deformed background. In order to distinguish between the left and right-moving $U(1)$ charges associated with translations along $y$, it is also convenient to introduce winding $w^{(y)}$ along this coordinate. Hence, the deformed target space coordinates satisfy the following boundary conditions
\eq{
(u,v) \sim (u + 2\pi w, v + 2\pi w), \qquad y \sim y + 2 \pi w^{(y)}. \label{windingbc}
}
For the class of tri-TsT backgrounds described in section \ref{se:tstbackgrounds}, the nonlocal boundary conditions satisfied by the $\hat X^\a$ coordinates can be read off from \eqref{nonlocalbd} and are given by (see \eqref{twistedbc} for the general formulae)
\eqsp{
\hat u (\tau, \s + 2\pi) &= \hat u(\tau, \s) + 2\pi \gamma^{(u)}, \qquad   \gamma^{(u)} = w + 2\check\mu_0  p_{(v)} + 2\check\mu_- p_{(y)}, \\
\hat v (\tau, \s + 2\pi) &= \hat v(\tau, \s) + 2\pi \gamma^{(v)}, \qquad   \gamma^{(v)} = w - 2 \check\mu_0 p_{(u)} - 2\check\mu_+ p_{(y)},  \\
\hat y (\tau, \s + 2\pi) &= \hat y(\tau, \s) + 2\pi \gamma^{(y)}, \qquad  \gamma^{(y)} =  w^{(y)} + 2\check\mu_+  p_{(v)} - 2\check\mu_- p_{(u)}, \label{twistedbcexp}
}
where $(\tau, \s)$ denote the worldsheet coordinates while $p_{(\a)}$ is the momentum conjugate to $X^{\a}$ which is defined in terms of the worldsheet Noether currents in \eqref{momentumdef}.

The worldsheet fields $\hat X^\a$ obey the same equations of motion and constraints as the undeformed fields $\tilde X^\a$, but satisfy twisted boundary conditions \eqref{twistedbcexp}. This allows us to carry out step 3, namely to generate new solutions satisfying nontrivial boundary conditions by a spectral flow transformation. Our background fields satisfy conditions \eqref{constraint} and \eqref{constraint2} which guarantee that the hatted fields given by
\eq{
\hat u = \til u + \gamma^{(u)} (\tau+\sigma) , \qquad \hat v = \til v - \gamma^{(v)} (\tau-\sigma), \qquad \hat y = \til y + \gamma^{(y)} \sigma, \label{specflowexp}
}
are solutions to the equations of motion provided that the tilded fields are also solutions to the equations of motion. This shift of coordinates induces a spectral flow transformation of the affine left and right-moving $SL(2,R) \times U(1)$ parts of the symmetry algebra of the worldsheet theory. Note that the equations of motion allow for one extra parameter in the shift of $\hat y$, namely $\hat y = \til y + \gamma^{(y)} (\sigma+\nu \tau)$ (see appendix \ref{ap:spectrum} for details). After the TsT transformations, this parameter changes the target space momentum and winding of the string along the $y$ coordinate. On the other hand, the identifications \eqref{coordsim2} and \eqref{windingbc} guarantee that these charges are quantized and hence unchanged by any continuous deformation. As a result, for the tri-TsT backgrounds constructed in section \ref{se:tstbackgrounds}, charge quantization requires $\nu = 0$. 

We can now follow step 4 and use \eqref{specflowexp} to relate the worldsheet stress tensor before and after the deformations. For general backgrounds and choices of the TsT coordinates, the zero modes of the deformed stress tensor are given in \eqref{L0} and \eqref{L0bar}. For our tri-TsT backgrounds, the zero modes of the stress tensor before $(\til L_0, \til {\bar L}_0)$ and after $(L_0,  {\bar L}_0)$ the TsT transformations are given by
\eqsp{
L_0& = \til L_0 + p_{(u)} \g^{(u)} + \frac{1}{2} p_{(y)} \g^{(y)} + \frac{k_4}{4} \big( \g^{(y)}\big)^2, \\
{\bar L}_0& = \til {\bar L}_0 - p_{(v)} \g^{(v)} - \frac{1}{2} p_{(y)} \g^{(y)} + \frac{k_4}{4} \big( \g^{(y)} \big)^2. \label{zeromodes2}
} 
In order to discuss the flow of the $U(1)$ charges, we also need to keep track of how the deformed chiral $U(1)$ currents \eqref{chiralcurrents} associated with linear combinations of translations and winding along $y$ are related to the undeformed currents. Plugging our background into the general expressions given in \eqref{undefchiralcurrents2}, we obtain
\eq{
\hat K_0 = \til K_0 + \frac{k_4}{2} \gamma^{(y)}, \qquad \hat{\bar K}_0 = \til{\bar K}_0 - \frac{k_4}{2} \gamma^{(y)}, \label{U1zeromodes}
}
where $(\til K_0, \til{\bar K}_0)$ are the zero modes before, and $(\hat K_0, \hat{\bar K}_0)$ are the zero modes after, the TsT transformations. It is natural to identify the zero modes $\hat K_0$ and $\hat{\bar K}_0$ with the deformed $U(1)$ charges $q_L(\check \mu_i)$ and $q_R(\check \mu_i)$ in the dual $T\bar T+T\bar J+J\bar T$-deformed CFT. Using the $\gamma$ parameters \eqref{twistedbcexp} and the expressions for the zero modes \eqref{U1zeromodes}, we find that the deformed $U(1)$ charges are given by
\eqsp{
q_L(\check\mu_i) & \equiv  \hat K_0  = \frac{1}{2} \big( p_{(y)} + k_4 w^{(y)} \big) + k_4 \check\mu_+  p_{(v)} - k_4 \check\mu_-    p_{(u)}, \\
q_R(\check\mu_i) & \equiv  \hat{\bar K}_0 =  \frac{1}{2} \big( p_{(y)} - k_4 w^{(y)} \big) - k_4 \check\mu_+  p_{(v)} + k_4 \check\mu_-   p_{(u)}. \label{QLQR0}
}

The last step in the derivation of the spectrum amounts to imposing the Virasoro constraints  before and after the deformation. From the dictionary relating the bulk and boundary coordinates, it is natural to identify the worldsheet Noether charges associated with $\ell^{-1}\p_u$ and $-\ell^{-1}\p_{v}$ with the left and right-moving energies $E_L$ and $E_R$ associated with $\p_x$ and $-\p_{\bar x}$ in the boundary theory. As a result, we have
\eq{
p_{(u)} = \ell E_L(\check\mu_i), \qquad p_{(v)} = - \ell E_R(\check\mu_i). \label{ELERdef}
}
Furthermore, it is convenient to express the zero modes of the deformed stress tensor in terms of the undeformed $U(1)$ charges which are given via \eqref{QLQR0} by
\eq{
q_L(0) & = \frac{1}{2} \big( p_{(y)} + k_4 w^{(y)}\big), \qquad q_R(0) = \frac{1}{2} \big( p_{(y)} - k_4 w^{(y)}\big). \label{undefqLqR}
}
Note that the momentum $p_{(y)}$ and the winding $w^{(y)}$ are integers since $y$ is a compact coordinate identified modulo $2\pi$. Consequently, the undeformed $U(1)$ charges \eqref{undefqLqR} are quantized and hence unchanged by the deformation, a fact that makes them useful in the parametrization of the deformed spectrum.

Finally, in terms of the energies $E_{L,R}(\check\mu_i)$ and the $U(1)$ charges $q_{L,R}(0)$, we find that the spectrum of strings winding on the TsT-transformed backgrounds constructed in section \ref{se:tstbackgrounds} can be written as
   \eqsp{
   E_{L}(0)  &= E_{L}(\check\mu_i) + \frac{2}{w} \Pi(\check\mu_i), \qquad   E_R(0)  = E_R(\check\mu_i)  +\frac{2}{w}   \Pi(\check\mu_i),
   \label{spectrum}
   }
where $\Pi(\check\mu_i)$ is given by
   \eq{
\!\! \! \Pi(\check\mu_i)  =  \check\mu_- q_R(0) E_L(\check\mu_i)  \! -\! \check\mu_+ q_L(0)  E_R(\check\mu_i) \! - \!  \ell \check\mu_0 E_L(\check\mu_i) E_R(\check\mu_i) \! + \! \frac{\ell k_4}{2} \big[ \check\mu_+ E_R(\check\mu_i) + \check\mu_- E_L(\check\mu_i)\big]^2\!. \label{Pidef}
   }
The expressions for the deformed energies \eqref{spectrum} are valid for strings winding on the spatial and internal circles of the tri-TsT backgrounds of section \ref{se:tstbackgrounds} for any values of the phase space parameters $T_u$, $T_v$, $\a_u$, and $\a_v$. In particular, this spectrum agrees with the spectrum computed in \cite{Chakraborty:2019mdf} for strings winding on the Ramond vacuum where all of the phase space parameters vanish.

The worldsheet spectrum is also characterized by the values of the deformed $U(1)$ charges $q_{L,R}(\check \mu_i)$, which can be written in terms of the deformed energies \eqref{ELERdef} and the undeformed $U(1)$ charges \eqref{undefqLqR} as
\eqsp{
q_L(\check\mu_i) & = q_L(0) - k_4 \check\mu_+   \ell  E_R(\check\mu_i) - k_4 \check\mu_-   \ell E_L(\check\mu_i), \\
q_R(\check\mu_i) & = q_R(0) + k_4 \check\mu_+  \ell  E_R(\check\mu_i) + k_4 \check\mu_-  \ell  E_L(\check\mu_i). \label{QLQR}
}
Note that both of these charges depend on the energies in such a way that the momentum along $y$, namely $q_L(\check\mu_i)  + q_R(\check\mu_i)$, is unchanged by the deformation. As a result, the deformed $U(1)$ charge $q_R(\check\mu_i)$ differs from the undeformed one even when the $T\bar J$ deformation parameter $\check\mu_-$ vanishes, contrary to what may have been otherwise expected.


\subsection{The spectrum of the symmetric orbifold} \label{se:matchingspectrum}

In this section we interpret the worldsheet spectrum in terms of the symmetric orbifold structure of single-trace $T\bar T + J\bar T + T\bar J$-deformed CFTs. In particular we will show that the spectrum of strings with winding $|w| = 1$ matches that of the untwisted sector of the symmetric orbifold, namely the spectrum of the $T\bar T + J\bar T + T\bar J$-deformed seed CFT.

\subsubsection{The twisted sector}

We begin by recalling that a single-trace deformation of the symmetric orbifold $\textrm{Sym}^p\, \mathcal M$ consists of deforming each copy of the seed $\M \to \M_{\mu}$ such that the deformed theory remains a symmetric orbifold, namely $\textrm{Sym}^p\, \mathcal M_\mu$.  As a result, each state in the single-trace $T\bar T + J\bar T + T\bar J$-deformed CFT can be thought of as a composite state that consists of $r$ single-particle states satisfying ${\textstyle \sum}_{a=1}^r n_a = p$ where $n_a$ is the length of a $Z_{n_a}$ cycle and $r$ is the total number of cycles that make up the state. In our conventions, $n_a = 1$ corresponds to a state in the untwisted sector of $\textrm{Sym}^p\, \mathcal M_\mu$, i.e.~a state in the $T\bar T + J\bar T + T\bar J$-deformed seed CFT. We can label the states in the deformed theory by the choice of $n_a$ coefficients $\{n_a\} \equiv \{n_1, n_2, \dots, n_p \}$ such that, for example, $\{1,1, \dots, 1\}$ describes a state in the untwisted sector while $\{p, 0, \dots, 0\}$ describes a state in the maximally-twisted sector of the symmetric orbifold.

The total energies and $U(1)$ charges of a state in $\textrm{Sym}^p\, \mathcal M_\mu$ can be written as
\eq{
E_{L,R}(\mu_i) = \sum_{a=1}^r E_{L,R}^{(n_a)}(\mu_i), \qquad q_{L,R}(\mu_i) = \sum_{a=1}^r q_{L,R}^{(n_a)}(\mu_i). \label{totalenergies}
}
where $E_{L,R}^{(n_a)}(\mu_i)$ and $q_{L,R}^{(n_a)}(\mu_i)$ are the energies and $U(1)$ charges of a single particle state. For single-trace $T\bar T + J\bar T + T\bar J$-deformed CFTs, we expect the energies $E_{L,R}^{(n_a)}(\mu_i)$ to satisfy the same formulae for the spectrum of strings with winding $w = - n_a$, where the minus sign follows from our conventions.\footnote{It would be interesting to confirm this directly from a field theoretical calculation.} Using \eqref{spectrum}, \eqref{QLQR}, and the dictionary \eqref{dictionary}, the energies and $U(1)$ charges of a twisted state in the deformed theory can be written as
 \eqsp{
   E_{L}^{(n_a)}(0)  &= E^{(n_a)}_{L}(\mu_i) - \frac{2}{n_a} \Pi^{(n_a)}(\mu_i), \qquad \quad E_{R}^{(n_a)}(0)  = E^{(n_a)}_{R}(\mu_i) -\frac{2}{n_a} \Pi^{(n_a)}(\mu_i), \\
 q_L^{(n_a)} (0) &=  q_L^{(n_a)} (\mu_i) + k_4 \mu_+ E_R^{(n_a)}(\mu_i) + k_4  \mu_- E_L^{(n_a)}(\mu_i), \\
q_R^{(n_a)} (0) &= q_R^{(n_a)} (\mu_i) - k_4  \mu_+ E_R^{(n_a)}(\mu_i) - k_4  \mu_- E_L^{(n_a)}(\mu_i),  \label{spectrum2}
   }
where $k_4$ is the $U(1)$ level of the undeformed seed and $\Pi^{(n_a)}(\mu_i)$ is given in \eqref{Pidef} with $E_{L,R} \to E^{(n_a)}_{L,R}$, $q_{L,R} \to q^{(n_a)}_{L,R}$, and $\check\mu_i $ is written in terms of $\mu_i$ using \eqref{dictionary}.

The general solution for the deformed energies of a twisted state can be written as 
\eq{
E_R^{(n_a)}(\mu_i) &= - \frac{B - \sqrt{B^2 + 4 A C}}{2A}, \qquad J^{(n_a)}(\mu_i) = J^{(n_a)}(0) \label{EJsols}
}
where $J= \ell (E_L -E_R)$ is the angular momentum, while $A$, $B$, and $C$ are defined by
\eqsp{
A & = 2 \mu_0 - k_4 (\mu_+ + \mu_-)^2, \\
C & = n_a \ell E^{(n_a)}_R(0) -  k_4^{-1} q_R^{(n_a)}(0)^2 +  k_4^{-1} \big[q_R^{(n_a)}(0) + \ell^{-1} k_4\mu_- J^{(n_a)}(0) \big]^2,\\
B& = n_a \ell  + 2 \mu_+ q_L^{(n_a)}(0) - 2 \mu_- q_R^{(n_a)}(0) + 2\ell^{-1} (\mu_0 - k_4 \mu_+ \mu_- - k_4 \mu_-^2) J^{(n_a)}(0).
 }
In particular, the spectrum of twisted states is real for states above the Ramond vacuum, namely for states satisfying $n_a \ell k_4 E_R^{(n_a)}(0) - q_R^2(0) \ge 0$, provided that
 \eq{
 A = 2  \mu_0 - k_4  (\mu_+ + \mu_-)^2 > 0.
 }
This is the same constraint \eqref{ctccondition} that guarantees the absence of closed timelike curves in the space of TsT transformed backgrounds, in agreement with the results obtained for the Ramond vacuum in \cite{Chakraborty:2019mdf}.


\subsubsection{Matching the spectrum in the untwisted sector}

Let us now consider the spectrum of single-particle states in the untwisted sector of the symmetric orbifold $\textrm{Sym}^p\, \mathcal M_\mu$. The latter should correspond to the spectrum of the $T\bar T + J\bar T +T \bar J$-deformed seed CFT $\M_\mu$. It is not difficult to verify that the spectrum of strings with one unit of winding --- obtained by setting $n_a = -w = 1$ in \eqref{spectrum2} or \eqref{EJsols} --- matches the spectrum of (double-trace) $T\bar T + J\bar T +T \bar J$-deformed CFTs on a cylinder of size $2\pi R = 2\pi\ell$~\cite{LeFloch:2019rut,Frolov:2019xzi}. The matching between the worldsheet and the field theory spectrum in the untwisted sector is a first indication of the correspondence between states in single trace $T\bar T + J\bar T + T\bar J$-deformed CFTs and backgrounds obtained from a sequence of TsT transformations of AdS$_3 \times S^3 \times T^4$ spacetimes. 

The NS vacuum of the deformed theory is made up of the ground states from each copy of the symmetric orbifold, hence it is state in the untwisted sector of $\textrm{Sym}^p\, \mathcal M_\mu$. As a result, the left and right-moving energies of the vacuum are given by
\eq{
E_{L,R}(\mu_i) = \sum_{a = 1}^p E_{L,R}^{(a)}(\mu_i) = p E_{L,R}^{(1)}(\mu_i), \label{totalenergies2}
}
where $E_{L,R}^{(1)}(\mu_i)$ are the energies of the vacuum in the deformed seed $\M_\mu$, the latter of which can be obtained from the undeformed energies $E_{L,R}^{(1)}(0)$ via \eqref{EJsols}. In the undeformed seed, the ground state is characterized by an energy and angular momentum that are given by
\eq{
\ell E^{(1)}(0) = -\frac{k}{2}, \qquad J^{(1)}(0) = 0.
}
Then, using \eqref{EJsols} and \eqref{totalenergies2}, we find that the energy and the angular momentum of the NS vacuum in single-trace $T\bar T + J\bar T + T\bar J$-deformed CFTs is given by
\eq{
E^{vac}(\mu_i) = - \frac{c}{6} \Bigg[ \frac{\ell - \sqrt{ \smash[b]{\ell^2 -2 k\mu_0 + k k_4 (\mu_+ + \mu_-)^2}}}{2k\mu_0 - k k_4 (\mu_+ + \mu_-)^2}\, \bigg], \qquad J^{vac}(\mu_i) = 0, \label{NSvacenergies2}
}
where we recall that $c = 6 kp$ is the total central charge of the undeformed symmetric orbifold. Since the NS vacuum has vanishing $U(1)$ charges before the deformation, the undeformed $U(1)$ charges $q^{vac}_{L,R}(0)$ vanish, and the deformed $U(1)$ charges read
\eq{
q^{vac} _L(\mu_i) &= - \frac{k_4}{2}  (\mu_+ + \mu_-)E^{vac}(\mu_i) , \qquad q^{vac}_R(\mu_i) = \frac{k_4 }{2}(\mu_+ + \mu_-) E^{vac}(\mu_i). \label{NSQLQR}
}
Note that the vacuum energy \eqref{NSvacenergies2} differs from the energy of the vacuum in the deformed seed $\M_\mu$ only by a factor of $p$. In addition, we note that 
\eq{
2 k \mu_0 - k k_4 (\mu_+ + \mu_-)^2 \le \ell^2 \label{NSconstraint2}
}
guarantees the energy and the deformed $U(1)$ charges of the vacuum are real and that this is the same constraint \eqref{NSconstraint} that leads to a real vacuum solution in the bulk side of the correspondence. 


\section{ Gravitational charges and thermodynamics} \label{se:thermodynamics}

In this section we study the thermodynamics of the TsT-transformed backgrounds described in section~\ref{se:tstbackgrounds}. We begin by computing the gravitational charges, entropy, and thermodynamic potentials of the charged tri-TsT black holes, the latter of which are shown to satisfy the first law of thermodynamics. We then derive the density of high energy states in single-trace $T\bar T + J\bar T + T\bar J$-deformed CFTs and show that this quantity matches the entropy of the black holes constructed in section \ref{se:tstbtz}. Finally, we compute the energies and $U(1)$ charges of the TsT-transformed background obtained in section \ref{se:NSvacuum} which match the charges of the NS vacuum in the dual $T\bar T + J\bar T + T\bar J$-deformed CFT.

\subsection{Thermodynamics of charged tri-TsT black holes} \label{se:tstthermo}

In this section we use the covariant formulation of charges to derive the energies and $U(1)$ charges of the charged TsT black hole \eqref{tstbtz}, and determine its thermodynamics potentials, entropy, and the first law of thermodynamics. For convenience, we carry out this analysis in three dimensions, where the computation of the gravitational charges and the analysis of the thermodynamics simplifies significantly. 

\subsubsection{Conserved charges} \label{se:tstcharges}

The charged tri-TsT black hole is a 7-parameter solution to the equations of motion of (the bosonic sector of) type IIB supergravity. We have written the black hole in \eqref{tstbtz}  in terms of three-dimensional variables, as the computation of its gravitational charges and thermodynamics simplifies in lower dimensions. In three dimensions, the TsT black hole is characterized by the metric, Kalb-Ramond field, gauge fields $A^{(1)}$ and $A^{(2)}$, as well as the dilatons $\Phi$ and $\omega$ given in \eqref{tstbtz}.  By construction, the backgrounds considered in this paper have fixed values of the magnetic and electric charges \eqref{QeQmBTZ}. Moreover, as argued in section~\ref{se:tstbackgrounds}, the variables $\l_0$, $\l_+$, and $\l_-$ parametrize the deformed theory, instead of the phase space of solutions, and are therefore fixed constants. As a result, the charged tri-TsT black holes are characterized by four independent charges corresponding to linear combinations of the energy and angular momentum, as well as two $U(1)$ charges associated with the gauge fields $A^{(1)}$ and $A^{(2)}$. 

In the covariant formulation of gravitational charges \cite{Wald:1993nt,Iyer:1994ys, Barnich:2001jy}, the infinitesimal charge associated with a symmetry generated by the Killing vector $\xi$ can be written as
  \eqsp{
    \d {\cal Q}_{\xi} & \equiv \frac{\ell^3}{4\pi \ell_s^4}  \int_{\p\ss} \bm \chi_{\xi},
  }
where $\p\ss$ is the boundary of a codimension-1 spacelike surface and the one-form $ \bm\chi_\xi$ is determined from the action of the theory (see appendix \ref{ap:charges} for details). The left and right-moving energies of the black hole \eqref{tstbtz}, denoted by $\Q_u$ and $\Q_v$, are the gravitational charges associated with the $\ell^{-1} \p_u$ and $- \ell^{-1}\p_{v}$ Killing vectors. Using the covariant formalism, we find that the left and right-moving energies are integrable in the space of solutions parametrized by $T_u$, $T_v$, $\a_u$, and $\a_v$ with fixed values of the deformation parameters. This is consistent with our interpretation of $\l_0$, $\l_+$, and $\l_-$ as deformation parameters instead of phase space variables. The left and right-moving energies of the charged tri-TsT black holes are given by
\eqsp{
\Q_{u}  = \frac{c \eta }{6 \ell} ( \a_u^2 + T_u^2) \big[ 1 - \a_{v} ( \l_+ + \l_-) +  \l_0 ( \a_{v}^2 + T_{v}^2) \big] , \\
\Q_{v}   = \frac{c \eta }{6\ell} ( \a_{v}^2 + T_{v}^2) \big[ 1 - \a_u ( \l_+ + \l_-) + \l_0( \a_u^2 + T_u^2) \big] . \label{tstbtzenergies}
}

In addition to the gravitational charges \eqref{tstbtzenergies}, the charged tri-TsT black holes have electric charges associated with the $A^{(1)}$ and $A^{(2)}$ gauge fields that are defined by
   \eqsp{
   Q^{(1)} &=  -\frac{\ell^3 \ell_4^2}{4\pi \ell_s^4}\int e^{-4\Phi - 2\omega} \star d A^{(1)} - e^{-8\Phi - 2\omega} A^{(2)}\! \wedge \star dB,  \\
     Q^{(2)} &= -\frac{\ell^3 \ell_4^2}{4\pi \ell_s^4}\int e^{-4\Phi}  \star d A^{(2)} - e^{-8\Phi - 2\omega} A^{(1)}\! \wedge \star dB,    \label{qqtildedef}
   }
where $\star$ denotes the Hodge dual in the three-dimensional Einstein frame. In ten dimensions, these charges correspond to the momentum and the electric charge along the $y$ coordinate, namely
\eq{
 Q^{(1)} = \mathcal Q_{y}, \qquad  Q^{(2)} = k_4 Q_e^{(y)},  \qquad  Q_e^{(y)} =   \frac{1}{(2\pi \ell_s)^6} \int_{C} e^{-2\Phi_{(10)}} \star_{10} d B_{(10)}, \label{10dU1charges}
}
where $\Q_{y}$ is the gravitational charge associated with the Killing vector $\p_y$, $Q_e^{(y)}$ is the electric charge along $y$, $C$ is the spatial surface transverse to $(r,y)$, and $\star_{10}$ is the Hodge dual in the ten-dimensional string frame. In addition, we note that the factor of $k_4$ in the definition of $Q^{(2)}$ in \eqref{10dU1charges} comes from our ansatz for the dimensional reduction \eqref{10dbackground}. 

It is convenient to write the $U(1)$ charges in terms of the following linear combinations
\eq{
Q_L \equiv \frac{1}{2}\big(Q^{(1)} - Q^{(2)}\big), \qquad Q_R \equiv \frac{1}{2} \big(Q^{(1)} + Q^{(2)}\big).\label{chiralchargebulk}
}
These definitions mimic the definitions of the undeformed $U(1)$ charges \eqref{undefqLqR} obtained in the derivation of the worldsheet spectrum. This follows from the fact that $\Q_y$ corresponds to the momentum conjugate to $y$ while $-Q_e^{(y)}$ corresponds to the winding along $y$, where the sign follows from our conventions on the worldsheet. Since the coordinate $y$ is compact, these $U(1)$ charges are quantized and preserved by any kind of continuous deformation. As a result, it is natural to identify the $U(1)$ charges \eqref{chiralchargebulk} with the undeformed left and right-moving $U(1)$ charges of the bulk spacetime.

For the charged tri-TsT black hole \eqref{tstbtz} the $U(1)$ charges are explicitly given by
\eqsp{
Q_L& = -  \frac{c\eta \ell_4}{6\ell} \a_u \big[1 -  \a_v ( \l_+ + \l_-) +  \l_0 (\a_v^2 + T_{v}^2)  \big] + \frac{\ell_4}{2} ( \l_- \Q_{u}   +  \l_+ \Q_{v})  ,   \\
Q_R & = + \frac{c\eta \ell_4}{6\ell} \a_v  \big[ 1 -  \a_u  ( \l_+ + \l_-) +   \l_0(\a_u^2 + T_u^2)   \big]  - \frac{\ell_4}{2} (\l_-  \Q_{u}  + \l_+ \Q_{v})  . \label{tstbtzU1charges} 
}
We observe that the $\a_u$ and $\a_v$ parameters are necessary to describe TsT-transformed backgrounds with independent $U(1)$ charges. Otherwise, when $\a_u = \a_v = 0$, the $U(1)$ charges become proportional to the left and right-moving energies, as considered previously in \cite{Apolo:2019yfj}. It is also important to note that these charges are still identified with the undeformed $U(1)$ charges of the black hole despite their dependence on the deformation parameters.\footnote{As discussed in \cite{Apolo:2019zai}, there is a subtlety in trying to relate the deformed and undeformed spectra directly from the expressions for the conserved charges computed in the bulk. In particular, \eqref{tstbtzU1charges} suggests that the undeformed $U(1)$ charges $Q_{L,R}$ depend on the deformation parameters, but this is ultimately a result of our choice of variables. Instead, in order to determine the nature of the deformed background in a gauge-independent way, we should consider the relationship between different physical observables.} 

Motivated by the definition of the deformed $U(1)$ charges \eqref{QLQR} obtained in the derivation of the perturbative spectrum, we can define the following $U(1)$ charges in the bulk
\eqsp{
Q_L^{\l_i} & \equiv Q_L - \frac{\ell_4 }{2} ( \l_- \Q_{u}   +  \l_+ \Q_{v}) =  -  \frac{c\eta \ell_4}{6\ell}  \a_u \big[1 -  \a_v ( \l_+ + \l_-) +  \l_0 (\a_v^2 + T_{v}^2)  \big],  \\
Q_R^{\l_i} & \equiv Q_R + \frac{\ell_4 }{2} ( \l_- \Q_{u}   +  \l_+ \Q_{v}) = + \frac{c\eta \ell_4}{6\ell}  \a_v \big[ 1 -  \a_u  ( \l_+ + \l_-) +   \l_0(\a_u^2 + T_u^2)   \big].  \label{tstbtzQLQR}
}
Although these charges lack a geometrical interpretation, they are still useful in the characterization of the tri-TsT black holes and can be identified with the deformed $U(1)$ charges of the $T\bar T+J\bar T+T\bar J$-deformed CFT. Note that the deformed $U(1)$ charges \eqref{tstbtzQLQR} are proportional to the spectral flow parameters $\a_u$ and $\a_v$. Hence, the tri-TsT black holes constructed in section \ref{se:specialtstbtz} are neutral with respect to the deformed $U(1)$ charges, while the tri-TsT black holes constructed in section \ref{se:tstbtz} are charged with respect to these charges.

The conserved charges \eqref{tstbtzenergies} and \eqref{tstbtzQLQR} allow us to express the tri-TsT black hole directly in terms of the physical charges $\Q_{u,v}$ and $Q_{L,R}^{\lambda_i}$. In order to accomplish this we note that the phase space parameters $T_u$, $T_v$, $\a_u$, and $\a_v$ can be written in terms of the gravitational charges as
\eqsp{
\a_u^2  + T_u^2 &= \frac{\ell \Q_{u}}{ \frac{c}{6}  + \l_0 \ell \Q_{v}  -  \frac{\ell \l_-}{\ell_4}  \big( Q_L^{\l_i}+ Q_R^{\l_i} \big) }, \qquad  \a_u = \frac{- Q_L^{\l_i}}{  \frac{c}{6}  + \l_0 \ell \Q_{v}  -  \frac{\ell \l_-}{\ell_4}  \big( Q_L^{\l_i}+ Q_R^{\l_i} \big)  }, \notag \\
\a_v^2 + T_{v}^2 &= \frac{\ell \Q_{v}}{ \frac{c}{6} + \l_0 \ell \Q_{u} + \frac{\ell \l_+}{\ell_4} \big( Q_L^{\l_i} +  Q_R^{\l_i}\big) }, \qquad \a_v = \frac{Q_R^{\l_i}}{ \frac{c}{6} + \l_0 \ell \Q_{u} + \frac{\ell \l_+}{\ell_4} \big( Q_L^{\l_i} +  Q_R^{\l_i}\big)  }. \notag
}
Interestingly, the factors appearing in the denominator of these equations are also featured in the entropy of the charged tri-TsT black holes described below.


\subsubsection{Thermodynamic potentials, entropy, and the first law} \label{se:tstentropy}

The charged TsT black hole \eqref{tstbtz} features a horizon at $r_h \equiv 2 T_u T_{v}$ whose location is not affected by the TsT transformations. In particular, the horizon generator for this background is given by the following Killing vector
\eq{
\xi_h  = \p_t - \Omega \p_\vp.
}
The inverse Hawking temperature $\beta$ and the angular potential $\Omega$ featured above can be written as
\eq{
\beta = \frac{1}{2}\bigg( \frac{1}{T_R} + \frac{1}{T_L}\bigg), \qquad \Omega = \frac{1}{2\b}\bigg( \frac{1}{T_R} - \frac{1}{T_L} \bigg), \label{betaomega}
}
where $T_{L,R}$ denote the left and right-moving temperatures which are given by
\eqsp{
\frac{1}{T_L} &= \frac{\ell\pi\big[T_{v} -\l_-( \a_v T_u + \a_u T_{v})  +  \l_0 T_u (\a_v^2 + T_{v}^2) \big]}{T_u T_{v}}, \\
\frac{1}{T_R} &= \frac{\ell\pi\big[ T_u - \l_+( \a_v T_u + \a_u T_{v})  + \l_0 T_{v}  (\a_u^2 + T_u^2)   \big]}{T_u T_{v}}. \label{tstbtzTLTR}
}
The temperatures \eqref{tstbtzTLTR} are thermodynamically conjugate to the left and right-moving energies \eqref{tstbtzenergies}. They can be alternatively determined from the thermal identification $(u,v) \sim (u + i/\ell T_L, v - i/\ell T_R)$ which guarantees that the Euclidean continuation of \eqref{tstbtz} is free of conical singularities. 

In addition to the temperatures \eqref{tstbtzTLTR}, there are two chemical potentials $\mu_{L}$ and $\mu_{R}$ that are conjugate to the $U(1)$ charges $Q_{L}$ and $Q_{R}$ in \eqref{tstbtzU1charges}. These chemical potentials can be determined from the values of the gauge fields at the horizon via (see e.g.~\cite{Compere:2007vx})
\eqsp{
\mu_L \equiv \xi_h^\mu \big(A_\mu^{(1)} -  A_\mu^{(2)}\big) \big|_{r_h},  \qquad \mu_R \equiv \xi_h^\mu \big(A_\mu^{(1)} + A_\mu^{(2)} \big)   \big|_{r_h}. \label{chempot}
} 
For the charged tri-TsT black hole \eqref{tstbtz} the chemical potentials read
\eqsp{
\mu_L & = -\frac{\pi \ell^2 \big[ 2\a_u T_{v} +  \l_+  T_u (\a_v^2 + T_{v}^2) - \l_-T_{v} (\a_u^2 + T_u^2)    \big]}{\ell_4 \beta T_u T_{v}}, \\
\mu_R & = +\frac{\pi \ell^2 \big[ 2\a_v T_u -  \l_+T_u  (\a_v^2 + T_{v}^2)   + \l_- T_{v} (\a_u^2 + T_u^2)   \big]}{ \ell_4 \beta T_u T_{v}}. \label{tstbtzmuLmuR}
}
In particular, note that $\mu_R - \mu_L$ is independent of the deformation parameters and  is thus unchanged by the sequence of TsT transformations.  In the ten-dimensional theory, the chemical potential $\frac{1}{2}(\mu_R + \mu_L)$ is conjugate to the momentum along $y$ while $\frac{1}{2}(\mu_R - \mu_L)$ is conjugate to the winding charge. Hence, the chemical potentials can also be determined from the ten-dimensional horizon generator $\xi_{h(10)}$ and from the value of the ten-dimensional $B$-field at the horizon, namely
\eq{
\xi_{h(10)} = \p_t - \Omega \p_\vp - \frac{1}{2}(\mu_R + \mu_L) \p_y, \qquad \frac{1}{2}(\mu_R - \mu_L) = \ell_4^{-2}\xi_{h(10)}^{\a} (\p_{y})^{\b} B_{(10)\a\b}\big|_{r_h}.
} 
The entropy of the charged tri-TsT black hole can be obtained from the area $\A$ of the horizon and is given by
\eqsp{
S_{BH} = \frac{\mathcal A}{4G_N^{(3)}} = \frac{\pi \eta c}{3} \Big\{ & T_u \big[ 1 - \a_v ( \l_+ + \l_-) + \l_0 (\a_v^2 + T_{v}^2)\big]  \\
&  \hspace{15pt} +T_{v}\big[ 1 - \a_u (\l_+ + \l_-) + \l_0 (\a_u^2 + T_u^2) \big] \Big\}, \label{tstbtzentropy} 
}
where $\eta$ is given by \eqref{etadef}, which we reproduce here for convenience,
\eq{
   \eta^{-1} &= (1 - \a_u \l_-)(1 - \a_v \l_+) - \big[\a_u \l_+ - \l_0 (\a_u^2 + T_u^2) \big] \big[ \a_v \l_- - \l_0 (\a_v^2 + T_{v}^2) \big]. \notag
}
In particular, it is not difficult to verify that the tri-TsT black hole satisfies the first law of thermodynamics, which can be written as
 \eq{
  \delta S_{BH} & =  \frac{1}{T_L} \d \Q_{u} + \frac{1}{T_R} \d \Q_{v} - \ell^{-1}\b (\mu_L \delta Q_L + \mu_R \delta Q_R). \label{firstlaw}
  }
Note that the contribution of the $\l_{\pm}$ parameters to the entropy \eqref{tstbtzentropy} is multiplied by the $\a_u$ and $\a_v$ variables. As a result, when $\a_u$ and $\a_v$ vanish, the entropy \eqref{tstbtzentropy} takes the same form as the entropy of the black holes dual to thermal states in single-trace $T\bar T$-deformed CFTs \cite{Apolo:2019zai}. This is also a feature of thermal states in $T\bar T + J\bar T + T\bar J$-deformed CFTs with vanishing values of the deformed $U(1)$ charges, as described in more detail in the next section.

The entropy \eqref{tstbtzentropy} takes a simple form in terms of the phase space variables but it is not suitable for comparison with the entropy in the field theory as it is not written in terms of the physical charges. Furthermore, the regime where \eqref{tstbtzentropy} is well defined is not manifest in terms of the phase space variables. For these reasons, it is more convenient to write the entropy in terms of the conserved charges such that
\eqsp{
\!\! S_{BH} &= 2\pi \bigg\{ \sqrt{ \ell \Q_{u} \big[\tfrac{c}{6} + \l_0 \ell \Q_{v} - \tfrac{\ell \l_-}{\ell_4} (Q_L + Q_R) \big] - \tfrac{\ell^2}{\ell_4^2} \big(Q_L - \tfrac{\ell_4}{2} \l_- \Q_{u} -\tfrac{\ell_4}{2} \l_+ \Q_{v} \big)^2} \\
& \hspace{16pt} + \sqrt{ \ell\Q_{v} \big[\tfrac{c}{6} + \l_0  \ell\Q_{u} + \frac{\ell\l_+}{\ell_4} (Q_L + Q_R) \big] - \tfrac{\ell^2}{\ell_4^2} \big(Q_R +\tfrac{\ell_4}{2} \l_-  \Q_{u} + \tfrac{\ell_4}{2} \l_+ \Q_{v} \big)^2 } \, \bigg\}. \label{tstbtzentropy2} 
}



\subsection{The entropy of single-trace $T\bar T + J\bar T + T\bar J$-deformed CFTs}
In this section we derive the asymptotic density of states in single-trace $T\bar T + J\bar T + T\bar J$-deformed CFTs following the lines of \cite{Apolo:2019zai} (see also \cite{Giveon:2017nie}).

The states in the symmetric product $\textrm{Sym}^p\, \mathcal M_\mu$ obtained from a single-trace $T\bar T + J\bar T + T\bar J$ deformation consists of composite states made of the product of twisted and untwisted states.\footnote{The results of this section rely on the assumption that the deformed theory  is a symmetric orbifold. As discussed in section \ref{se:tst}, the symmetric orbifold structure has been conjectured to exist, before the deformation, only for the long string sector of the bulk theory.} As described in section \ref{se:matchingspectrum}, these states can be labeled by a set of integers $\{n_a\} = \{n_1, n_2, \dots, n_p\}$ that correspond to the lengths of each of the $Z_{n_a}$ cycles that make up the state. In a symmetric orbifold the log of the density of $\{n_a\}$ states at fixed values of the total energies $E_{L,R}(\mu_i)$ and deformed $U(1)$ charges $q_{L,R}(\mu_i)$ is given by
\eq{
S^{\{n_a\}}(E_{L,R}, q_{L,R}) = \sum_{a=1}^r S^{(n_a)} \big(E_{L,R}^{(n_a)}, q_{L,R}^{(n_a)} \big), \label{derentropy1}
}
where $\exp S^{(n_a)} (E_{L,R}^{(n_a)}, q_{L,R}^{(n_a)})$ is the density of single particle states with energies $E_{L,R}^{(n_a)}(\mu_i)$ and $U(1)$ charges $q_{L,R}^{(n_a)}(\mu_i)$. The latter are related to the total energies and charges of a state in the symmetric orbifold via \eqref{totalenergies}.

Since the energy levels do not cross under the $T\bar T + J\bar T + T\bar J$ deformation, the density of states is preserved by the deformation. As a result, the Cardy formula for each twisted sector can be written in terms of the deformed energy using \eqref{spectrum2} such that 
\eq{
S^{(n_a)}  &= 2\pi\bigg\{ \sqrt{ n_a k \ell E_L^{(n_a)} (0) - \tfrac{k}{k_4} q_L^{(n_a)}(0)^2} + \sqrt{ n_a k \ell E_R^{(n_a)} (0) - \tfrac{k}{k_4} q_R^{(n_a)}(0)^2} \, \bigg\}  \notag \\
\begin{split}
& = 2\pi \bigg\{ \sqrt{k E_L^{(n_a)}(\mu_i)  \Big[ n_a \ell +  2\mu_0 E_R^{(n_a)}(\mu_i) - 2\mu_- \big[ q_L^{(n_a)}(\mu_i) + q_R^{(n_a)}(\mu_i)\big] \Big] - \tfrac{k}{k_4} q_L^{(n_a)}(\mu_i)^2}  \\
&\hspace{15pt} + \sqrt{k E_R^{(n_a)}(\mu_i)  \Big[ n_a \ell + 2\mu_0 E_L^{(n_a)} (\mu_i) + 2\mu_+ \big[ q_L^{(n_a)}(\mu_i)  + q_R^{(n_a)}(\mu_i)  \big] \Big] - \tfrac{k}{k_4} q_R^{(n_a)}(\mu_i)^2}\, \bigg\}. \label{derentropy2}
\end{split}
} 
The entropy formula \eqref{derentropy2} is valid in the Cardy regime where
\eq{
E_L^{(n_a)} (0) - \frac{q_L^{(n_a)}(0)^2 }{n_a k_4} \gg \frac{k}{n_a}.  \label{cardycondition}
}
Furthermore, it is not difficult to verify that the following partition of the energies and $U(1)$ charges extremizes the entropy \eqref{derentropy2} 
\eq{
E_{L,R}^{(n_a)}(\mu_i) = \frac{n_a}{p} E_{L,R}(\mu_i),  \qquad q_{L,R}^{(n_a)}(\mu_i) = \frac{n_a}{p} q_{L,R}(\mu_i). \label{partition}
} 
Using the partition of the energies and the $U(1)$ charges \eqref{partition}, we find that the total entropy \eqref{derentropy1} is independent of the choice $\{ n_a \}$ of the $Z_{n_a}$ cycles and is given by
\eq{
\!\!\!\! S(E_{L,R}, & q_{L,R} )  =S^{\{n_a\}}(E_{L,R}, q_{L,R}) \notag \\
 \begin{split}
 &= 2\pi \bigg \{ \!\sqrt{ \frac{c}{6} E_L(\mu_i)\Big[ \ell + \frac{2\mu_0}{p} E_R(\mu_i) - \frac{2\mu_-}{p} \big[q_L(\mu_i) + q_R(\mu_i)\big] \Big] - \tfrac{k}{k_4} q_L(\mu_i)^2}  \\
&\hspace{23pt} + \sqrt{ \frac{c}{6} E_R(\mu_i) \Big[ \ell + \frac{2 \mu_0}{p} E_L(\mu_i) + \frac{2\mu_+}{p} \big[q_L(\mu_i)+ q_R(\mu_i)\big] \Big] - \tfrac{k}{k_4} q_R(\mu_i)^2} \,\bigg\}, \end{split} \label{entropyqft}
}
where $c = 6pk$ is the total central charge of the symmetric orbifold before the deformation.
The regime of validity of the deformed Cardy formula \eqref{derentropy2} depends on the twisted sector. In particular, it is clear  from \eqref{cardycondition} that maximally twisted states have the lowest threshold and dominate the entropy at lower energies (see \cite{Apolo:2019zai} for more details). In addition, using \eqref{partition} we find that the total energies and $U(1)$ charges of the states contributing to \eqref{entropyqft} satisfy
\eqsp{
 E_{L,R}(0)  &= E_{L,R}(\mu_i) - \frac{2}{p} \Pi(\mu_i), \\
q_L(0)&=  q_L (\mu_i) + k_4 \mu_+ E_R (\mu_i) + k_4 \mu_- E_L (\mu_i)  , \\
 q_R(0)    &= q_R  (\mu_i) - k_4 \mu_+ E_R (\mu_i)  - k_4 \mu_- E_L(\mu_i), \label{spectrum3}
}
where $\Pi(\mu_i)$ is given in \eqref{Pidef} with $\check\mu_i $ expressed in terms of $\mu_i$ via the dictionary \eqref{dictionary}.

The entropy \eqref{entropyqft} generalizes the expression found in \cite{Apolo:2019zai} for single-trace $T\bar T$-deformed CFTs by including the contributions from the $J\bar T$ and $T\bar J$ deformations. In particular, up to corrections proportional to the deformation parameters, the entropy takes a similar form as Cardy's formula for a CFT with additional $U(1)$ charges.

We note that the entropy \eqref{entropyqft} has been derived in the microcanonical ensemble where the energies and $U(1)$ charges are held fixed. We can express the entropy in the canonical ensemble in terms of the conjugate temperatures and chemical potentials defined by
\eq{
\frac{1}{T_{L,R}} \equiv \frac{\p S}{\p E_{L,R}(\mu_i)}, \qquad \mu_{L,R}  \equiv \frac{\p S}{\p q_{L,R}(0)}. \label{thermopotqft}
}
In this way we can find the relationship between the deformed thermodynamic potentials and the deformed charges, expressions which can be compared to the ones found for the tri-TsT black holes in section \ref{se:tstthermo}. In particular, from the entropy \eqref{entropyqft} it is not difficult to derive the Hagedorn temperature $T_H$ of single-trace $T\bar T + J\bar T + T\bar J$-deformed CFTs. For fixed values of the undeformed $U(1)$ charges $q_{L,R}(0)$, the Hagedorn temperature can be obtained by letting $2E_{L,R}(\mu_i) \to E(\mu_i)\to \infty$ in \eqref{entropyqft} such that
\eq{
S(E_{L,R}, q_{L,R}) \sim \beta_H E(\mu_i), \qquad \beta_H = \frac{1}{T_H} = 2\pi \sqrt{2 k \mu_0 - k k_4(\mu_+ + \mu_-)^2}. \label{hagedorn}
}
We observe that the Hagedorn temperature is real provided that \eqref{NSconstraint2} is satisfied. This is to be expected since \eqref{NSconstraint2} guarantees the existence of real vacuum energies in single-trace $T\bar T + J\bar T + T\bar J$-deformed CFTs. Up to conventions, \eqref{hagedorn} agrees with the expression for double-trace $T\bar T + J\bar T + T\bar J$-deformed CFTs with fixed $U(1)$ charges derived in~\cite{Chakraborty:2020xyz}. This is compatible with the fact that the Hagedorn temperature is not an extensive quantity and should be given by the Hagedorn temperature of the seed theory.

It is also interesting to note that \eqref{entropyqft} differs from the entropy of double-trace $T\bar T + J\bar T + T\bar J$-deformed CFTs by the additional factors of $p$ accompanying the deformation parameters. In particular, since $p$ counts the number of copies making up the symmetric orbifold, when $p = 1$ we recover the entropy of double-trace $T\bar T + J\bar T + T\bar J$-deformed CFTs. 


\subsection{Matching the thermodynamics} \label{se:machingthermo}
In this section we show that the asymptotic density of states in single-trace $T\bar T + J\bar T + T\bar J$-deformed CFTs matches the entropy of the charged tri-TsT black holes computed in the previous section. We will also show that the neutral black holes \eqref{neutraltstbtz} constructed in section \ref{se:specialtstbtz} correspond to a special class of thermal states whose deformed $U(1)$ charges vanish and whose thermodynamics share several similarities with thermal states in $T\bar T$-deformed CFTs.

The holographic dictionary \eqref{dictionary} implies that the bulk and boundary charges are identified via
\eq{ 
\Q_{u,v}  \leftrightarrow E_{L,R}(\mu_i) ,\qquad Q_{L,R} \leftrightarrow q_{L,R}(0). \label{dictionary2}
}
In addition, the flow equation \eqref{QLQR} guarantees that the bulk $U(1)$ charges \eqref{tstbtzQLQR} are identified with the deformed $U(1)$ charges, namely $Q_{L,R}^{\l_i} = q_{L,R}(\mu_i)$. It is then straightforward to see that the entropy of the charged tri-TsT black holes \eqref{tstbtzentropy2} matches the asymptotic density of states \eqref{entropyqft}, namely
\eq{
S_{BH}  = S (E_{L,R}, q_{L,R}).
}
Similarly, the thermodynamic potentials \eqref{tstbtzTLTR} and \eqref{tstbtzmuLmuR} of the black hole match the corresponding expressions derived from the field theory side via \eqref{thermopotqft}. We thus conclude that the charged tri-TsT black holes \eqref{tstbtz} are dual to thermal states in $T\bar T + J\bar T + T\bar J$-deformed CFTs with arbitrary values of the deformed energies and $U(1)$ charges.

As a consistency check we note that if we assume that the entropy \eqref{tstbtzentropy2} of the tri-TsT black holes is unchanged before and after the deformation, we can reproduce the relationship between the deformed and undeformed energies of a maximally twisted state in single-trace $T\bar T + J \bar T + T\bar J$-deformed CFTs given in \eqref{spectrum3}. We can obtain the deformed energies from \eqref{tstbtzentropy2} because the angular momentum and the undeformed $U(1)$ charges are quantized, and hence preserved by the deformation. The reason \eqref{tstbtzentropy2} only reproduces the spectrum of maximally twisted states is that they dominate the entropy at low energies.

It is important to note that the matching of the entropy holds for all values of the deformation parameters, including the $T\bar T$ and $J\bar T$ limits. In particular, when the $\l_\pm$  parameters vanish and the $U(1)$ charges are turned off, our tri-TsT black holes, their perturbative spectrum, and the matching of their entropy with the field theory side reproduces the $T\bar T$ results obtained in \cite{Apolo:2019zai}. On the other hand, in the $J\bar T$ case where $\l_0 = \l_- = 0$, our results generalize the finite-temperature analysis of \cite{Apolo:2019yfj} in the following ways: ($i$) by constructing a more general class of black hole solutions with arbitrary values of the deformed $U(1)$ charges; and ($ii$) by showing that their entropy still matches the density of high-energy states in single-trace $J\bar T$-deformed CFTs.

We conclude with comments on the entropy of the neutral tri-TsT black holes \eqref{neutraltstbtz}. Recall that the latter are TsT-transformed backgrounds characterized by vanishing values of the $\a_u$ and $\a_v$ parameters. This means that the $U(1)$ charges vanish before the TsT transformation, i.e.~that the undeformed background is just the BTZ black hole. From the values of the deformed $U(1)$ charges given in \eqref{tstbtzQLQR}, we see that these charges also vanish for this class of backgrounds. As a result, the entropy of the neutral tri-TsT black holes becomes\footnote{One may wonder how \eqref{entropyttbar} is compatible with the fact that the neutral tri-TsT black holes have nonvanishing undeformed $U(1)$ charges \eqref{tstbtzU1charges} and chemical potentials \eqref{tstbtzmuLmuR}. This can be understood by noting that the charges and chemical potentials satisfy $Q_L = -Q_R$ and $\mu_L = \mu_R$, so that their contribution to the first law of thermodynamics \eqref{firstlaw} vanishes.}
\eq{
S_{BH}\big|_{\a_u = \a_v = 0} & =  2\pi \bigg \{ \sqrt{ \frac{c}{6} E_L \Big( 1 + \frac{2\mu_0}{p} E_R  \Big)}  + \sqrt{ \frac{c}{6} E_R \Big( 1 + \frac{2 \mu_0}{p} E_R \Big) } \,\bigg\}. \label{entropyttbar}
}
In addition, we note that when the $\a_u$ and $\a_v$ parameters vanish, the energies \eqref{tstbtzenergies} and the physical temperatures \eqref{tstbtzTLTR} of the tri-TsT black holes also reduce to the energies and temperatures characterizing the $T\bar T$-deformed backgrounds constructed in \cite{Apolo:2019zai}. As a result, we find that the neutral black holes describe special states in the dual $T\bar T + J\bar T + T\bar J$-deformed CFT that mimic the behavior of thermal states in $T\bar T$-deformed CFTs. 


\subsection{Matching the ground state} \label{se:NSenergies}

Let us now revisit the ground state geometry constructed in section \eqref{se:NSvacuum}. Therein, we argued that the NS vacuum can be obtained from analytic continuation of the charged tri-TsT black hole, where the values of the phase space variables are determined by requiring a smooth non-rotating background free of conical defects or singular gauge fields. Plugging the vacuum parameters \eqref{r0sol} into the gravitational charges \eqref{tstbtzenergies} we find that the deformed energy and angular momentum of this background are respectively given by
\eq{
\Q_{t}^{vac}  = -\frac{c}{3 \ell} \bigg[ \frac{2 - \sqrt{4 -4\l_0 + (\l_+ + \l_-)^2}}{4\l_0 - (\l_+ + \l_-)^2}\,\bigg], \qquad \Q_{\vp}^{vac} = 0, \label{NSvacenergies}
}
where $\Q_{t}^{vac}$ and $\Q_{\vp}^{vac}$ are the gravitational charges associated with the Killing vectors $\ell^{-1} \p_t$ and $\p_\vp$. Similarly, plugging \eqref{r0sol} into \eqref{tstbtzU1charges}, we find that the undeformed $U(1)$ charges $Q_{L,R}$ vanish, which is consistent with the fact that the vacuum must have vanishing $U(1)$ charges before the deformation. As a result, the deformed $U(1)$ charges defined in \eqref{tstbtzQLQR} are given by
\eq{
Q_L^{\l_i, vac}  =  - \frac{\ell_4}{4} (\l_+ + \l_-) \Q_{\p_t}^{vac}, \qquad Q_R^{\l_i, vac}  =   \frac{\ell_4}{4} (\l_+ + \l_-) \Q_{\p_t}^{vac}.  \label{NSvaccharges}
}
Using the holographic dictionary \eqref{dictionary2}, the above energy, angular momentum, and deformed $U(1)$ charges match the corresponding quantities for the NS vacuum in single-trace $T\bar T + J\bar T + T\bar J$-deformed CFTs given in \eqref{NSvacenergies2} and \eqref{NSQLQR}.

The matching of the vacuum energies provides additional evidence, beyond the spectrum and the thermodynamics, that the charged tri-TsT backgrounds \eqref{tstbtz} describe a general class of backgrounds that are dual to states in single-trace $T\bar T + J\bar T + T\bar J$-deformed CFTs.

 
 \bigskip

\section*{Acknowledgments}
We are grateful to Soumangsu Chakraborty, Amit Giveon, Peng-Xiang Hao, David Kutasov, Wen-Xin Lai, and Eva Silverstein for helpful discussions. The work of LA and WS was supported by NFSC Grant No.~11735001. The work of LA was also supported by NFSC Grant No.~11950410499 and by the European Research Council under the European Union's Seventh Framework Programme (FP7/2007-2013), ERC Grant
agreement ADG 834878. 



\bigskip

\appendix

\section{Spectrum for multiple TsT transformations} \label{ap:spectrum}

In this appendix we derive general expressions for the spectrum of strings winding on TsT-transformed backgrounds supported by NS-NS flux.

Let us consider IIB string theory on a TsT-transformed background with NS-NS flux. The worldsheet action can be written as
  \eq{
  \!\!  S_{\textrm{WS}} =- \frac{1}{4\pi\ell_s^{2}}\!  \int \! d^2z \big( \sqrt{-\g} \g^{ab} {G}_{\a\b} + \e^{ab}  {B}_{\a\b} \big)  \p_a {X}^{\a} \p_b {X}^{\b}  = \frac{1}{2\pi\ell_s^2} \! \int \! d^2z   {M}_{\a\b} \p  {X}^{\a} \bp  {X}^{\b} , \label{action}
  }
where $G_{\a\b}$ and $B_{\a\b}$ denote the ten-dimensional string frame metric and Kalb-Ramond field, $M_{\a\b} = G_{\a\b} + B_{\a\b}$, $z = \tau + \s$ and $\bar{z} = \tau - \s$ are the worldsheet coordinates, $\p \equiv \p_z$ and $\bp \equiv \p_{\bar z}$, and the worldsheet metric and Levi-Civita symbol satisfy $\g_{z\bar{z}} = - 1/2$ and $\e_{z\bar{z}} = 1$. A special feature of TsT transformations is that the equations of motion and Virasoro constraints of the deformed theory \eqref{action} can be mapped to those of an undeformed theory with twisted boundary conditions by a nonlocal change of coordinates~\cite{Frolov:2005dj,Rashkov:2005mi,Alday:2005ww}. This fact continues to be true for backgrounds obtained from multiple TsT transformations. In this case, the nonlocal change of coordinates mapping the deformed and undeformed theories can be written in matrix notation as
  \eqsp{
  \p \hat{X} = \p X  M   \hat M^{-1} = \p X - 2\ell^{-2}_s \textstyle \sum_i \check\mu_i \, \p X   M  \GG_{i},\\
   \bp \hat{X} = \hat M^{-1} M \bp X = \bp X - 2\ell^{-2}_s \textstyle \sum_i \check\mu_i  \,\GG_{i} M\bp X, \label{nonlocal}
  }
where $\hat X^\a$ denote the twisted target space coordinates of the undeformed background $\hat M_{\a\b}$. 

The change of coordinates \eqref{nonlocal} introduces twisted boundary conditions on the  coordinates involved in the TsT transformations. In order to see this, let $X^m$ and $X^{\bar m}$ denote the target space coordinates satisfying winding and/or twisted boundary conditions, and let $X^i$ denote the rest of the coordinates which are assumed to satisfy trivial boundary conditions. Using the nonlocal map \eqref{nonlocal} we find that the $\hat X^m$, $\hat X^{\bar m}$, and $X^i$ coordinates satisfy
\eqsp{
\hat  X^m (\sigma + 2\pi) &= \hat X^m(\sigma) + 2\pi \gamma^{(m)}, \qquad \gamma^{(m)} = w^{(m)} +  2 \textstyle\sum_i \check\mu_i (\GG_i)_{m\bar{n}} p_{(\bar n)},\\
\hat X^{\bar m}(\sigma + 2\pi) &= \hat X^{\bar m}(\sigma) + 2\pi \gamma^{(\bar m)}, \qquad \gamma^{(\bar m)} = w^{(\bar m)} +  2\textstyle\sum_i  \check\mu_i (\GG_{i})_{\bar{m}{n}} p_{(n)}, \\
\hat X^i (\s + 2\pi) &= \hat X^i (\s), \label{twistedbc}
}
where $w^{(\a)}$ denotes the winding along $X^\a$ while $p_{(\a)}$ is the momentum conjugate to $X^{\a}$. The latter is defined by
  \eq{
  p_{(\a)} = \frac{1}{2\pi} \oint j^{\tau}_{(\a)} = \frac{1}{2\pi} \oint  j^z_{(\a)} + j^{\bar{z}}_{(\a)}, \label{momentumdef}
  }
where $j_{(\a)}$ is the worldsheet Noether current generating shifts of $X^{\a}$ that is given by
\eq{
 {j}_{(\a)}& \equiv  {j}^z_{(\a)} \p + {j}^{\bar z}_{(\a)} \bp = \ell_s^{-2} \big( {M}_{\a\b} \bp X^{\b} \p + \p X^{\b} {M}_{\a\b}  \bp \big). \label{noethervector} 
  }
In particular, using \eqref{nonlocal} it is not difficult to verify that the Noether currents of the deformed and undeformed backrounds satisfy $j_{(\a)} = \hat j_{(\a)}$.

It is also instructive to write the winding number as a conserved charge on the worldsheet. The Noether current associated with the gauge transformation $\delta B = d \Lambda$ reads
\eq{ 
\theta_{w}^{(\alpha)} =  \bp X^\alpha\p- \p X^\alpha\bp.
}
The conserved charge associated with this current corresponds to the winding number along the $X^\alpha$ direction and is given by
\eq{ 
q_{w}^{(\alpha)} \equiv \frac{1}{2\pi} \oint  \theta_{w}^{(\alpha), \tau} = \frac{1}{2\pi} \oint \p_\sigma X^\alpha = w^{(\alpha)}.
}

Once the boundary conditions of the $\hat X^{\a}$ coordinates have been determined, the worldsheet spectrum can be obtained from spectral flow of the underlying symmetry algebra. Following~\cite{Apolo:2019zai}, let us assume a general undeformed background $\hat M_{\a\b}$ satisfying
  \eq{
  \p_{[\a}  \hat M_{|m| \b]} = 0, \qquad \p_{[\a} { \hat M}_{\b] \bar{m}} = 0. \label{constraint}
  }
This constraint implies the existence of chiral and anti-chiral currents associated with shifts of the $\hat X^m$ and $\hat X^{\bar m}$ coordinates that are respectively defined by
  \eqsp{
\hat h_{(m)} &= \ell_s^{-2} \hat G_{m\a} \p \hat X^{\a} \bp  = \frac{1}{2} \hat j_{(m)} -  \frac{1}{2\ell_s^2} \hat M_{m\a} \hat \theta_w^{(\alpha)} , \\
\hat {\bar h}_{(\bar m)} &=  \ell_s^{-2} \hat G_{\bar{m}\a} \bp \hat X^{\a} \p = \frac{1}{2} \hat j_{(\bar m)} -  \frac{1}{2\ell_s^2} \hat M_{\a \bar m} \hat \theta_w^{(\alpha)}.\label{chiralcurrents}
  }
Consequently, the chiral currents are equivalent to the Noether currents associated with a combination of a translations and gauge transformations.

The boundary conditions \eqref{twistedbc} can then be implemented by a shift of the target space coordinates such that
\eq{
\hat X^{m} = \tilde X^{m} + \gamma^{(m)} z, \qquad \hat X^{\bar m} = \tilde X^{\bar m} - \gamma^{(\bar m)} \bar z, \label{specflow}
}
where $\tilde X^{m}$ and $\tilde X^{\bar m}$ satisfy trivial boundary conditions and are interpreted as the target space coordinates of the undeformed theory. The shift \eqref{specflow} is responsible for a spectral flow transformation of the symmetry algebra generated by the currents \eqref{chiralcurrents}. Furthermore, \eqref{specflow} is guaranteed to be a solution to the equations of motion provided that $\til X^{\a}$ satisfies the equations of motion and that the constraint \eqref{constraint} holds. 

A subtlety arises when the undeformed background features a $T^n$ factor such that one or more of the target space coordinates corresponds to a free field in the undeformed theory. For these coordinates, which we denote by $\hat X^{\check a}$, the corresponding components of the undeformed background field $\hat M_{\a\b}$ satisfy a stronger condition than \eqref{constraint}, namely
\eq{
\p_{\a} \hat M_{\check a \b} = 0, \qquad \p_{\a} \hat M_{\b \check a} = 0, \label{constraint2}
}
which implies the existence of chiral and anti-chiral currents $\hat h_{(\check a)}$ and $\hat{\bar h}_{(\check a)}$ associated with shifts of the same coordinate $\hat X^{\check a}$. Consequently, for this kind of coordinates the twisted and/or winding boundary conditions \eqref{twistedbc} can be implemented by the following generalization of the spectral flow transformation
\eq{
\hat X^{\check a} = \til X^{\check a} +  \g^{(\check  a)} (\s + \nu \tau). \label{specflow2}
}

The $\nu$ variable affects the relationship between the chiral currents of the undeformed theory and the Noether currents of the deformed one. This can be seen by writing the undeformed chiral currents $\til {h}_{(\check a)}$ and $\til {\bar h}_{(\check a)}$ in terms of the deformed ones $\hat {h}_{(\check a)}$ and $\hat {\bar h}_{(\check a)}$. The former are defined in \eqref{chiralcurrents} with all hats replaced by tildes, namely
  \eq{
\til h_{(m)} &= \ell_s^{-2} \hat G_{m\a} \p \til X^{\a} \bp, \qquad  \til {\bar h}_{(\bar m)} =  \ell_s^{-2} \hat G_{\bar{m}\a} \bp \til X^{\a} \p. \label{chiralcurrents2}
  }
 Using the shift of coordinates \eqref{specflow2} we then find
\eq{
 \til h_{(\check a)} &=  \hat  h_{(\check a)} -  \frac{1}{2}(1 + \nu) \ell_s^{-2} \hat G_{\check a \check a'} \gamma^{(\check a')} \bp, \qquad \til {\bar h}_{(\check a)} =  \hat  {\bar h}_{(\check a)} +  \frac{1}{2}(1 - \nu) \ell_s^{-2} \hat G_{\check a \check a'} \gamma^{(\check a')} \p. \label{undefchiralcurrents}
}
Note that the deformed chiral currents  $\hat h_{(\check a)}$ and $\hat {\bar h}_{(\check a)}$ depend on the Noether currents \eqref{noethervector} and the $\g^{(\a)}$ parameters characterizing the TsT transformations but are independent of $\nu$. The physical implication of this is that different choices of $\nu$ correspond to different choices of circles in the deformed background along which the conjugate momenta are quantized. The reason for this is that the undeformed chiral currents \eqref{undefchiralcurrents} must be independent of the deformation parameters. This is possible for different values of $\nu$ provided that linear combinations of the momenta featured in \eqref{undefchiralcurrents} are quantized and hence preserved by the deformation. As a result, different choices of $\nu$ lead to slightly different spectra due to the corresponding changes in the global properties of the deformed spacetime. In particular, in terms of the zero mode charges $\hat K_0 = \frac{1}{2\pi} \oint \hat h_{\check a}$ and $\hat{\bar K}_0 = \frac{1}{2\pi} \oint \hat{\bar h}_{\check a}$, the relationship between the deformed and undeformed chiral currents \eqref{undefchiralcurrents} becomes
\eq{
 \til K_0 &=  \hat  K_0 -  \frac{1}{2}(1 + \nu) \ell_s^{-2} \hat G_{\check a \check a'} \gamma^{(\check a')}, \qquad \til {\bar K}_0 =  \hat  {\bar K}_0 +  \frac{1}{2}(1 - \nu) \ell_s^{-2} \hat G_{\check a \check a'} \gamma^{(\check a')}, \label{undefchiralcurrents2}
}
where $\til K_0$ and $\til{\bar K}_0$ are the undeformed zero mode charges.

Let us now derive general formulae for the spectrum of strings winding on backgrounds obtained from multiple TsT transformations where some of the TsT coordinates may be taken from $T^n$ factors of the undeformed spacetime. In this case, the spectral flow transformations associated with the TsT transformations can be implemented in different ways via \eqref{specflow} or \eqref{specflow2}, leading to more general expressions for the spectrum than the ones obtained in \cite{Apolo:2019zai}. 

We first note that under the shifts of coordinates \eqref{specflow} and \eqref{specflow2}, the left-moving component of the deformed stress  tensor $ \hat{T} \equiv \hat{T}_{zz} = - \ell_s^{-2} \hat{G}_{\a\b} \p \hat{X}^{\a} \p \hat{X}^{\b}$ transforms as
\eqsp{
\hat{T}  &= \tilde T - 2 \ell_s^{-2} \hat G_{m\a} \p \tilde X^{\a} \gamma^{(m)} - (1 + \nu) \ell_s^{-2} \hat G_{ \check a\a} \p \tilde X^{\a} \gamma^{(\check a)} -  \ell_s^{-2} \hat G_{m m'} \gamma^{(m)} \gamma^{(m')}  \\
& \phantom{=} -  (1 + \nu)  \ell_s^{-2} \hat G_{m \check a} \gamma^{(m)} \gamma^{(\check a)} -  \frac{1}{4}(1 + \nu) ^2 \ell_s^{-2} \hat G_{\check a \check a'} \gamma^{(\check a)} \gamma^{(\check a')}  , 
 }
  where $\tilde T$ denotes the stress tensor in the undeformed theory. Next, using the definition of the chiral currents \eqref{chiralcurrents} together with eqs.~\eqref{specflow} and \eqref{specflow2} we find that
  \eq{
  \!\!\!\! \!\! \begin{split}
  \frac{1}{2\pi} \oint  2 \til G_{m\a} \p \til X^{\a} &= \ell_s^{2}p_{(m)} + \big( \hat M_{mm'} - 2 \hat G_{mm'} \big) \g^{(m')}   + \big[ \hat M_{m \check a}  - (1 + \nu)   \hat G_{m \check a} \big] \g^{(\check a)} \\
  & \hspace{15pt} + \hat M_{m \bar m} \g^{(\bar m)} + \frac{1}{2\pi} \oint \hat M_{m i} \p_\s \hat X^i, \label{tilG1}
 \end{split} \\
 \!\!\!\! \!   \begin{split}
  \frac{1}{2\pi} \oint 2  \til G_{\check a\a} \p \til X^{\a} &= \ell_s^{2} p_{(\check a)} +  \big( \hat M_{\check a m}  - 2  \hat G_{\check a m}  \big) \g^{(m)} + \big[ \hat M_{\check a \check a'} - (1 + \nu)   \hat G_{\check a \check a'}   \big]  \g^{(\check a')} + \hat M_{\check a \bar m} \g^{(\bar m)} \label{tilG2}
 \end{split}
  }
  where we used the fact that $\hat M_{mm'}$, $\hat M_{m\bar m}$, $\hat M_{\bar m\bar m'}$,  $\hat M_{\check a \a}$, and $\hat M_{\a \check a}$ are all constant, a result that follows from the constraints \eqref{constraint} and \eqref{constraint2} imposed on the undeformed background. The equations above imply that the zero mode $L_0 = \hat L_0 = -\frac{1}{2\pi} \oint \hat T$ of the left-moving component of the stress tensor satisfies
\eq{
\!\!\!\! L_0 &= \til L_0 + p_{(m)} \gamma^{(m)}  + \frac{1}{2}(1 + \nu)  p_{(\check a)} \gamma^{(\check a)} + \ell_s^{-2} \hat M_{m \bar m} \gamma^{(m)} \gamma^{(\bar m)}+ \frac{1}{2} (1 - \nu) \ell_s^{-2} \hat M_{m \check a} \g^{(m)} \g^{(\check a)} \notag \\
\!\!\!\! &\phantom{=} + \frac{1}{2} (1 + \nu)  \ell_s^{-2} M_{\check a \bar m} \g^{(\check a)} \g^{(\bar m)}  + \frac{1}{4}(1 - \nu^2) \ell_s^{-2}  \hat M_{\check a \check a'} \g^{(\check a)} \g^{(\check a')} + \frac{\g^{(m)}}{2\pi \ell_s^2} \oint \hat M_{m i} \p_\s \hat X^i, \label{L0}
}
where $\til L_0$ denotes the zero mode before the deformation. Similarly, following the steps described above, we find that the right-moving zero mode of the stress tensor $\bar L_0 = \hat {\bar L}_0 = -\frac{1}{2\pi} \oint \hat {\bar T}$ is related to its undeformed value $\til {\bar L}_0$ by
\eq{
\!\!\!\!  {\bar L}_0 &= \til {\bar L}_0 - p_{(\bar m)} \gamma^{(\bar m)} - \frac{1}{2}(1 - \nu) p_{(\check a)} \gamma^{(\check a)} + \ell_s^{-2} \hat M_{m \bar m} \gamma^{(m)} \gamma^{(\bar m)} + \frac{1}{2}(1 + \nu)  \ell_s^{-2} \hat M_{\check a \bar m} \g^{(\check a)} \g^{(\bar m)} \notag \\
\!\!\!\! & \phantom{=} + \frac{1}{2}(1 - \nu) \ell_s^{-2} \hat M_{m \check a} \g^{(m)} \g^{(\check a)} + \frac{1}{4}(1 - \nu^2)  \ell_s^{-2}  \hat M_{\check a \check a'} \gamma^{(\check a)} \gamma^{(\check a')} + \frac{\g^{(\bar m)}}{2\pi \ell_s^2} \oint \hat M_{i \bar m} \p_\s \hat X^i.    \label{L0bar}
}
Finally, imposing the Virasoro constraints before and after the deformation, we can express the spectrum of the deformed theory in terms of the undeformed one.

Note that eqs.~\eqref{L0} and \eqref{L0bar} generalize the expressions for the zero modes of the stress tensor derived in \cite{Apolo:2019zai} in two ways: first, by including the effect of multiple TsT transformations; and second, by allowing the TsT coordinates to be taken from a $T^n$ factor of the undeformed spacetime. In particular, we note that when $\nu = 1$, \eqref{specflow2} reduces to the first equation in \eqref{specflow}, while it reduces to the second equation in \eqref{specflow} when $\nu = -1$. In both of these cases, the shift of coordinates affects only one component of the stress tensor such that \eqref{L0} and \eqref{L0bar} take the same general form derived in \cite{Apolo:2019zai}. In contrast, for any other values of $\nu$ the spectral flow transformation induced by \eqref{specflow2} affects both of these zero modes. As a result, the worldsheet spectrum depends on the choice of $\nu$, in addition to the choice of the TsT/winding coordinates, and the choice of the undeformed background.


\section{More tri-TsT backgrounds} \label{ap:tstbackgrounds}

In this appendix we describe two additional kinds of TsT-transformed backgrounds where the internal $U(1)$ currents in the $T\bar T+J\bar T+T\bar J$ deformation are associated with two isometries of $T^4$ or two isometries of $S^3$. We also derive the spectrum of strings winding on these backgrounds and show that it differs slightly from the spectrum derived in section \ref{se:spectrum}.

\subsection{$J$ and $\bar J$ associated with two isometries of $T^4$} \label{ap:case2}

Let us first consider the case where the $J$ and $\bar J$ currents in the $T\bar T+J\bar T+T\bar J$ deformation are associated with two different isometries of $T^4$. This case corresponds to the second entry of table \ref{table1}. 

\subsubsection{Neutral black holes} \label{ap:case2neutral}

We begin by considering the simpler class of neutral tri-TsT black holes obtained by performing the following sequence of TsT transformations on the undeformed BTZ$\,\times S^3 \times T^4$ background
\eq{
& \textrm{TsT}_{(u,v;\, \check\mu_0)}, \qquad \textrm{TsT}_{(y,v;\, \check\mu_+)},  \qquad \textrm{TsT}_{(u,y_8;\, \check\mu_-)}.  \label{tstrecipeap1}
}
The resulting TsT-transformed background is more complicated than \eqref{neutraltstbtz}, where $J$ and $\bar J$ were taken from the same isometry of $T^4$, and can be written in ten dimensions as
 \eq{
\!\!\!\! ds^2_{(10)} &=  \frac{\ell^2 dr^2}{4(r^2 - 4 T_u^2 T_{v}^2)} + \ell^2_4 {\textstyle\sum}_{i=9}^{10}\, dy_i^2 + d\Omega_3^2 + h \Big[ \ell^2 r du d{v} + \ell^2 (1 + \l_+^2 T_{v}^2) T_u^2 du^2     \notag \\
 & \phantom{=} +  \ell^2 (1 + \l_-^2 T_u^2) T_{v}^2 d{v}^2  - \ell \ell_4 (r + 2 \l_0 T_u^2 T_{v}^2) (\l_+ du dy + \l_- d{v} dy_8) + \ell_4^2 \l_+ \l_- r dy dy_8  \notag \\
 & \phantom{=} +   \ell_4^2 (1 + \l_0 r + \l_-^2 T_u^2 + \l_0^2 T_u^2 T_{v}^2) dy^2  +   \ell_4^2 (1 + \l_0 r + \l_+^2 T_{v}^2 + \l_0^2 T_u^2 T_{v}^2) dy_8^2  \Big],  \label{neutraltstbtz2} \\
\!\!\!\! B_{(10)}& = B_{\Omega_3} + \frac{h}{2} \Big\{ (r + 2 \l_0 T_u^2 T_{v}^2) (\ell^2 d{v} \wedge du + \ell_4^2 \l_+ \l_- dy \we dy_8 ) + \ell \ell_4\big[ \l_+ r  \, du \wedge dy     \notag \\
 &\phantom{=}  - \l_- r \, d{v} \we dy_8 - 2\l_- T_u^2 (1+\l_+^2 T_{v}^2) du \we dy_8  + 2 \l_+ T_{v}^2 (1 + \l_-^2 T_u^2) d{v} \we dy\Big] \Big\}   ,  \notag  \\
\!\!\!\! e^{2\Phi_{(10)}} & = \frac{k_4^2 k  h}{\eta p}, \notag
 }
 where we used the definition of the $\l_i$ parameters introduced in \eqref{dictionarylambdas}, while the quantization parameter $\eta$ and the flow function $h$ are given by
 \eq{
 \eta^{-1} &= 1 - \l_0^2 T_u^2 T_{v}^2, \\
 h ^{-1} &= 1 + \l_0 r+  \l_-^2 T_u^2 + \l_+^2 T_{v}^2 + \l_+^2 \l_-^2 T_u^2 T_{v}^2 + \l_0^2 T_u^2 T_{v}^2. \label{shdef2}
 }

There are several similarities between \eqref{neutraltstbtz2} and the netural tri-TsT background \eqref{neutraltstbtz} constructed in section \ref{se:specialtstbtz}. We first note that both backgrounds satisfy the same identification of coordinates as the undeformed BTZ$\,\times S^3 \times T^4$ spacetime, namely \eqref{coordsim1} -- \eqref{coordsim3}. Similarly, the quantization parameter $\eta$, which is determined by requiring the magnetic and electric charges to remain quantized after the TsT transformations, is the same for both of these backgrounds. Importantly, both \eqref{neutraltstbtz2} and \eqref{neutraltstbtz} describe neutral black holes (with a horizon at $r_h = 2 T_u T_v$) whose deformed $U(1)$ charges vanish.  Although \eqref{neutraltstbtz2} cannot be used to describe the NS vacuum for this class of TsT-transformed backgrounds --- we will discuss how this can be done in the next section --- we note that when $T_u = T_v = 0$ in \eqref{neutraltstbtz2} we obtain the Ramond vacuum found in \cite{Chakraborty:2019mdf}. 

There is one important difference between the tri-TsT black holes \eqref{neutraltstbtz2} and  \eqref{neutraltstbtz} regarding the range of parameters for which the metric is free of pathologies. We find that the black holes \eqref{neutraltstbtz2} are free of curvature singularities and CTCs provided that
\eq{
\l_0 > 0, \label{singcondition2}
}
which is more constraining than the condition \eqref{singcondition1} where $\l_0$ can be negative. When $\l_0 < 0$, there is a curvature singularity at $h^{-1} = 0$ which corresponds to the following value of the radial coordinate
\eq{
r^*_h = \frac{1 +  \l_-^2 T_u^2 + \l_+^2 T_{v}^2 + \l_+^2 \l_-^2 T_u^2 T_{v}^2 + \l_0^2 T_u^2 T_{v}^2}{| \l_0 |}.
}
This is also the location beyond which CTCs appear in the deformed background \eqref{neutraltstbtz2}.  The difference between \eqref{singcondition1} and \eqref{singcondition2} is related to the different spectra of $T\bar T + J\bar T + T\bar J$-deformed CFTs when the $J$ and $\bar J$ currents are taken from the same or different $U(1)$ symmetries of the undeformed theory. This is described in more detail in section \ref{ap:case2spectrum} where we derive the spectrum of strings winding on \eqref{neutraltstbtz2}. In particular, we note that \eqref{singcondition2} guarantees that the spectrum of both neutral and charged thermal states in the dual $T\bar T + J\bar T + T\bar J$-deformed CFT is always real.


\subsubsection{Charged black holes} \label{se:case2charged}

In order to construct charged black holes for this class of TsT-transformed backgrounds, we first need to turn on a left and right-moving $U(1)$ charges along the $\p_{y}$ and $\p_{y_8}$ isometries of the BTZ$\,\times S^3 \times T^4$ spacetime. This can be accomplished by a spectral flow transformation that is given by
  \eq{
   \!\!\! y &\to y - \frac{\ell}{\ell_4} \a_u u,  \quad  y_8 \to y_8  - \frac{\ell}{\ell_4} \a_{v} v, \quad  B_{(10)} \! \to B_{(10)} \!  + \ell \ell_4 \big( \a_u du \we dy  - \a_{v} d{v} \we dy_8\big). \label{cc2}
  }
Note that we impose the same identifications of coordinates \eqref{coordsim1} -- \eqref{coordsim3}  after the gauge transformations~\eqref{cc2}, which guarantees that the new solution differs from the BTZ$\,\times S^3 \times T^4$ background. The resulting background features left and right-moving energies that are given by \eqref{ELERbtzU1}, and the same left and right-moving $U(1)$ charges \eqref{qLqRbtzU1} (although the latter are now associated with the isometries along $y$ and $y_8$). The reason why the change of coordinates \eqref{cc2} yields a black hole with $U(1)$ charges can be understood from a three-dimensional perspective, since \eqref{cc2} turns on additional gauge fields associated with translations along the $y$ and $y_8$ coordinates in the three-dimensional theory. In addition, we note that the gauge transformation \eqref{cc2} preserves the same boundary conditions on $\hat M_{\alpha\beta} = \hat G_{(10)\alpha\beta} + \hat B_{(10)\alpha\beta}$ as the undeformed BTZ$\,\times S^3 \times T^4$ background, namely
\eq{
\hat M_{uy} =  \hat M_{y_8 v}  = \hat M_{uv} = 0. \label{bfieldbc2}
}

Performing the sequence of TsT transformations \eqref{tstrecipeap1} on the charged background described above yields black hole solutions that are dual to general thermal states in the class of $T\bar T + J \bar T + T \bar J$-deformed CFTs under consideration. These black holes are 7-parameter solutions to the supergravity equations of motion that are characterized by three deformation parameters $(\l_0, \l_+, \l_-)$ and four phase space variables $(T_u, T_v, \a_u, \a_v)$ parametrizing its mass, angular momentum, and two $U(1)$ charges. In particular, after analytic continuation of $(T_u, T_v)$, and an appropriate choice of $(\a_u, \a_v)$, the resulting background can be used to describe the NS vacuum in this class of $T\bar T + J \bar T + T \bar J$-deformed CFTs.


\subsubsection{Perturbative spectrum} \label{ap:case2spectrum}

The spectrum of strings winding on backgrounds obtained from a sequence of TsT transformations was derived in appendix \ref{ap:spectrum}. The main ingredients in this derivation are the undeformed background and the directions along which we perform the TsT transformations. 

For the charged tri-TsT backgrounds considered in this section, the undeformed background is described by the charged BTZ$\,\times S^3 \times T^4$ background obtained from \eqref{10dbackground} and \eqref{btz} together with the spectral flow transformation \eqref{cc2}. Let us recall that TsT transformations induce twisted boundary conditions on the target space coordinates of the undeformed background, the latter of which are characterized by the so-called $\gamma$ parameters \eqref{twistedbc}. For the sequence of TsT transformations \eqref{tstrecipeap1}, the $\gamma$ parameters are given by
\eqsp{
\g^{(u)} &= w + 2 \check \mu_0 p_{(v)} + 2 \check \mu_- p_{(y_8)}, \qquad \g^{(y)} = w^{(y)} + 2 \check \mu_+ p_{(v)}, \\
\g^{(v)} &= w - 2 \check \mu_0 p_{(u)} - 2 \check \mu_+ p_{(y)}, \qquad \g^{(y_8)} = w^{(y_8)} - 2\check \mu_- p_{(u)}, \label{gammaap1}
}
where $w$ denotes the winding along the spatial circle $\vp = \frac{1}{2}(u + v)$, while $w^{(y)}$ and $w^{(y_8)}$ denote the winding along the $y$ and $y_8$ coordinates. Note that in contrast to \eqref{twistedbcexp}, we must impose additional twisted boundary conditions along the $\hat y_8$ coordinate, which follows from the fact that this is one of the TsT directions in \eqref{tstrecipeap1}.

The zero modes $(L_0, \bar L_0)$ of the worldsheet stress tensor for this class of TsT-transformed backgrounds are given in terms of the undeformed values $(\til L_0, \til{\bar L}_0)$ by
\eqsp{
L_0& = \til L_0 + p_{(u)} \g^{(u)} + \frac{1}{2} \Big( p_{(y)} \g^{(y)} +  p_{(y_8)} \g^{(y_8)} \Big)+ \frac{k_4}{4} \Big[\big( \g^{(y)}\big)^2 + \big( \g^{(y_8)}\big)^2\Big], \\
{\bar L}_0& = \til {\bar L}_0 - p_{(v)} \g^{(v)} - \frac{1}{2} \Big( p_{(y)} \g^{(y)} + p_{(y_8)} \g^{(y_8)} \Big) -\frac{k_4}{4} \Big[\big( \g^{(y)}\big)^2 + \big( \g^{(y_8)}\big)^2\Big],\label{apzeromodes1}
}
where we used \eqref{L0} and \eqref{L0bar} with $\nu = 0$, which is compatible with the identification of coordinates \eqref{coordsim2}. The momenta featured in \eqref{apzeromodes1} are related to the deformed energies $E_{L,R}(\check\mu_i)$ and the undeformed $U(1)$ charges $q_L(0)$ and $q'_R(0)$ of the string via
\eq{
p_{(u)} &= \ell E_L(\check\mu_i) , \qquad\qquad\qquad \,\,\,\, p_{(v)} = - \ell E_R(\check\mu_i), \\
 q_L(0) &= \frac{1}{2}\big( p_{(y)} + k_4 w^{(y)} \big), \qquad\, q'_R(0) = \frac{1}{2} \big( p_{(y_8)} - k_4 w^{(y_8)} \big).
}
Thus, imposing the Virasoro constraints before and after the deformation, we find that the spectrum of strings winding on this class of tri-TsT backgrounds can be written as
   \eqsp{
   E_{L}(0)  &= E_{L}(\check\mu_i) + \frac{2}{w} \Pi_2(\check\mu_i), \qquad   E_R(0)  = E_R(\check\mu_i)  +\frac{2}{w}   \Pi_2(\check\mu_i),
   \label{spectrumap1}
   }
where $\Pi_2(\check\mu_i)$ is given by
   \eq{
\!\! \! \Pi_2(\check\mu_i)  =  \check\mu_- q'_R(0) E_L(\check\mu_i)  \! -\! \check\mu_+ q_L(0)  E_R(\check\mu_i) \! - \!  \ell \check\mu_0 E_L(\check\mu_i) E_R(\check\mu_i) \! + \! \frac{\ell k_4}{2} \big[ \check\mu_+^2 E_R(\check\mu_i)^2 + \check\mu_-^2 E_L(\check\mu_i)^2\big]. \label{Pidefap1}
   }
Additionally, the deformed $U(1)$ charges for strings on these backgrounds can be obtained from the zero modes of the chiral currents given in~\eqref{undefchiralcurrents} and satisfy
\eq{
q_L(\check \mu_i) = q_L (0) -  k_4 \check\mu_+ \ell E_R(\check\mu_i) , \qquad q'_R (\check \mu_i) = q'_R(0) + k_4 \check \mu_- \ell E_L(\check\mu_i). \label{U1chargesap1}
}    
The deformed $U(1)$ charges allow us to write the deformed energies in a more compact way where the effect of the $J\bar T$ and $T\bar J$ deformations is captured entirely by the deformed $U(1)$ currents
\eq{
\!\!\!\!\! w \ell E_{L,R}(0) + \frac{q_L(0)^2 + q'_R(0)^2}{k_4} = w \ell E_{L,R}(\check\mu_i)+ \frac{q_L(\check\mu_i)^2 + q'_R(\check\mu_i)^2}{k_4} - 2 \ell^2 \check\mu_0 E_L(\check \mu_i) E_R (\check\mu_i).  
}  
   
The spectrum described by \eqref{spectrumap1} and \eqref{U1chargesap1} is valid for strings winding on TsT-transformed backgrounds obtained from the charged BTZ$\,\times S^3 \times T^4$ background via \eqref{tstrecipeap1}, and includes deformed versions of the NS vacuum and charged rotating black holes. The spectrum \eqref{spectrumap1} differs slightly from the $T\bar T + J\bar T + T\bar J$ spectrum given in \eqref{spectrum} where the $J$ and $\bar J$ currents are taken from the same isometry of $T^4$. From the field theory side, this is a result of having $J$ and $\bar J$ currents that are the chiral and antichiral components of different $U(1)$ currents in the deformed theory. We also note that \eqref{spectrumap1} reduces to the spectrum derived for the Ramond vacuum in \cite{Chakraborty:2019mdf} using a different approach.


\subsection{$J$ and $\bar J$ associated with two isometries of $S^3$} \label{ap:case3}

Let us now consider the case where the $J$ and $\bar J$ currents are associated with two isometries of $S^3$ (the fourth entry in table \ref{table1}). This case corresponds to a $T\bar T + J\bar T + T\bar J$ deformation where the left and right-moving $U(1)$ currents $J$ and $\bar J$ are part of a larger $SU(2)_L \times SU(2)_R$ symmetry group before the deformation. 

\subsubsection{Neutral black holes} \label{ap:case3neutral}

In this case the deformed neutral black holes can be obtained by performing the following sequence of TsT transformations on the undeformed BTZ$\,\times S^3 \times T^4$ background
\eq{
& \textrm{TsT}_{(u,v;\, \check\mu_0)}, \qquad \textrm{TsT}_{(\psi,v;\, 2\check\mu_+)},  \qquad \textrm{TsT}_{(u,\phi;\, 2\check\mu_-)}.  \label{tstrecipeap2}
}
The factors of 2 accompanying the deformation parameters $\check\mu_\pm$ are necessary to account for the normalization of the Noether currents associated with translations along $\psi$ and $\phi$, which follows from the identification of coordinates given in \eqref{coordsim1}. The TsT transformation \eqref{tstrecipeap2} of the BTZ$\,\times S^3 \times T^4$ background is described by the following ten-dimensional metric, $B$-field, and dilaton
  \eq{
\!\! ds^2_{(10)} &= \frac{\ell^2 dr^2}{4(r^2 - 4 T_u^2 T_{v}^2)} + \ell^2_4{\textstyle\sum}_{i=7}^{10}\, dy_i^2 + \frac{\ell^2}{4} d\t^2 + \frac{\ell^2 h}{4} \Big\{ (4r + 8\l_+ \l_- \cos\t \,T_u^2 T_v^2) du d{v}  \notag \\
 & \hspace{15pt}+ 4(1 + \l_+^2 T_v^2) T_u^2 du^2  + 4 (1 + \l_-^2 T_u^2) T_v^2 dv^2 - 2(r + 2 \l_0 T_u^2 T_v^2)( \l_+ du d\psi + \l_- dv d\phi) \notag \\
  & \hspace{15pt} - 4 \cos\t (r +  \l_0 T_u^2 T_v^2)(\l_+ du d\phi + \l_- dv d\psi) - 4 \cos\t\,( \l_- T_u^2 du d\psi + \l_+ T_v^2 dv d\phi)  \notag \\
  & \hspace{15pt}+   (1 + \l_0 r + \l_-^2 T_u^2 + \l_0^2 T_u^2 T_v^2) d\psi^2  +   (1 + \l_0 r + \l_+^2 T_v^2 + \l_0^2 T_u^2 T_v^2) d\phi^2 \notag \\
 & \hspace{15pt} + \big[ 2(1+ \l_0 r + \l_0^2 T_u^2 T_v^2) \cos\t + \l_+ \l_- r \big] d\psi d\phi \Big\}, \label{sbackground3}  \\
\!\! B_{(10)}  & = \frac{\ell^2 h}{4} \Big\{ 2(r + 2 \l_0 T_u^2 T_v^2 - 2 \l_+ \l_- T_u^2 T_v^2\cos\t) dv \wedge du + (\l_+ r - 2 \l_- T_u^2 \cos\t) du \we d\psi  \notag  \\
  & \hspace{15pt}  - (\l_- r - 2 \l_+ T_v^2\cos\t) dv \we d\phi + 2 \big[ \l_+ (r + \l_0 T_u^2 T_v^2) \cos\t  - \l_- T_u^2 (1 + \l_+^2 T_v^2) \big] du \we d\phi \notag  \\
  & \hspace{15pt} - 2 \big[ \l_- (r + \l_0 T_u^2 T_v^2) \cos\t  -  \l_+ T_v^2 (1 + \l_-^2 T_u^2) \big] dv \we d\psi  \notag  \\
 & \hspace{15pt} -\big[ (1 + \l_0 r + \l_0^2 T_u^2 T_v^2)\cos\t - \l_+ \l_-(\tfrac{r}{2} + \l_0 T_u^2 T_v^2) \big] d\psi \we d\phi \Big\}   \notag    \\
\!\! e^{2\Phi_{(10)}} & = \frac{k_4^2 k h}{\eta p}, \notag
 }
where the deformation parameters $\l_i$ are related to $\check\mu_i$ via $\l_i = 2 k \check\mu_i$, while the quantization parameter $\eta$ and the flow function $h$ satisfy
\eqsp{
\eta^{-1} &= 1 - \l_0^2 T_u^2 T_{v}^2 -  \l_+^2 T_{v}^2-  \l_-^2 T_u^2 + \l_+^2 \l_-^2 T_u^2 T_{v}^2, \\
h ^{-1} &= 1 +  \big(\l_0 - 2 \l_+ \l_- \cos\t \big)(r + \l_0 T_u^2 T_{v}^2)  +  \l_+^2 T_{v}^2 +  \l_-^2 T_u^2 + \l_+^2 \l_-^2 T_u^2 T_{v}^2.  \label{etaap2} 
}
Unlike the neutral backgrounds \eqref{neutraltstbtz} and \eqref{neutraltstbtz2}, we see that in this case the flow function acquires dependence on $\theta$ in addition to its dependence on the radial coordinate. Furthermore, we note that a real value of the dilaton requires $\eta^{-1} > 0$, which leads to constrains on the thermodynamic potentials of the dual $T\bar T + J\bar T + T\bar J$-deformed CFT.

For this class of TsT-transformed backgrounds, the identification of coordinates is different after the TsT transformations and is given by \eqref{coordsim2} and \eqref{coordsim3} together with
\eq{
(v, \psi) \sim ( v - 2\pi\l_+ , \psi + 4\pi), \quad  (u,v,\psi, \phi) \sim (u - \pi \l_- , v - \pi \l_+ , \psi + 2 \pi, \phi + 2\pi). \label{coordsim4}
}
The identification of coordinates \eqref{coordsim4} differs from \eqref{coordsim1} and is necessary for a smooth background geometry. The identification \eqref{coordsim4} affects the definition of spacelike surfaces and, consequently, the values of the conserved charges. In particular, \eqref{coordsim4} guarantees that the magnetic charge in this class of backgrounds is always given by $Q_m = k$ for any values of the temperatures. Similarly, the quantization parameter in \eqref{etaap2} is sensitive to the global properties of the spacetime and guarantees that the electric charge is also preserved after the deformation and given by $Q_e = p$. Relatedly, the background \eqref{sbackground3} features several compact $U(1)$ isometries associated with the circles described by \eqref{coordsim2}, \eqref{coordsim3}, and \eqref{coordsim4}. Translations along these circles are generated by the Killing vectors
\eq{
\p_u + \p_{v},  \qquad 2\p_\psi - \l_+  \p_{v}, \qquad 2 \p_\phi -  \l_-  \p_u, \qquad \p_{y_i}, \label{killingsap2}
}
and their gravitational charges --- which correspond to the angular momentum and several internal $U(1)$ charges --- are quantized and hence preserved by the deformation. 

The backgrounds obtained via the sequence of TsT transformations \eqref{tstrecipeap2} are dual to states in $T\bar T + J\bar T + T\bar J$-deformed CFTs where the $J$ and $\bar J$ currents belong to an $SU(2)_L \times SU(2)_R$ symmetry group before the deformation. In particular, when the $T_u$ and $T_v$ parameters are real and nonvanishing, the tri-TsT backgrounds \eqref{sbackground3} describe rotating black hole solutions that are dual to a special class of thermal states in $T\bar T + J\bar T + T\bar J$-deformed CFTs characterized by vanishing values of the deformed $U(1)$ charges. On the other hand, when $T_u = T_v = 0$, the background \eqref{sbackground3} describes the Ramond vacuum of the dual field theory.

In analogy with previous cases, the black holes in this class of TsT-transformed backgrounds are free of curvature singularities and CTCs provided that $h^{-1} > 0$. This means that the deformation parameters must satisfy
\eq{
\l_0 - 2 |\l_+ \l_-| > 0, \label{singcondition3}
}
where the expectation value is needed since the value of $h^{-1}$ depends on $\t$. When \eqref{singcondition3} is violated, there is a curvature singularity at $h^{-1} = 0$ which is located at a $\t$-dependent value of the radial coordinate
\eq{
r^*_h = \frac{1 +  \big(\l_0 -  2\l_+ \l_- \cos\t \big) \l_0 T_u^2 T_{v}^2  +  \l_+^2 T_{v}^2 +  \l_-^2 T_u^2 +  \l_+^2 \l_-^2 T_u^2 T_{v}^2}{\big| \l_0 - 2 \l_+ \l_- \cos\t \big |}.
}
Closed timelike curves also appear in the region $r > r^*_h$ when \eqref{singcondition3} is not satisfied.

It is interesting to note that at fixed $\theta =\pi/2$, the finite-temperature generalization of \eqref{sbackground3} is locally equivalent to the black hole \eqref{neutraltstbtz2} considered in section \ref{ap:case2} up to a constant shift of the dilaton. This can be seen by exchanging the $(\psi,\phi)$ coordinates with $(2y_1, 2y_2)$. The global properties of these spacetimes differ, however, due to the identification \eqref{coordsim4}. This is the reason why these backgrounds are locally equivalent only up to a shift of the dilaton, since the latter is sensitive to the global properties of the spacetime through the quantization parameter $\eta$.


\subsubsection{Charged black holes} \label{ap:case3charged}

We can construct charged TsT black holes for this class of $T\bar T + J\bar T + T\bar J$ deformations by turning on $U(1)$ charges associated with the $\psi$ and $\phi$ coordinates of the 3-sphere. This can be achieved by performing the following spectral flow transformation on the BTZ$\,\times S^3 \times T^4$ background before the deformation
  \eq{
   \psi &\to \psi  - \a_u u,  \quad \phi \to \phi  - \a_{v} {v}, \quad B_{(10)} \to B_{(10)}  + \ell^2 \big( \a_u du \we d\psi  - \a_{v} d{v} \we d\phi\big). \label{cc3}
  }
As discussed previously, we must also impose the identification of coordinates \eqref{coordsim1} -- \eqref{coordsim3} after the gauge transformation \eqref{cc3}. The resulting background is globally inequivalent to the BTZ$\,\times S^3 \times T^4$ background and features the same left and right-moving energies given in \eqref{ELERbtzU1}, and the same values of the left and right-moving $U(1)$ charges shown in \eqref{qLqRbtzU1}, although the latter are now associated with translations along the $\psi$ and $\phi$ coordinates. In addition, we note that the spectral flow transformation \eqref{cc3} preserves the following boundary conditions on $\hat M_{\alpha\beta} = \hat G_{(10)\alpha\beta} + \hat B_{(10)\alpha\beta}$
\eq{
\hat M_{u\psi} = \hat M_{\phi v}  = \hat M_{uv} = 0. \label{bfieldbc3}
}

The tri-TsT holes obtained from the sequence of TsT transformations \eqref{tstrecipeap2} describe charged thermal states in single-trace $T\bar T + J\bar T +T \bar J$-deformed CFTs where the $J$ and $\bar J$ currents are part of a larger $SU(2)_L \times SU(2)_R$ symmetry group before the deformation. These black holes generalize the special solutions described in section \ref{ap:case3neutral} and feature arbitrary values of the deformed $U(1)$ charges. In particular, in the pure $J\bar T$ case where $\l_0$ and  $\l_-$ vanish, these solutions generalize the neutral black holes constructed in \cite{Apolo:2019yfj}. In addition, we note that the NS vacuum of the dual $T\bar T + J\bar T + T\bar J$-deformed CFT can be obtained by an appropriate choice of the phase space parameters.


\subsubsection{Perturbative spectrum} \label{ap:case3spectrum}

Let us conclude this section by deriving the spectrum of strings winding on this class of TsT-transformed backgrounds. The $\gamma$ parameters characterizing the twisted boundary conditions induced by the sequence of TsT transformations \eqref{tstrecipeap2} are given by
\eqsp{
\g^{(u)} &= w + 2 \check \mu_0 p_{(v)} + 4 \check \mu_- p_{(\phi)}, \qquad \g^{(\psi)} =  4 \check \mu_+ p_{(v)}, \\
\g^{(v)} &= w - 2 \check \mu_0 p_{(u)} - 4 \check \mu_+ p_{(\psi)}, \qquad \g^{(\phi)} = -4\check \mu_- p_{(u)}.\label{gammaap2}
}
These $\gamma$ parameters take a similar form as \eqref{gammaap1}, except that ($i$) we have not included winding for the $\psi$ and $\phi$ coordinates since it is not necessary to distinguish between chiral and antichiral currents ($\psi$ leads to a chiral current while $\phi$ leads to antichiral one); and ($ii$) there is a rescaling of $\check\mu_\pm$ which, as described previously, is necessary to account for the normalization of the Noether currents generating translations along $\psi$ and $\phi$. Using \eqref{L0} and \eqref{L0bar}, the deformed zero modes of the worldsheet stress tensor are given by
\eq{
L_0& = \til L_0 + p_{(u)} \g^{(u)} + p_{(\psi)} \g^{(\psi)}, \qquad {\bar L}_0= \til {\bar L}_0 - p_{(v)} \g^{(v)} - p_{(\phi)} \g^{(\phi)} ,\label{apzeromodes2}
}
where $(\til L_0, \til{\bar L}_0)$ denote the zero modes before the deformation. 

It is convenient to write the spectrum in terms of the (un)deformed energies and the undeformed $U(1)$ charges. For the class of TsT transformations \eqref{tstrecipe2}, the undeformed $U(1)$ charges are given by
\eqsp{
\frac{q_L(0)}{2} &\equiv \frac{1}{\pi} \oint \til h_{(\psi)} = \frac{1}{\pi} \oint \Big[  \hat  h_{(\psi)} -  \ell_s^{-2} \hat G_{\psi \psi} \gamma^{(\psi)}  \Big] =  p_{(\psi)} - k \check\mu_+ p_{(v)} , \\
\frac{q'_R(0)}{2} &\equiv \frac{1}{\pi} \oint \til{\bar h}_{(\phi)} = \frac{1}{\pi} \oint \Big[  \hat{\bar h}_{(\phi)} + \ell_s^{-2} \hat G_{\phi \phi} \gamma^{(\phi)}  \Big] =  p_{(\phi)} - k \check\mu_- p_{(u)}, \label{qLqRap2}
}
where the deformed $\hat h_{(\a)}$ and undeformed $\til h_{(\a)}$ currents are defined in \eqref{chiralcurrents} and \eqref{chiralcurrents2}. Note that these charges are preserved by the deformation despite their dependence on the $\check \mu_\pm$ parameters. This follows from the fact that these linear combinations of the momenta are quantized as can be seen from the Killing vectors generating the compact $U(1)$ isometries of the background \eqref{killingsap2}. Using $p_{(u)} = \ell E_L(\check\mu_i)$, $p_{(v)} = - \ell E_R(\check\mu_i)$, and imposing the Virasoro constraints before and after the deformation, we then obtain
 \eqsp{
   E_{L}(0)  &= E_{L}(\check\mu_i) + \frac{2}{w} \Pi_3(\check\mu_i), \qquad   E_R(0)  = E_R(\check\mu_i)  +\frac{2}{w}   \Pi_3(\check\mu_i),
   \label{spectrumap2}
   }
where $\Pi_3(\check\mu_i)$ is given by
   \eq{
\!\! \! \Pi_3(\check\mu_i)  =  \check\mu_- q'_R(0) E_L(\check\mu_i)  \! -\! \check\mu_+ q_L(0)  E_R(\check\mu_i) \! - \!  \ell \check\mu_0 E_L(\check\mu_i) E_R(\check\mu_i) \! + \! 2\ell k \big[ \check\mu_+^2 E_R(\check\mu_i)^2 + \check\mu_-^2 E_L(\check\mu_i)^2\big]. \label{Pidefap2}
   }
In addition, we find that the deformed $U(1)$ charges, which correspond to the zero modes of the chiral and antichiral currents $\hat h_{(\psi)}$ and $\hat{\bar h}_{(\phi)}$, read
\eq{
q_L(\check\mu_i)  = q_L(0) - 4 k \check \mu_+ E_R(\check\mu_i), \qquad q'_R(\check\mu_i)  = q'_R(0) + 4 k \check \mu_- E_L (\check\mu_i). \label{U1defap2}
}

The spectrum described in \eqref{spectrumap2} and \eqref{U1defap2} is valid for any tri-TsT background obtained from the charged BTZ$\,\times S^3 \times T^4$ background described by \eqref{10dbackground} and \eqref{btz} together with the spectral flow transformation \eqref{cc3}. This spectrum agrees with the spectrum derived in appendix \ref{ap:case2} up to an identification of the $U(1)$ level that is given by $k_4 \to 4 k$. This identification follows from the aforementioned normalization of the $U(1)$ currents which implies that the $U(1)$ level in this case is $4k$. The fact that the spectrum derived in this appendix agrees with the one derived in appendix \ref{ap:case2} is to be expected, as both of the TsT transformations \eqref{tstrecipeap1} and \eqref{tstrecipeap2} associate the $J$ and $\bar J$ currents in the $T\bar T + J\bar T + T\bar J$ deformation with two different isometries of the background spacetime. This contrasts the TsT transformation used in the main body of the paper \eqref{tstrecipe2} where both the $J$ and $\bar J$ currents are associated with the same isometry of $T^4$.


\section{Gravitational charges} \label{ap:charges}

The charged TsT black hole described by \eqref{10dbackground} and \eqref{tstbtz} is a solution to the equations of motion of (the bosonic sector of) ten-dimensional type IIB supergravity. We have written the black hole in terms of three-dimensional variables, as the computation of its gravitational charges and thermodynamics simplifies significantly in lower dimensions. In three dimensions, the TsT black hole is characterized by the metric, Kalb-Ramond field, gauge fields $A{(1)}$ and $A{(2)}$, as well as the dilatons $\Phi$ and $\omega$ given in \eqref{tstbtz}. These fields are solutions to the equations of motion of the dimensionally reduced action which, in the Einstein frame, can be written as \eqsp{
  \!\! I_{3} &= \frac{\ell^3}{4\pi \ell_s^4} \int d^3x\sqrt{|g|} \bigg\{ R  + 4 \ell^{-2} e^{4\Phi+\omega} - 4 \p_\mu \Phi \p^{\mu} \Phi - \frac{1}{2} \p_\mu \omega \p^{\mu} \omega - 2 \p_\mu \Phi \p^\mu \omega \\
   &\hspace{60pt} - \frac{\ell_4^2 }{4} e^{-4\Phi-2\omega} F^{(1)}_{\mu\nu} F^{(1)\mu\nu} - \frac{\ell_4^2 }{4} e^{-4\Phi} F^{(2)}_{\mu\nu} F^{(2)\mu\nu} - \frac{e^{-8\Phi-2\omega}}{12} H_{\mu\nu\alpha} H^{\mu\nu\alpha} \bigg\}, \label{sugraaction3d}
    }
where the field strengths $F^{(1)}$, $F^{(2)}$, and $\til H$ are respectively given by
  \eq{
  F^{(1)} = d A^{(1)}, \qquad F^{(2)} = d A^{(2)}, \qquad H = d B - \ell_4^2 d A^{(2)}\! \wedge A^{(1)}. \label{fieldstrengths}
  }
In particular, we note that $H$ contains a Chern-Simons term, namely $d A^{(2)}\! \we A^{(1)}$, and that the action \eqref{sugraaction3d} is invariant under the following gauge transformations 
\eq{
\delta  A^{(2)} = d   \lambda^{(2)}, \qquad \delta B = d \Lambda, \qquad (\delta A^{(1)}, \delta B) = (d\lambda^{(1)}, \ell_4^2 A^{(2)}\! \wedge d\lambda^{(1)} ).
}

In the covariant formulation of gravitational charges \cite{Wald:1993nt,Iyer:1994ys, Barnich:2001jy}, the infinitesimal charge associated with a symmetry generated by the Killing vector $\xi$ is given by
  \eqsp{
    \d {\cal Q}_{\xi} &\equiv \a_d \int_{\p\ss}  \bm \chi_\xi,
  }
where $\bm \chi_\xi = \frac{1}{(d-2)!} \chi_\xi^{\mu\nu} \epsilon_{\mu \nu \a_1 \dots \a_{d-2}} dx^{\a_1} \we \dots \we dx^{\a_{d-2}}$ is a $(d-2)$-form determined from the $d$-dimensional action, $\a_d$ is the overall constant in front of the action, and $\p\ss$ is the boundary of a codimension-1 spacelike surface. For the three-dimensional action \eqref{sugraaction3d} we have $\a_3 = \ell^3/4\pi\ell_s^4$ while the one-form $\bm\chi_\xi$ is defined by
  \eq{
 \bm \chi_{\xi} =  \bm k_{\xi,g}  + \bm k_{\xi,B} + \bm k_{\xi,A^{(1)}} + \bm k_{\xi,A^{(2)}}  + \bm k_{\xi,\Phi} + \bm k_{\xi, \om} ,
  }
where $\bm  k_{\xi, f}$ denote the contributions of the background fields to the gravitational charge. The components $k_{\xi, f}^{\mu\nu}$ of $\bm  k_{\xi, f}$ can be succinctly written as\footnote{Note that in these expressions we have omitted terms that vanish when $\xi$ is an exact Killing vector but which otherwise contribute in the analysis of asymptotic symmetries.}
\eq{
 k^{\mu\nu}_{\xi, g} & =  |g|^{-1/2} \d (|g|^{1/2}\nabla^{\nu} \xi^{\mu})+ \xi^\nu \nabla^{\mu} \d g^\a{}_\a - \xi^\nu \nabla_\a \d g^{\a \mu}, \\ 
 \begin{split}
k^{\mu\nu}_{\xi, B} & =  \frac{1}{2} |g|^{-1/2} \d \big[ |g|^{1/2}e^{-8\Phi -2\om} H^{\nu\mu\b} \xi^{\a} \big( B_{\a\b} - \ell_4^2 A_\a^{(2)} A_\b^{(1)}\big)\big] \\
& \hspace{12pt} +  \frac{1}{2} e^{-8\Phi -2\om} \xi^\nu H^{\mu \a \b} \big( \d B_{\a \b} - 2 \ell_4^2 \d A_\a^{(2)} A_\b^{(1)} \big), 
\end{split} \\
k^{\mu\nu}_{\xi, A^{(1)}} &= \frac{\ell_4^2}{2} |g|^{-1/2} \d \big( |g|^{1/2} e^{-4\Phi -2\om} F^{(1)\nu \mu} A_\a^{(1)} \xi^\a  \big) + \ell_4^2e^{-4\Phi -2\om} \xi^\nu F^{(1)\mu \a} \d A_\a^{(1)} ,  \\
k^{\mu\nu}_{\xi, A^{(2)}} &=  \frac{\ell_4^2}{2}   |g|^{-1/2} \d \big( |g|^{1/2}  e^{-4\Phi}  F^{(2)\nu \mu}  A_\a^{(2)} \xi^\a  \big) + \ell_4^2 e^{-4\Phi} \xi^\nu F^{(2)\mu \a} \d A_\a^{(2)} , \\
k^{\mu\nu}_{\xi, \Phi} &= 2 \xi^\nu \nabla^\mu (4\Phi + \om) \d \Phi, \\
k^{\mu\nu}_{\xi, \om} &= \xi^\nu \nabla^\mu (2\Phi + \om) \d \om, 
}
where the variation $\d$ of any background field $f_{\mu_1 \dots \mu_n}$ is defined as $\d f_{\mu_1 \dots \mu_n} \equiv (\d T_u \p_{T_u} +  \d T_{v} \p_{T_{v}} +  \d \a_u \p_{\a_u} +  \d \a_{v} \p_{\a_{v}}) f_{\mu_1 \dots \mu_n} $ and $\d f^{\mu_1 \dots \mu_m}{}_{\mu_{m+1}\dots \mu_n} \equiv g^{\mu_1 \nu_1} \dots g^{\mu_m \nu_m} \d f_{\nu_1 \dots \nu_m \mu_{m+1} \dots \mu_n}$.


\section{The charged BTZ black hole} \label{ap:btzU1}

In this appendix we describe in more detail the conserved charges and the thermodynamics of the charged BTZ black hole \eqref{btzU1}. In particular, we point out that the covariant formalism is automatically compatible with holographic renormalization in the three-dimensional theory and does not require the addition of counterterms.

We first note that the presence of the Chern-Simons term in the definition of the three-form $H$ in \eqref{fieldstrengths} suggests that the gauge fields $A^{(1)}$ and $A^{(2)}$ are dual to linear combinations of chiral and antichiral currents in the dual CFT. This can be made manifest by noting that $\sqrt{|g|} e^{-8\Phi - 2\om} H$ is required to be a constant such that the on-shell variation of the action \eqref{sugraaction3d} satisfies $\d I_3 = -(p k_4/2\pi) \int_{\p \M} \epsilon^{\mu\nu} A_\mu^{(1)} \delta A_{\nu}^{(2)} + \dots$ where $\p \M$ denotes the asymptotic boundary and we have ignored the variation of the other fields.\footnote{The fact that $\sqrt{|g|} e^{-8\Phi - 2\om} H$ is a constant follows from the fact that this combination is proportional to the electric charge in higher dimensions.} With this in mind, the energies and $U(1)$ charges of the charged BTZ black hole can be written as
\eq{
\ell \Q_{u}  &= \frac{c}{6 } T_u^2 + \frac{Q_L^2}{p k_4 }  , \qquad \qquad\,\, \ell \Q_{v}  =  \frac{c}{6 }  T_{v}^2   + \frac{Q_R^2}{p k_4  } , \label{appELER} \\
 Q_L & = \frac{p k_4}{2} \big(A^{(1)} - A^{(2)} \big), \qquad  Q_R = -\frac{p k_4}{2} \big(A^{(1)} + A^{(2)} \big), \label{appqLqR}
}
where $ p k_4$ is the total level of two affine $U(1)$ symmetry algebras whose zero mode charges are given by $Q_L$ and $Q_R$. In particular, we note that \eqref{appELER} is the Sugawara-like combination of charges we expect to find in a theory with additional $U(1)$ symmetries. Furthermore, the $U(1)$ charges \eqref{appqLqR} are compatible with the holographic dictionary of the AdS/CFT correspondence associated with the presence of the aforementioned Chern-Simons term. 

In terms of the conserved charges \eqref{appELER} and \eqref{appqLqR}, the entropy of the charged BTZ black hole \eqref{btzU1} is given by
\eq{
S_{BH} = \frac{\pi c}{3}( T_u + T_{v}) =  2\pi \bigg[ \sqrt{\frac{c}{6}\bigg(\ell\Q_{u}  - \frac{Q_L^2}{p k_4}\bigg)} + \sqrt{\frac{c}{6}\bigg(\ell \Q_{v}  - \frac{Q_R^2}{p k_4}\bigg)}\, \bigg] = S_{Cardy}. \label{microbtzU1}
}
The entropy \eqref{microbtzU1} matches the Cardy formula for the density of high energy states in two-dimensional CFTs with total central charge $c = 6pk$ and a pair of left and right-moving $U(1)$ currents with total level $p k_4$. This provides evidence that the charged BTZ black hole \eqref{btzU1} is dual to a charged thermal state in the symmetric orbifold CFT described in section \ref{se:innerlayer}. Furthermore, \eqref{microbtzU1} suggests that this is the appropriate background on which to perform the sequence of TsT transformations \eqref{tstrecipe2}.

It is important to note that a boundary counterterm is necessary to make the two linear combinations of the gauge fields $\tfrac{1}{2}\big(A^{(1)} + A^{(2)} \big)$ and $\tfrac{1}{2}\big( A^{(1)} - A^{(2)} \big)$ dual to two chiral currents in the dual CFT \cite{Kraus:2006wn}. This boundary counterterm leads to a modification of the Brown-York stress tensor in the bulk, and consequently, to a modification of the charges obtained from integrals of the stress tensor. More generally, in holographic renormalization it is crucial to take into account all of the contributions of the counterterms that are compatible with the desired boundary conditions. The covariant formalism automatically takes these contributions into account, which are essential to obtain the right expressions for the energy and match the entropy to Cardy's formula in the microcanonical ensemble. 

Finally, we note that the $\a_u\a_{v} d{v}\we du$ term in the $B$-field of the charged black hole \eqref{btzU1} does not affect any of its charges as long as the dilaton is a fixed constant, i.e.~as long as the dilaton is not a phase space variable. It is not clear how to justify the presence of this term from a purely three-dimensional perspective although it is necessary to satisfy the boundary condition \eqref{bfieldbc} on the higher-dimensional fields. Note that the supergravity equations of motion require the dilaton to be a constant that can in principle vary in the phase space of solutions. Although the $\a_u \a_{v} d{v}\we du$ term does not affect the gravitational charges, it does guarantee that the latter are integrable if the dilaton is allowed to vary. A BTZ black hole with a variable dilaton is not physically desirable since, for example, it renders the electric charge counting the number of NS1 branes supporting the background a non-integer. Nevertheless, the charged tri-TsT black holes \eqref{tstbtz} do require constant shifts of the dilaton that depend on the phase space variables. In this case, we find that the $\a_u \a_{v} d{v}\we du$ term in the undeformed background \eqref{btzU1} is essential to render the gravitational charges of the deformed background \eqref{tstbtz} integrable. 



\ifprstyle
	\bibliographystyle{apsrev4-1}
\else
	\bibliographystyle{JHEP}
\fi

\bibliography{jtbar}



\end{document}
